\documentclass[a4paper,11pt]{article}  
\pdfoutput=1

\usepackage{graphicx} 
\usepackage{bbold}
\usepackage{mathtools}
\usepackage{amsmath}
\numberwithin{equation}{section} 
\usepackage{amsfonts}
\usepackage{amssymb}
\usepackage{slashed} 
\usepackage{color}
\usepackage{xcolor}
\usepackage{tabularx}
\usepackage{hyperref}
\usepackage{bm}
\usepackage{cite}  
\usepackage{mathrsfs}
\usepackage{framed}

\def\eq#1{{Eq.~(\ref{#1})}}
\def\eqs#1#2{{Eqs.~(\ref{#1})--(\ref{#2})}}

\def\Tr{\mbox{Tr}\,}

\def\di{\mbox{d}}

\colorlet{grayline}{gray!70}
\definecolor{blueline}{rgb}{0,0.27,0.55}
\definecolor{DarkGray}{gray}{0.4}
\definecolor{Gray}{gray}{0.6}
\definecolor{oucrimsonred}{rgb}{0.6, 0.0, 0.0}
\definecolor{persianblue}{rgb}{0.11, 0.22, 0.73}
\definecolor{forestgreen}{rgb}{0.13,0.35,0.13}
 \hypersetup{colorlinks, citecolor=forestgreen, linkcolor=forestgreen, urlcolor=forestgreen}
\def\hhref#1{\href{http://arxiv.org/abs/#1}{#1}} 
\newcommand{\be}{\begin{equation}}
\newcommand{\ee}{\end{equation}}
\newcommand{\bea}{\begin{eqnarray}}
\newcommand{\eea}{\end{eqnarray}}
\newcommand{\nn}{\nonumber}

\newcommand*\xbar[1]{%
  \hbox{\;%
    \vbox{%
      \hrule height 0.5pt 
      \kern0.5ex
      \hbox{%
        \kern-0.25em
        \ensuremath{#1}%
        \kern-0.07em
      }%
    }%
  }%
} 
\newcommand{\com}[1]{}
\newcommand{\gsim}{\lower.7ex\hbox{$\;\stackrel{\textstyle>}{\sim}\;$}}
\newcommand{\lsim}{\lower.7ex\hbox{$\;\stackrel{\textstyle<}{\sim}\;$}} 

\newcommand{\bc}{\begin{center}}
\newcommand{\ec}{\end{center}}

\newcommand{\hk}{{\bf \hat{k}}}
\newcommand{\hr}{{\bf \hat{r}}}
\newcommand{\hn}{{\bf \hat{n}}}
\newcommand{\hp}{{\bf \hat{p}}}
\newcommand{\lambdap}{\lambda^{\prime}}
\newcommand{\lambdaA}{\lambda_{1}}
\newcommand{\lambdaB}{\lambda_{2}}
\newcommand{\lambdaAp}{\lambda_1^{\prime}}
\newcommand{\lambdaBp}{\lambda_2^{\prime}}
\newcommand{\mup}{\mu^{\prime}}
\newcommand{\nup}{\nu^{\prime}}
\newcommand{\ct}{\cos{\Theta}}
\newcommand{\st}{\sin{\Theta}}
\newcommand{\cmb}{\mathscr{C}_2}

\newcommand{\mVV}{m_{\scriptscriptstyle{V\!V}}}
\newcommand{\mWW}{m_{\scriptscriptstyle{W\!W}}}
\newcommand{\mZZ}{m_{\scriptscriptstyle{Z\!Z}}}

\newcommand{\sW}{\sin \theta_W}

\newcommand{\ssW}{s_{\W}}
\newcommand{\ccW}{c_{\W}}

\newcommand{\U}{\scriptscriptstyle{U}}

\newcommand{\Ct}{c_{\Theta}}
\newcommand{\St}{s_{\Theta}}
\newcommand{\gAb}{\bar{g}_A^{u}}
\newcommand{\gAbD}{\bar{g}_A^{d}}
\newcommand{\gAbb}{\bar{g}_A^{u 2}}
\newcommand{\gVb}{\bar{g}_V^{u}}
\newcommand{\gVbD}{\bar{g}_V^{d}}
\newcommand{\gVbb}{\bar{g}_V^{u 2}}

\newcommand{\gVV}{g_V^{q 2}}
\newcommand{\gAA}{g_A^{q 2}}
\newcommand{\gVVVV}{g_V^{q 4}}
\newcommand{\gAAAA}{g_A^{q 4}}
\newcommand{\htuu}{\tilde{h}^{u\bar u}}
\newcommand{\htud}{\tilde{h}^{u \bar d}}
\newcommand{\htdd}{\tilde{h}^{d\bar d}}

\newcommand{\ftuu}{\tilde{f}^{u\bar u}}
\newcommand{\gtuu}{\tilde{g}^{u\bar u}}
\newcommand{\ftdd}{\tilde{f}^{d\bar d}}
\newcommand{\gtdd}{\tilde{g}^{d\bar d}}

\newcommand{\gtud}{\tilde{g}^{u\bar d}}
\newcommand{\ftud}{\tilde{f}^{u\bar d}}
\newcommand{\Aud}{A^{u\bar d}}

\newcommand{\Aee}{A^{\ell\bar \ell}}
\newcommand{\htee}{\tilde{h}^{\ell\bar \ell}}
\newcommand{\ftee}{\tilde{f}^{\ell\bar \ell}}
\newcommand{\gtee}{\tilde{g}^{\ell\bar \ell}}

\newcommand{\fWW}{f_{\scriptscriptstyle{W\!W}}}
\newcommand{\fZZ}{f_{\scriptscriptstyle{Z\!Z}}}
\newcommand{\fWZ}{f_{\scriptscriptstyle{W\!Z}}}
\newcommand{\DW}{{\rm D}_{\W\!\W}}
\newcommand{\DZ}{{\rm D}_{\Z\!\Z}}
\newcommand{\DV}{{\rm D}_{\W\!\Z}}
\newcommand{\W}{\scriptscriptstyle{W}}
\newcommand{\Z}{\scriptscriptstyle{Z}}
\newcommand{\V}{\scriptscriptstyle{V}}
\newcommand{\betaW}{\beta_{\W}}
\newcommand{\betaZ}{\beta_{\Z}}
\newcommand{\betaV}{\beta_{\V}}

\newcommand{\Aqq}{A^{q\bar q}}
\newcommand{\htqq}{\tilde{h}^{q\bar q}}
\newcommand{\ftqq}{\tilde{f}^{q\bar q}}
\newcommand{\gtqq}{\tilde{g}^{q\bar q}}
\newcommand{\gV}{g_V^{q}}
\newcommand{\gA}{g_A^{q}}

\begin{document}
\thispagestyle{empty}
\begin{center}
{\color{oucrimsonred}\Large {\bf 
   Bell inequalities and quantum entanglement \\[0.3cm] 
    in weak gauge boson production at the LHC and future colliders
}}

\vspace*{1.5cm}
{\color{DarkGray}
  {\bf M. Fabbrichesi$^{a}$,}
{\bf   R. Floreanini$^{a}$,}
{\bf E. Gabrielli$^{{b,a,c,d}}$ and} 
 {\bf L. Marzola$^{{d}}$}
}\\

\vspace{0.5cm}
{\small 
{\it  \color{DarkGray} (a)
INFN, Sezione di Trieste, Via Valerio 2, I-34127 Trieste, Italy}
\\[1mm]
  {\it \color{DarkGray}
    (b) Physics Department, University of Trieste, Strada Costiera 11, \\ I-34151 Trieste, Italy}
  \\[1mm]  
  {\it \color{DarkGray} 
   (c) CERN, Theoretical Physics Department, Geneva, Switzerland}
    \\[1mm]
  {\it \color{DarkGray}
(d) Laboratory of High-Energy and Computational Physics, NICPB, R\"avala 10, \\ 10143 Tallinn, Estonia}
}
\ec

 \vskip0.5cm
\bc
{\color{DarkGray}
\rule{0.7\textwidth}{0.5pt}}
\ec
\vskip1cm
\bc
{\bf ABSTRACT}
\ec

\vspace*{5mm}

\noindent
Quantum entanglement of weak interaction gauge bosons produced at colliders can be explored by computing the corresponding polarization density matrix. To this end, we  consider the Higgs boson decays $H\to W  W^*$ and $H\to Z  Z^*$, in which  $W^*$ and $Z^*$ are off-shell states, and the $WW$, $WZ$ and $ZZ$ di-boson production in proton collisions. The polarization density matrix of the di-boson state is determined by the amplitude of the production process and can be experimentally reconstructed from the angular distribution of the momenta of the   final states into which the gauge bosons decay. We show that a suitable instance of the Bell inequality is violated in $H\to Z  Z^*$ to a degree that can be tested  at the LHC with future data. The same Bell inequality is violated in the production of $WW$ and $ZZ$ boson pairs for invariant masses above 900 GeV and scattering angles close to $\pi/2$ in the center of mass frame.  LHC data  in this case  are not sufficient  to establish the violation of the Bell inequality. We also analyze the prospects for detecting Bell inequality violations in di-boson final states at future $e^+e^-$ and muon colliders. A further observable that provides a lower bound on the amount of polarization entanglement in the di-boson system is computed for each of  the examined processes. 
The analytic expressions for the polarization density matrices  are  presented in full in an Appendix. We also provide the unitary matrices required in the optimization procedure necessary in testing the Bell inequalities. 

  \vskip 3cm

\bc 
{\color{DarkGray} \vbox{$\bowtie$}}
\ec


\section{Introduction\label{sec:intro}}

The most natural way to generate entanglement~\cite{Horodecki:2009zz} between two quantum systems is through their mutual interaction; 
any interaction dynamics involving the degrees of freedom of both systems is bound to create quantum correlations  and yield  detectable effects measurable through suitable quantum observables.

An instance of such an interaction   is high-energy collisions: they  give rise to quantum entanglement among the elementary particles partaking in a scattering process---thereby  providing the possibility to study and test entanglement in a novel setting. This opportunity   has indeed recently drawn some interest and has been explored in a series of papers~\cite{Caban:2008qa,Afik:2020onf,Fabbrichesi:2021npl,Barr:2021zcp,Gong:2021bcp,Severi:2021cnj,Larkoski:2022lmv,Afik:2022kwm,Barr:2022wyq,Aguilar-Saavedra:2022uye,Aguilar-Saavedra:2022wam,Aoude:2022imd,Fabbrichesi:2022ovb,Severi:2022qjy,Ashby-Pickering:2022umy,Altakach:2022ywa}.
Most of these analyses have focused on distinguishing quantum mechanics from alternative local and deterministic theories through the exploration of  Bell inequalities~\cite{Bell1,Bell2,Redhead,Bertlmann,Brunner} in the energy range probed at the Large Hadron Collider (LHC). We continue these studies by analyzing  possible quantum correlations in the polarization states of weak interaction gauge bosons produced at colliders.  

Massive gauge bosons act as their own polarimeters and their spin polarizations can be reconstructed from the angular distribution of the final leptons or jets (when generated by down-type quarks).  We assume that the polarization density matrix of the two bosons can be fully reconstructed (albeit with  limited efficiency) from the angular distributions of their decays into final states. The actual uncertainty affecting such a reconstruction---as well as the effect of backgrounds, unfolding and of the detector---can only be estimated through dedicated numerical simulations.

In the following, we  analyze the production of gauge bosons via the resonant Higgs boson decays $H\to W  W^*$ and $H\to Z  Z^*$, where $W^*$ and $Z^*$  denote off-shell states, and study the $WW$, $WZ$ and $ZZ$ production proceeding from proton-proton collisions in a manner reminiscent of the Drell-Yan mechanism. The density matrix of the di-boson state is determined analytically for each process from the corresponding amplitudes and depends on the kinematic variables characterizing the scattering process. Once the density matrix is known, it is possible to test
the (Collins-Gisin-Linden-Massar-Popescu) CGLMP inequality~\cite{CGLMP,CGLMP2}---a Bell inequality optimized for three-level systems as those describing the polarizations of massive spin-1 particles--- on the available kinematic configurations. We also compute an observable used as proxy for the entanglement in processes characterized by massive spin-1 final states. The corresponding operator yields a lower bound for the entanglement of the two boson system in each of the analyzed cases, thereby serving as a witness of the entanglement in their polarizations. 

We find that the CGLMP inequality is violated in di-boson Higgs decays and that such a violation in the case $H\to Z  Z^*$ case could be tested  at the LHC  with future data. For the $WW$, $WZ$ and $ZZ$ production in proton collisions, instead, the CGLMP inequality is violated only in the $WW$ and $ZZ$ channels for invariant masses above 900 GeV and scattering angles close to $\pi/2$ in the center of mass frame. Due to the small number of events expected in this kinematic region, it is difficult  to assert the CGLMP inequality violation with sufficient accuracy. A better significance will be achieved at future lepton colliders, as we also show in the following.

The theoretical uncertainty of our results is at most of order 10\%  in the continuum $WW$, $WZ$ and $ZZ$ Drell-Yan processes, due to the implied QCD next-to-leading order (NLO) contributions~\cite{Denner:2020bcz}. We checked that uncertainties in the parton distribution functions (PDF) are negligible. Smaller theoretical uncertainties of order of a few percent are expected for the $WW^*$ and $ZZ^*$ processes in the resonant Higgs decay region, due to NLO EW corrections. \cite{Boselli:2015aha}

We estimate the overall uncertainty in the event selection by considering the efficiency in the identification of the lepton final states and their momenta reconstruction---which conservatively we take to be 70\% for each  lepton---and a  distribution of the observables due to uncertainty in the off-shell gauge boson mass.  When $W$-bosons are in the final states, we add a systematic error to take into account the intrinsic difficulty of the physical analysis that requires dedicated algorithms to reconstruct the neutrino momenta. Though naive, our estimates still indicate the most promising processes for this kind of studies at the considered collider machines. It is our hope that this preliminary study encourages the experimental collaborations to assess the power of current and future machine to probe quantum entanglement through full simulations. Since our results are analytic and the related uncertainties, as mentioned, only include a guess on the efficiency of the reconstruction, these numerical simulations are paramount for a realistic estimate of the uncertainty. Numerical simulations were performed at the parton level in~\cite{Barr:2021zcp} and~\cite{Aguilar-Saavedra:2022wam}, for the Higgs boson decays, and in~\cite{Ashby-Pickering:2022umy} for the di-boson production from quarks. We compare these numerical results with ours when discussing our findings.

The polarization of weak gauge bosons has been studied in the literature in terms of helicity amplitudes~\cite{helicity}. Our approach differs in that we derive the full density matrix of the di-boson system as required for studying the  presence of entanglement.

The paper is organized as follows. Sec.~\ref{sec:met} introduces the  entanglement witness we utilize, the Bell and CGLMP inequalities and shows how the polarization density matrix of interest can be built from the amplitude of the underlying scattering process. In Sec.~\ref{sec:hvv} we apply our methodology to two massive gauge bosons originated in a Higgs boson decay. The entanglement of a di-boson system created at the LHC or at future colliders is  investigated in Sec.~\ref{sec:DY}. We conclude in Sec.~\ref{sec:sum} by briefly summarizing our findings. The correlation coefficients of all the considered density matrices are listed in~\ref{sec:a2}. These expressions can be useful for future studies and, to the best of our knowledge, have not been reported in the literature before.

\section{Methods and tools\label{sec:met}}

In this section we introduce the observables used in the study of entanglement and violation of Bell inequalities. We also briefly review the calculation of  the polarization density matrix for a system formed by one or two massive gauge bosons, as well as the procedure to reconstruct it from the momenta of the leptons emitted in their decay process.

\subsection{Observables\label{sec:obs}}

We start by discussing quantum correlations in the context of three-level systems---in short, \textit{qutrits}---implemented, for instance, by the possible polarizations of massive gauge bosons. We use a specific Bell inequality optimized for this system and a suitable measure of entanglement to probe the correlations encoded in the polarization density matrix of the two qutrits.

\subsubsection{Entanglement}

Quantifying the entanglement content of the state of a quantum system is generally challenging as the complexity of the problem increases with the system dimensionality~\cite{Horodecki:2009zz}. For pure states--- systems described by a vector in the Hilbert space, or equivalently, by a density matrices that is a projector--- the problem can be addressed by considering their Schmidt decomposition~\cite{Horodecki:2009zz}, but already for the simple case of bipartite systems only partial answers are available. No general rule is applicable to mixed states, in which case one can only rely on so-called {\it entanglement witnesses}: quantities that give conditions sufficient to establish the presence of entanglement in the system. A computable example of such a witness is connected to {\it concurrence},
a reliable entanglement measure for bipartite two-level systems---that is, consisting of two qubits~\cite{Bennett:1996gf}.

Consider a bipartite quantum system comprising two subsystems of equal dimensionality, $A$ and $B$, described by a normalized pure state $|\Psi\rangle$ and density matrix $|\Psi\rangle\langle\Psi|$. The concurrence of the system is then defined as~\cite{Rungta}
\be
{\cal C}[|\Psi\rangle]=\sqrt{2\left( 1-{\rm Tr}\big[(\rho_r)^2\big]\right)}\ ,\qquad r=A\ {\rm or}\ B\ ,
\label{C_psi}
\ee
where $\rho_r$ is the reduced density matrix obtained by tracing over the degrees of freedom of either subsystem: \textit{e.g.} for $r=A$ one has $\rho_A={\rm Tr}_B\big[|\Psi\rangle\langle\Psi|\big]$. Any mixed state $\rho$ of the bipartite system can be decomposed into a set of pure states $\{|\Psi_i\rangle\}$,
\be
\rho=\sum_i p_i\, |\Psi_i\rangle\langle\Psi_i|\ ,\qquad p_i\geq0\ ,\qquad  \sum_i p_i=1\;
\ee
its concurrence is then defined by means of the concurrence of the pure states appearing in the decomposition through an optimization process:
\be
{\cal C}[\rho]=\underset{\{|\Psi\rangle\}}{\rm inf} \sum_i p_i\, {\cal C}[|\Psi_i\rangle]\ ,
\label{C_rho}
\ee
where the infimum is taken over all the possible decompositions of $\rho$ into pure states. Clearly, for a pure state (\ref{C_psi}) the concurrence vanishes if and only if the state is separable, that is: $|\Psi\rangle=|\Psi_A\rangle\otimes |\Psi_B\rangle$. As the same holds for mixed states~\cite{Buchleitner}, the concurrence appears to be a good entanglement detector. Unfortunately, the optimization problem appearing in (\ref{C_rho}) makes the evaluation of the concurrence a very hard mathematical task with a simple analytic solution only when $A$ and $B$ are two-level systems. Any approximation or numerical computation of ${\cal C}[\rho]$ only holds as an upper bound and thus cannot serve to reliably distinguish between entangled and separable states, or to give an estimate of a state entanglement content.

Lower bounds on ${\cal C}[\rho]$ for a generic density matrix $\rho$ can  be analytically computed and, if non-vanishing, unequivocally signal the presence of entanglement. One of these bounds is easily computable,  yielding~\cite{Mintert}
\be
\big({\cal C}[\rho]\big)^{2} \geq \cmb[\rho]\ ,
\ee
where
\be
{\color{Gray}
\boxed{\color{ black}
\cmb [\rho] = 2 \,\text{max}\, \Big( 0,\, \Tr[\rho^{2}] - \Tr[(\rho_A)^2],\, \Tr[\rho^{2}] - \Tr[(\rho_B)^2]  \Big) \, ,
\label{C_bound}
}}
\ee
with $\rho_A={\rm Tr}_B[\rho]$ and $\rho_B={\rm Tr}_A[\rho]$ being the reduced density matrices. A non-vanishing value of $\cmb$ then implies a concurrence larger than zero, thus witnessing the entanglement of the density matrix $\rho$. 

Interestingly enough, an upper bound for ${\cal C}[\rho]$ has also been obtained~\cite{Zhang}; explicitly, one finds
\be
\big({\cal C}[\rho]\big)^{2} \leq 2 \,\text{min}\, \Big(1 - \Tr[(\rho_A)^2],\  1 - \Tr[(\rho_B)^2] \Big)\, .
\ee
The maximum value for the concurrence is obtained for a totally symmetric and maximally entangled pure state. For two qutrits this is 
\be
|\Psi_+\rangle=\frac{1}{\sqrt{3}} \sum_{i=1}^3 |i\rangle\otimes|i\rangle\ ,
\label{max-ent}
\ee
with $\{|i\rangle\}$ an orthonormal basis in the $A$- or $B$-Hilbert space,
resulting in ${\cal C}[|\Psi_+\rangle]=2/\sqrt{3}$. Accordingly, $ \cmb$ is at most equal to 4/3.

The concurrence lower bound (\ref{C_bound}) will play the role of entanglement witness in our study of the spin polarization states formed with two massive gauge bosons.

If the bipartite state of interest is a pure state, it is possible to quantify its entanglement by computing the \emph{entropy of entanglement}: 
\be
{\color{Gray}
\boxed{\color{ black}
\mathscr{E}[\rho] = - \Tr[\rho_A \log \rho_A] =  - \Tr[\rho_B \log \rho_B] \, , \label{E}
}}
\ee
given by the von Neumann entropy~\cite{Horodecki:2009zz} of either of the two component subsystems $A$ or $B$ with reduced density matrix $\rho_A$ and $\rho_B$, respectively.  Whereas the concurrence of a bipartite pure state is only an entanglement monotone, the von Neumann entropy is a true entanglement measure satisfying $0\leq \mathscr{E}[\rho] \leq \ln d$, where $d=3$ for a two-qutrit system. The first equality holds if and only if the bipartite state is separable, the second inequality saturates if the bipartite state is maximally entangled.

\subsubsection{Bell inequalities}

Local deterministic theories provide descriptions of a physical system that match the results of quantum mechanics for the averages of relevant system observables. Yet, in view of the deterministic and locality assumptions,
these stochastic classical models are bound to satisfy a set of inequalities  known as Bell inequalities~\cite{Bell1,Bell2,Redhead,Bertlmann,Brunner}, 
which are instead violated by the statistical predictions of quantum mechanics. An experimental determination of any Bell inequality is thus able to discriminate between these classical local models and quantum mechanics.

Whereas an essentially unique Bell inequality can be formulated~\cite{Clauser:1978ng} in the case of a bipartite system made of two qubits, different Bell inequalities can be found in the literature for systems of higher dimensionality. Among these, the CGLMP inequality~\cite{CGLMP,CGLMP2} is an optimal generalization of the qubit inequality for systems made of two qutrits.

In order to explicitly write this inequality, consider again the two components $A$ and $B$ of the two qutrit system. For the qutrit $A$, select two spin measurement settings, $\hat{A}_1$ and $\hat{A}_2$, which correspond to the projective measurement of two spin-1 observables having
each three possible outcomes $\{0,1,2\}$. Similarly, the measurement settings and corresponding observables for the other qutrit $B$ are $\hat{B}_1$ and $\hat{B}_2$. Then, denote by $P(A_{i}=B_{j}+k)$
the probability that the outcome $A_{i}$ for the measurement of $\hat{A}_i$ and $B_j$ for the measurement of $\hat{B}_j$, with $i$, $j$ either 1 or 2, differ by $k$ modulo 3. One can then construct the combination:
\bea
{\cal I}_{3}& =& P(A_{1}=B_{1}) +  P(B_{1}=A_{2}+1) +  P(A_{2}=B_{2}) +  P(B_{2}=A_{1}) \nn \\
& & -P(A_{1}=B_{1}-1) - P(A_{1}=B_{2}) -P(A_{2}=B_{2}-1) -P(B_{2}=A_{1}-1)\ . 
\label{CGLMP}
\eea

For deterministic local models, this quantity satisfies the following generalized Bell inequality,
\be
{\cal I}_{3}\leq 2\ ,
\label{CGLMP_inequality}
\ee
which instead can be violated by computing the above joint probabilities using the rules of quantum mechanics. Given a state $\rho$ of the two-qutrit system, the above probabilities are computed in quantum mechanics
as expectation values of suitable projector operators; for instance, the probability of the outcome
$A_1=B_1=1$, when measuring $\hat{A}_1$ and $\hat{B}_1$, is given by 
$P(A_1=B_1=1)={\rm Tr}[\rho\, ({\cal P}_{A_1=1}\otimes{\cal P}_{B_1=1})]$, where {\it e.g.} ${\cal P}_{A_1=1}$
projects onto the subspace of the $A$-Hilbert space where $\hat{A}_1$ assumes the value~1. Therefore, in quantum mechanics, ${\cal I}_{3}$ in (\ref{CGLMP}) can be similarly expressed as an expectation value
of a suitable Bell operator $\cal B$:
\be
{\color{Gray}
\boxed{\color{black}
{\cal I}_{3}={\rm Tr}\big[\rho\, {\cal B}\big].
}}
\ee
The explicit form of $\cal B$ depends on the choice of the four measured operators $\hat{A}_i$, $\hat{B}_i$, $i\in\{1,2$\}. Hence, given the two-qutrit state $\rho$, it is possible to enhance the violation of the Bell inequality (\ref{CGLMP_inequality}) through a specific choice of these operators. We remark that the numerical value of the observable is bound to be less than or equal to 4.

For the case of the maximally entangled state in (\ref{max-ent}), $\rho=|\Psi_+\rangle\langle\Psi_+|$, the problem of finding an optimal choice of measurements has been solved~\cite{CGLMP}. By working in the single spin-1 basis formed by the eigenstates of the $S_3$ spin operator~\eqref{eq:s3} 
with eigenvalues $\{1,0,-1\}$, the Bell operator takes the following explicit form (see~\cite{Acin:2002zz},
though there it is  written in the so-called computational basis):
\be
{\cal B} =  \begin{pmatrix} 
  0 & 0 & 0 & 0 & 0 & 0 & 0 & 0 & 0  \\
  0 & 0 & 0 & -\dfrac{2}{\sqrt{3}} & 0 & 0& 0 & 0 & 0  \\
  0 & 0 &0 & 0 & -\dfrac{2}{\sqrt{3}} & 0 &2 & 0 & 0  \\
  0 &  -\dfrac{2}{\sqrt{3}} & 0 & 0 & 0 & 0 & 0 &0 & 0  \\
  0& 0 & -\dfrac{2}{\sqrt{3}} & 0 & 0 & 0 & -\dfrac{2}{\sqrt{3}} & 0 &0  \\
  0 & 0 & 0 & 0 & 0 & 0 & 0 &  -\dfrac{2}{\sqrt{3}} & 0  \\
  0 & 0 & 2 & 0 & -\dfrac{2}{\sqrt{3}} & 0 &0& 0 & 0  \\
  0 & 0 & 0 & 0 & 0 &  -\dfrac{2}{\sqrt{3}} & 0 & 0 & 0  \\
  0 & 0 & 0 & 0 & 0& 0 & 0 & 0 & 0  \\
\end{pmatrix} \ .
\label{B}
\ee
It should be noticed that, perhaps surprisingly, the maximal violation of (\ref{CGLMP_inequality}) obtained with $\cal B$ is for a density matrix which is \textit{not} maximally entangled~\cite{Acin:2002zz}, making it evident that entanglement theory in higher dimensions is rather intricate.

Within the choice of measurements leading to the Bell operator in (\ref{B}), there is still the freedom of modifying the measured observables through local unitary transformations, which effectively corresponds to local changes of basis. Correspondingly, the Bell operator undergoes the change:
\be
{\cal B} \to (U\otimes V)^{\dag} \cdot {\cal B}\cdot (U\otimes V)\ , \label{uni_rot}
\ee
where $U$ and $V$ are independent three-dimensional unitary matrices.
In the following we make use of this freedom to maximize the value of ${\cal I}_3$ for any given density matrix $\rho$; as the gauge boson polarization states depend on the relevant kinematic variables, this optimization procedure is to be performed independently for each point in phase space. We give the explicit forms of the matrices that maximize the observable ${\cal I}_3$ for the processes analyzed as they can be useful in future numerical simulations.


\subsection{Density matrix for one spin-1 particle}
Let us start by defining the reference frame we use to describe the polarization of a spin-1 particle at rest. To this purpose we introduce a set of three orthonormal (three-)vectors,
$\left\{ \hn, \hr, \hk \right\}$, forming a right-handed system: $\hn=\hr\times \hk$. The normalized helicity eigenvectors $\psi_{\pm,0}$ of the massive spin-1 particle of mass $M$, corresponding respectively to eigenvalues $\lambda=\pm 1,0$, are
\bea
\psi_{\pm}=-\frac{1}{\sqrt{2}}\left(\pm {\hat {\bf n}}+i\, {\hat {\bf r}}\right)\quad \text{and} \quad \psi_{0}={\hat {\bf k}}\,,
\label{psi}
\eea
having chosen the ${\hat {\bf k}}$ direction as the direction of quantization.

In order to describe the helicity of the spin-1 particle in a more general reference frame and in a covariant manner, we first promote the three basis vectors to four-vectors by extending them with a null temporal component and then perform a Lorentz boost along the $-{\hat {\bf k}}$ direction. As a result, in the new frame the spin-1 particle acquires a velocity $\beta=\sqrt{1-M^2/E^2}$ along the positive ${\hat {\bf k}}$ direction and
possesses a 4-momentum $p^{\mu}=E(1,{\hat {\bf k}}\beta)$, where $E$ is the particle energy in this frame. By construction, the boosted basis vectors 
\bea
n^{\mu}_{1}=\left(0,\, \hn\right), ~ n^{\mu}_{2}=\left(0,\, \hr\right),~
n^{\mu}_3=\frac{E}{M}(\beta,\, \hk)\, ,
\label{ni}
\eea
are orthogonal to the four-vector $n^{\mu}_0= E/M(1,\, \hk\beta)$ (proportional to the particle momentum) and with it form an orthonormal \textit{vierbein} $n^{\mu}_m$. The label $m\in\{0,1,2,3\}$ indicates the vector: $g_{\mu \nu}\,n_m^{\mu} n_{n}^\nu  =-\delta_{mn}$ with $g_{\mu\nu}=\text{diag}(1,-1,-1,-1)$ being the Minkowski metric.

The wave vector $\varepsilon^{\mu}(p,\lambda)$
of a spin-1 particle can then be expressed in a covariant form as a linear combination of the three reference vectors $\left(n^{\mu}_1,n^{\mu}_2,n^{\mu}_3\right)$ orthogonal to the particle momentum
\bea
\varepsilon^{\mu}(p,\lambda)=-\frac{1}{\sqrt{2}}|\lambda|\left(\lambda \, n_1^{\mu}+i \, n_2^{\mu}\right)
+\Big(1-|\lambda| \Big)n_3^{\mu}\, ,
\label{eps}
\eea
giving the standard representation of the spin-1 wave vector in the helicity $\lambda$ basis. It can be easily checked that in the particle rest frame, where $(\beta\to 0)$, the equation above reduces to Eq.~(\ref{psi}).

From Eq.~(\ref{eps}) we can construct the covariant helicity projector operator of a spin-1 particle with four-momentum $p$, mass $M$ and polarization $\varepsilon_{\mu}(p,\lambda)$~\cite{Choi:1989yf}
\bea
\mathscr{P}^{\mu\nu}_{\lambda \lambdap}(p) &=&\varepsilon^{\mu}(p,\lambda)^{\star}\varepsilon^{\nu}(p,\lambdap) \nn\\&=&
\frac{1}{3}\left(-g^{\mu\nu}+\frac{p^{\mu}p^{\nu}}{M^2}\right)
\delta_{\lambda\lambdap}-\frac{i}{2M}
\epsilon^{\mu\nu\alpha\beta}p_{\alpha} n_{i\,\beta} \left(S_i\right)_{\lambda\lambdap}-\frac{1}{2}n_i^{\mu}n_j^{\nu} \left(S_{ij}\right)_{\lambda\lambdap}\, ,
\label{proj}
\eea
where $S_i$, $i\in\{1,2,3\}$, are the spin-1 representations\footnote{Explicit matrix representations are given in Appendix~\ref{sec:a1} on the basis where the eigenstates of $S_3$ read 
\begin{equation}
  |+\rangle = 
  \begin{pmatrix}
  1 \\ 0 \\ 0  
  \end{pmatrix},
  \quad
  |0\rangle = 
  \begin{pmatrix}
  0 \\ 1 \\ 0  
  \end{pmatrix},
  \quad
  |-\rangle = 
  \begin{pmatrix}
  0 \\ 0 \\ 1  
  \end{pmatrix}\,,
\end{equation}
corresponding to the eigenvalues $+1$, $0$ and $-1$, respectively.} of the $SU(2)$ generators and $\epsilon^{\mu\nu\alpha\beta}$ is the fully antisymmetric Levi-Civita tensor with $\varepsilon^{0123}=1$. The matrices $S_{ij}$ are defined as
\bea
S_{ij}= S_iS_j+S_jS_i-\frac{4}{3} \mathbb{1}\, \delta_{ij} \, ,
\label{Sij}
\eea
with $i,j\in\{1,2,3\}$ and $\mathbb{1}$ being the $3\times 3$ unit matrix. The covariant relation~(\ref{proj}) can be verified by substituting the expression for $\varepsilon^{\mu}(p,\lambda)$ in Eq.~(\ref{eps}) with $n_i^{\mu}$ given as in Eq.~(\ref{ni}) and the $(S_i)_{\lambda\lambda^{\prime}}$ and $(S_{ij})_{\lambda\lambda^{\prime}}$ matrix elements as provided in Appendix~\ref{sec:a2}. 
 
Consider now the probability amplitude ${\cal M}$ for the production of a massive spin-1 particle of momentum $p$ and helicity $\lambda$, given by
\bea
{\cal M}(\lambda)&=&{\cal M}_{\mu} \varepsilon^{\mu\star}(p,\lambda)\, .
\label{Mpol}
\eea
Then, the polarization density matrix of a massive spin-1 particle can be written in the helicity basis as 
\bea
\rho(\lambda,\lambdap)&=&\frac{{\cal M}(\lambda) {\cal M}^{\dag}(\lambdap)}{
|\xbar{{\cal M}}|^2}\, \label{Mrho}
\eea
where the $|\xbar{{\cal M}}|^2 =\sum_{\lambda}{\cal M}^{\dag}(\lambda){\cal M}(\lambda)$ is the unpolarized square amplitude (the sum in \eq{Mrho} over possible internal degrees of freedom of initial state particles is understood). By using the expression in Eq.~(\ref{Mpol}) we have that
\bea
\rho(\lambda,\lambdap)&= &\frac{{\cal M}_{\mu} {\cal M}^{\dag}_{\nu}\, \mathscr{P}^{\mu\nu}_{\lambda\lambdap}(p)}{|\xbar{{\cal M}}|^2}
 \label{rho1}
\eea
where the expression for the covariant projector is given in Eq.~(\ref{proj}).

The relation above provides a simple way to compute the polarization density matrix of one massive spin-1 particle starting from the amplitudes $\mathcal{M}$ of the related production process. As all 3$\times$3 matrices, $\rho$ can be decomposed on the basis formed by the eight Gell-Mann matrices $T^a$ (see Appendix~\ref{sec:a1}) and the unit matrix as follows
\bea
\rho(\lambda,\lambdap)=\left(\frac{1}{3}\,\mathbb{1}
+\sum_{a=1}^8 v^a T^a\right)_{\lambda\lambdap} \label{rhoi}
\eea
where the $T^a$ satisfy the orthogonality condition $\Tr[T^a T^b]= 2\, \delta^{ab}$.

The coefficients $v^a$, which depend on the kinematic variables of the process, are scalar quantities and can be easily obtained by projecting the $\rho$ matrix on the Gell-Mann basis:
  \bea
  v^a&=&\frac{1}{2}\Tr\left[\rho \, T^a\right]\, .
  \eea
Expressions for $S_i$ and $S_{ij}$, $i,j\in\{1,2,3\}$, in terms of the Gell-Mann matrices are given in Appendix~\ref{sec:a1}.


\subsection{Density matrix for two spin-1 particles\label{sec:rho}}

We first analyze the case of a pair of spin-1 particles with same mass $M$, and then generalize the kinematics to the case of two particles with different masses. 

Consider the production of a pair $V_1 V_2$ via the Drell-Yan topology initiated by quark-antiquark fusion
\bea
\bar{q}(p_{1})\, q(p_{2}) \to V_1(k_1,\lambda_1)\,  V_2(k_2,\lambda_2)
\eea
where $p_{i}$ are the momenta of initial state quarks and $k_{i}$ and $\lambda_{i}$ ($i\in\{1,2\}$) the four-momenta and helicities of $V_1$ and $V_2$, respectively. For processes initiated in proton-proton collisions we can assume massless quarks.

\begin{figure}[h!]
\begin{center}
\includegraphics[width=4.5in]{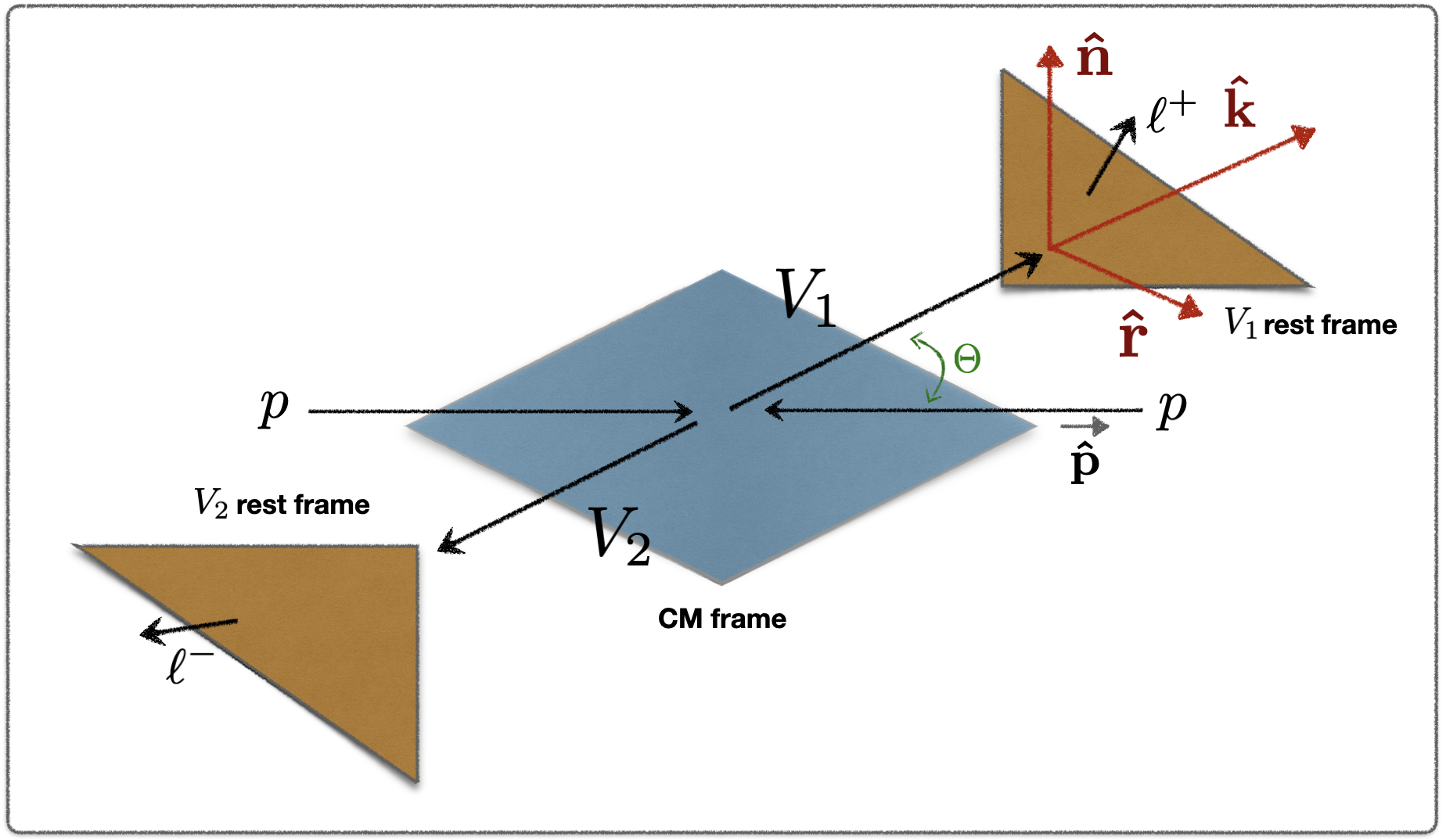}
\caption{\small Unit vectors and momenta in the CM system for the weak gauge bosons production $p\, p \to V_1 V_2$ as utilized in the text. Notice the definition of the scattering angle and the direction of the unit vector $\mathbf{\hat p}$.
\label{fig:coordinates} 
}
\end{center}
\end{figure}

Without the loss of generality we choose to work in the center of mass (CM)
frame, where the orientation of the basis unit vectors $\{\hn,\hr,\hk\}$ relative to the quark beam is illustrated in Fig.~\ref{fig:coordinates}. In this frame, the $\hk$ direction is taken to coincide with the axis defined by the three-momenta of the two spin-one particles produced, with ${\bf k}_1$ indicating the positive verse. Then, taking the 3-momentum of the antiquark along the $\hp$ direction, so that $p_1=E(1,\hp)$ and the scattering angle $\Theta$ matches the angle between the vectors $\hp$ and $\hn$. The remaining unit vectors composing the orthogonal $\{\hn,\hr,\hk\}$ system are then given by 
\bea
\hr&=&\frac{1}{\st}\left(\hp-\ct\hk\right)\, , ~~~~
\hn\,=\,\frac{1}{\st}\left(\hp\times\hk\right)\, .
\label{basis}
\eea

We define the spin eigenstates for each particle in its own rest frame as in Eq.~(\ref{psi}). The corresponding expressions for the CM frame are then obtained via a Lorentz boost by $-\beta$, for $V_1$, and $+ \beta$ for $V_2$, where $\beta=\sqrt{1-4M^2/s}$ and $s$ is the squared of the CM energy.  We report below the form taken in this frame by the initial and final states momenta 
\bea
p^{\mu}_1=E\,(1,\,\hp)\, ,~~p^{\mu}_2=E\,(1,\,-\hp)\,
,~~k^{\mu}_1=E\, (1,\,\beta\hk)\, ,~~k^{\mu}_2=E\,(1,\,-\beta\hk)\,.
\label{momenta}
\eea
Given the polarization vectors $\varepsilon^{\mu}(k_1,\lambdaA)$ and $\varepsilon^{\nu}(k_2,\lambdaB)$, associated respectively with $V_1$ and $V_2$, the corresponding polarization bases $n_i^{\mu}(1)$ and $n_i^{\mu}(2)$ ($i\in\{1,2,3\}$, \textit{cf.} \eqs{eps}{ni}) are given by
\bea
&&n^{\mu}_1(1)=n^{\mu}_1(2)=(0,\,\hn)\, ,~~n^{\mu}_2(1)=n^{\mu}_2(2)=(0,\,\hr)\, ,  \nn\\
&&n^{\mu}_3(1)=\gamma \,(\beta,\,\hk)\, , ~~~
n^{\mu}_3(2)=\gamma\, (-\beta,\,\hk)\, ,
\label{spinvect}
\eea
where $\gamma=1/\sqrt{1-\beta^2}$ is the Lorentz factor. The above vectors satisfy the normalization conditions
\bea
n_i^{\mu}(1) \,n_j(1)_{\mu}&=&
n_i^{\mu}(2)\, n_j(2)_{\mu}=-\delta_{ij}\, , \nonumber\\
n_3^{\mu}(1)\, n_{3\mu}(2)&=&-\gamma^2(\beta^2+1)\, , \nonumber\\
n_1^{\mu}(1)\, n_2(2)_{\mu}&=&
n_2^{\mu}(1)\, n_1(2)_{\mu}=0\, .
\label{normcond}
\eea

In case of production of two different gauge bosons, as well as for the decay of the Higgs boson into a pair of weak interaction gauge bosons (one necessarily off-shell\footnote{We model the off-shell particle as an on-shell gauge boson with a reduced mass for the purpose of computing the amplitude of the process.}), the above relations generalize as follows.

Let $k$ be the common magnitude of the momenta of the produced particles in the CM frame, $\sqrt{s}=E_1+E_2$ the total energy in the same frame, $M$ the heaviest mass of the two spin-1 particles and $f M$ the lightest one, with $0<f<1$. The four-vectors $k_{1}$ and $k_{2}$ are then given by 
\bea
k^{\mu}_1=(E_1,\, k\,\hk)\, ,~~k^{\mu}_2=(E_2,\, -k\,\hk)\, ,
\eea
where 
\bea
k&=&\frac{1}{2\sqrt s}\sqrt{s^2-2(1+f^2)s\,M^2+(1-f^2)^2M^4}\, , \nonumber\\
E_{1}&=&\frac{\sqrt s}{2}\left[1+(1-f^2)\frac{M^2}{s}\right]\,, \qquad E_{2}=\frac{\sqrt s}{2}\left[1-(1-f^2)\frac{M^2}{s}\right]\, ,
\eea
with corresponding velocities $\beta_{1,2}=k/E_{1,2}$. The expression of the vectors
$n_{1,2}^{\mu}(1)$ and $n_{1,2}^{\mu}(2)$ remain the same as in Eq.~(\ref{spinvect}), while here
\bea
n_{3}^{\mu}(1)=\gamma_1\, (\beta_1,\, \hk)
\, ,~~ n_{3}^{\mu}(2)=\gamma_2\, (-\beta_2,\, \hk)\, , \label{n3}
\eea
where $\gamma_{1,2}=1/\sqrt{1-\beta_{1,2}^2}$ are the corresponding Lorentz factors. The normalization conditions remain the same as in the degenerate case, Eq.~(\ref{normcond}), but with scalar product  
\be
n_3^{\mu}(1)\, n_{3\mu}(2)=-\gamma_1\gamma_2 \, (\beta_1\beta_2+1)\, .
\ee

Turning now to the computation of the polarization density matrix for two spin-1 bosons of arbitrary non vanishing masses, the matrix element ${\cal M}(\lambdaA,\lambdaB)$ for the related production amplitude can be written as
\bea
{\cal M}(\lambdaA,\lambdaB)&=&{\cal M}_{\mu\nu}
\varepsilon^{\mu\star}(k_1,\lambdaA)\varepsilon^{\nu\star}(k_2,\lambdaB)\, .
\eea
The polarization density matrix is accordingly defined as
\bea
\rho(\lambdaA,\lambdaAp,\lambdaB,\lambdaBp)&=&\frac{
{\cal M}(\lambdaA,\lambdaB) {\cal M}^\dag (\lambdaAp,\lambdaBp)}{|\xbar{{\cal M}}|^2}\, ,
\label{rho2}
\eea
where, as usual, $|\xbar{{\cal M}}|^2$ stands for the unpolarized square amplitude and a sum over the possible internal degrees of freedoms of initial state particles is understood.

By using the covariant expression for the spin-1 projectors
$\mathscr{P}^{\mu\nu}_{\lambda\lambda^{\prime}}(k)$ defined in \eq{proj}, we can rewrite the
the density matrix in Eq.~(\ref{rho2}) as
\bea
\rho(\lambdaA,\lambdaAp,\lambdaB,\lambdaBp)=\frac{{\cal M}_{\mu\nu}{\cal M}^{\dag}_{\mu^{\prime}\nu^{\prime}}
\mathscr{P}^{\mu\mu'}_{\lambdaA\lambdaAp}(k_1)
\mathscr{P}^{\nu\nu'}_{\lambdaB\lambdaBp}(k_2)}{|\xbar{{\cal M}}|^2}\, .
\eea

In the case at hand, $\rho(\lambdaA,\lambdaAp,\lambdaB,\lambdaBp)$ can be decomposed on the basis of the 9$\times$9 matrices formed by the tensor products $\{\mathbb{1}\otimes\mathbb{1}, \,\mathbb{1}\otimes T^a, \,T^a\otimes \mathbb{1}, \,T^a\otimes T^b\}$, with $T^{a}$ again the 3$\times$3 Gell-Mann matrices. In particular, we have\footnote{We use the abbreviation: $[A\otimes B]_{ii^{\prime} jj^{\prime}}= A_{ii^{\prime}}B_{jj^{\prime}}$.}
\bea
\label{eq:rhone}
\rho(\lambdaA,\lambdaAp,\lambdaB,\lambdaBp)= \Big(\frac{1}{9}\left[\mathbb{1}\otimes
    \mathbb{1}\right]+
    \sum_a f_a \left[T^a\otimes \mathbb{1}\right]+\sum_a g_a \left[\mathbb{1}\otimes T^a\right] 
    +\sum_{ab} h_{ab}  \left[T^a\otimes T^b\right]\Big)_{\lambdaA\lambdaAp\lambdaB\lambdaBp}.
\label{rho}
\eea

The eight components of $f_a$ and $g_a$, as well as the 64 elements of $h_{ab}$, can be obtained by projecting $\rho$ on the desired subspace basis via 
\bea
f_a=\frac{1}{6}\,\Tr\left[\rho \left(T^a \otimes \mathbb{1}\right)\right]\, , ~~
g_a=\frac{1}{6}\,\Tr\left[\rho \left(\mathbb{1}\otimes T^a\right)\right]\, , ~~
h_{ab}=\frac{1}{4}\,\Tr\left[\rho \left(T^a \otimes T^b\right)\right]\, .
\label{fgh}
\eea
All the terms computed via Eq.~(\ref{fgh}) are Lorentz scalars which depend only on the energy $E$, the velocity $\beta$ and the scattering angle $\Theta$ in the CM frame.

It is possible to compute the observable quantifying the entanglement in the gauge boson system once the coefficients $f_a$, $g_a$ and $h_{ab}$ are known. The lower bound $\cmb$, introduced in Section~\ref{sec:obs} as an entanglement witness, can be written in terms of the coefficients in \eq{fgh} as
\bea
\cmb &=& 2\max \Big[ -\frac{2}{9}-12 \sum_a f_{a}^{2} +6 \sum_a g_{a}^{2} + 4 \sum_{ab} h_{ab}^{2}\, ,\nn \Big.\\
& & \Big. -\frac{2}{9}-12  \sum_a g_{a}^{2} +6 \sum_a f_{a}^{2} + 4 \sum_{ab} h_{ab}^{2},\, 0 \Big]\,,
\eea
which is the expression we use throughout this work. 

Likewise, the observable ${\cal I}_3$ can be written in terms of the  coefficients $h_{ab}$  as 
\be
{\cal I}_3 = 4 \Big(h_{44} + h_{55} \Big)
 - \frac{4 \sqrt{3} }{3} \Big[ h_{61} + h_{66} + h_{72} +  h_{77} +  h_{11} + h_{16} + h_{22} +  h_{27} \Big]\,. \label{ii3}
\ee
\eq{ii3} is valid prior to performing the unitary rotation in \eq{uni_rot} of the $\cal B$ matrix that maximizes the value of the corresponding expectation value. Such a rotation might bring a dependence also on the coefficients $f_{a}$ and $g_{a}$, beside changing the number and the weights of the various coefficients $h_{ab}$.

\subsection{Reconstructing the correlation coefficients from the data \label{sec:qp}}

The actual processes observed at colliders are
\be
p\;p \to V_1 + V_2 + X \to \ell^+ \ell^- \;\ell^+\ell^-  \; (\text{or}\;\;  \ell^\pm\; j_s j_c + E^{\text{miss}}_{T})+ \text{jets}  \, ,
\ee
with missing energy $E^{\text{miss}}_{T}$ due to the possible presence of neutrinos in the final state. 
These processes include the production of the gauge bosons through  the resonant Higgs boson channel, as well as via quark fusion, and include the consequent decays into the final leptons (for the $Z$s) or the jets of interest (for the $W$s)---plus the jets originating from $X$ spectator quarks. 

The spin 1 gauge bosons act as their own polarimeters. For instance, in the decay $W^{+}\to \ell^{+}  \nu_{\ell}$ the lepton $\ell^{+}$ is produced in the positive helicity state while the neutrino $\nu_{\ell}$ in the negative helicity state. The polarization of the $W^{+}$ is therefore measured to be $+1$ in the direction of the lepton  $\ell^{+}$. The opposite holds for  the decay $W^{-}\to \ell^{-}  \bar \nu_{\ell}$ and the polarization of the $W^{-}$ is therefore  measured to be $-1$ in the direction of the lepton $\ell^{-}$. In both the cases, the momenta of the final leptons (see Fig.\ref{fig:coordinates}) provide a measurement of the gauge boson polarizations. The same is true for final jets from $d$ and $s$ quarks.
These momenta are the only information that we need to extract from  the  numerical simulation or the actual data. 

How do we go about reconstructing the correlation coefficients $h_{ab}$, $f_a$ and $g_a$ of the density matrix starting from the momenta of the final leptons? This problem has been recently discussed in~\cite{Ashby-Pickering:2022umy}, which we mostly follow in the remainder of this section. 
 
The cross section we are interested in can be written as~\cite{Rahaman:2021fcz}
\be 
  \frac{1}{\sigma}\frac{\di \sigma}{\di\Omega^{+}\,\di\Omega^{-}}
  = \left( \frac{3}{4 \pi} \right)^2
 \Tr  \Big[ \rho_{V_1V_2} \left(\Pi_+ \otimes \Pi_-\right)\Big]\, , \label{x-sec-leptons}  
\ee
in which the angular volumes $\di \Omega^\pm= \sin \theta^\pm \di \theta^\pm\,\di \phi^\pm$ are written in terms of the spherical coordinates  (with independent polar axes) for the momenta of the final charged leptons in the respective rest frames of the decaying particles. The dependence on the invariant mass $m_{VV}$ and scattering angle $\Theta$ in \eq{x-sec-leptons} is implied. The density matrix $\rho_{V_1V_2}$ in \eq{x-sec-leptons} is that for the production of two gauge bosons given in \eq{rho}. 

The density matrices $\Pi_\pm$ describe the polarization of the decaying gauge bosons. The final leptons are taken to be massless---for their masses are negligible with respect to that of the gauge boson. They are projectors  in the case of the $W$-bosons because of their chiral coupling  to leptons. These matrices can be computed by rotating to an arbitrary polar axis the  spin $\pm 1$ states of the weak gauge bosons taken in the $z$ direction and are given,  in the Gell-Mann basis, as
\be
\Pi_\pm=\frac{1}{3}\,\mathbb{1}
+\sum_{i=1}^8 \mathfrak{q}^a_\pm\, T^a \, , \label{pi_{f}}
\ee
where the functions $\mathfrak{q}^a_\pm$ can be written in terms of the respective spherical coordinates, as reported in \eq{Q} of  Appendix~\ref{sec:Aqp}, for the decay of $W$-bosons.\footnote{The  functions in \eq{Q}, are the Wigner's $Q\; symbols$ for the case of a spin 1 particle.}
 
We can define another set of functions
\be
\mathfrak{p}_\pm^n =   \sum_{m} (\mathfrak{m}^{-1}_{\pm})_{m}^{n} \,\mathfrak{q}_\pm^m \label{p}
\ee
orthogonal to those in \eq{Q}:
\be
\left(\dfrac{3}{4\, \pi} \right) \int \mathfrak{p}_\pm^n\, \mathfrak{q}_\pm^m \, \di \Omega^\pm = \delta^{nm} \, .
\ee
In \eq{P}, $\mathfrak{m}^{-1}$ is the inverse of the matrix 
\be
(\mathfrak{m}_{\pm})^{nm}= \left(\dfrac{3}{8\, \pi}  \right)\int \frak{q}_\pm^n \, \mathfrak{q}_\pm^m \, \di \Omega^\pm \, ,
\ee
which is assumed to exist.
The explicit form of the functions $\mathfrak{p}_\pm^n$ are given in Appendix~\ref{sec:Aqp} \eq{P} .

The functions in \eq{P}  can be used to extract the correlation coefficients  $h_{ab}$ from the bi-differential cross section in \eq{x-sec-leptons} through the projection
\bea
h_{ab} &=&  \frac{1}{\sigma} \int \int \frac{\di \sigma}{\di\Omega^{+}\,\di\Omega^{-}} \, \mathfrak{p}_+^a \, \mathfrak{p}_-^b \,\di \Omega^+ \di \Omega^-\, .\label{hh}
\eea
The correlation coefficients  $f_a$ and $g_a$ can be obtained in similar fashion by projecting the single differential cross sections:
\bea
f_{a} &=&  \frac{1}{\sigma}  \int  \frac{\di \sigma}{\di\Omega^{+}} \, \mathfrak{p}_+^a \, \di \Omega^+ \, ,\nn\\
g_{a} &=&  \frac{1}{\sigma}  \int  \frac{\di \sigma}{\di\Omega^{-}} \, \mathfrak{p}_-^a \, \di \Omega^- \, . \label{ffgg}
\eea

The density matrices $\Pi_\pm$ are not projectors in the case of the $Z$-bosons because  the coupling between $Z$-bosons and  leptons
\be
\mathcal{L}\supset
-i \frac{g}{\cos \theta_{W}} \Big[ g_{L} (1-\gamma^{5}) \gamma_{\mu} + g_{R} (1+\gamma^{5}) \gamma_{\mu}\Big]\, Z^{\mu}
\ee
contains both right- and  left-handed components, whose strengths are controlled by the coefficients $g_{L} =-1/2 + \sin^2 \theta_{W}$  and $g_{R}= \sin^2 \theta_{W}$. In this case, one must introduce a generalized form of the functions in \eq{Q} which is defined as the following linear combinations
\be
\tilde{\mathfrak{q}}^{n} = \dfrac{1}{g_{R}^2 + g_{L}^{2}} \Big[ g_{R}^2 \, \mathfrak{q}^{n}_{+}+  g_{L}^2 \, \mathfrak{q}_{-}^{n}\Big]\, ,
\ee
and define from these the corresponding orthogonal functions $\tilde{\mathfrak{p}}^{n}$ to be used in \eq{fgh}. They are the same for both the $\pm$ coordinate sets and given by
\be
\tilde{\mathfrak{p}}^{n} = \sum_{m}\mathfrak{a}^{n}_{m} \mathfrak{p}_+^m \, , \label{p_f}
\ee
where the matrix $\mathfrak{a}^{n}_{m}$ is given in \eq{Anm} in Appendix~\ref{sec:Aqp}. The \eqs{hh}{ffgg} can be used after replacing the functions $\mathfrak{p}_\pm^m$ with $\tilde{\mathfrak{p}}^{n} $ and including
a symmetry factor of 1/2 for the $f_{a}$ and $g_{a}$ coefficients and 1/4 for the $h_{ab}$ in the case of identical final states, namely for the $ZZ$ case.

 \eqs{hh}{ffgg} provide the means to reconstruct the correlation functions of the density matrix from the distribution of the lepton momenta and thus allow to infer the expectation values of the observables ${\cal I}_3$ and  $\cmb$ from the data. 
In a numerical simulation, or working with actual events, one extracts from  each single event the coefficient of the combinations of trigonometric functions indicated in \eq{P} in \ref{sec:Aqp}; that coefficient is the corresponding entry of the correlation matrix in \eqs{hh}{ffgg}. Running this procedure over all events gives an average value and its standard deviation.

The analysis outlined in this Section is experimentally rather challenging because both the CM frame of the collision  and the rest frame of the gauge bosons must be determined as precisely as possible to compute the correlation coefficients $h_{ab}$, $f_a$ and $g_a$ with reasonable uncertainties.

\subsection{Estimating the uncertainty}

We model the uncertainty in the value of our observables as a  Gaussian dispersion--- controlled by the number of events---in the determination of the kinematical variables. The number of events is modulated by the efficiency in the identification of the final charged leptons or  jets.

To this random error, we add, in the case of the $WW$ final states, a systematic error that takes into account  the significant uncertainty in the reconstruction of the gauge boson momentum from the missing momentum of the neutrinos. This reconstruction comes from the kinematical constraints together with dedicated algorithms.  Estimates show that the distribution of the differences between the true and the reconstructed momentum over the true momentum has a standard deviation ranging from 30\%, for simple kinematical reconstructions with smearing effects of the detector, to 3\% in more advanced machine learning algorithms (see, for instance, \cite{Choi:2009hn,Betchart:2013nba,D0:2009lmx,Grossi:2020orx,Leigh:2022lpn}). We take for this  uncertainty  a conservative benchmark value  of 30\% (and show for the $WW$ continuum  how the determination changes for a smaller value) and add it in quadrature to the Gaussian error coming from the number of available events.

We run from 1000 (Higgs decay) to 10000 (Drell-Yan-like production) pseudo experiments as we vary the kinematical variables and compute for each of these values the observable ${\cal I}_3$.  The distribution so obtained is skewed because the observable is computed near its maximum value and the random variation can only reduce this value.

The  significance of the violation of the Bell inequality can be defined as ${\cal Z} = \Phi^{-1}(1-p)$
where
\be
\Phi(x) = \frac{1}{2}\, \left[ 1 + \text{erf} \left(\frac{ x}{\sqrt{2}} \right) \right]\,,
\ee
and the $p$-value $p$ refers to the null hypothesis that the Bell inequality is not violated.
The value of  ${\cal Z}$ assigns a statistical significance to the separation between the  distribution we obtain for the values of  ${\cal I}_3$ and the value 2, above which  the Bell identity is violated. 

Values of the significance larger than 5 requires a very large number of pseudo experiments to be performed in order to find the actual value. For this reason, when this is the case, we do an extrapolation and quote a lower bound.


\section{Di-boson production in Higgs boson decays 
\label{sec:hvv}}
Consider the decay
\be
H\to V(k_1,\lambda_1)\, V^*(k_2,\lambda_2)\, ,
\label{HVV}
\ee
with $V\in\{W,Z$\}, and $V^*$ regarded as an off-shell vector boson. In the following, we treat the latter as an on-shell particle characterized by a mass 
\be
M_{V^*}= f M_V
\ee
reduced by a factor $0<f<1$ with respect to the original mass $M_V$.
\begin{figure}[h!]
\begin{center}
\includegraphics[width=3.8in]{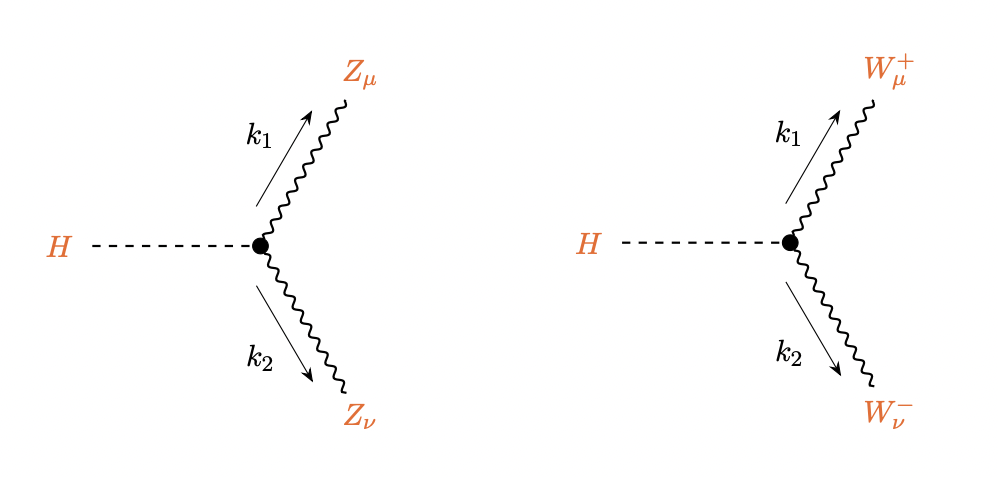}
\caption{\small Feynman diagrams for the decay of the Higgs boson into a pair of gauge bosons.
\label{fig:HtoVV} 
}
\end{center}
\end{figure}
The amplitude of the Higgs boson decay (\ref{HVV}) is given by 
\bea
{\cal M}_H(\lambdaA,\lambdaB)=g \,  M_V\, \xi_V \, g_{\mu\nu}
\varepsilon^{\mu\star}(k_1,\lambdaA)\varepsilon^{\nu\star}(k_2,\lambdaB)
\, ,
\label{MHVV}
\eea
where $g$ is the weak coupling, $\xi_W=1$, and $\xi_Z=1/(2c_W)$, with $c_W=\cos{\theta_W}$ and $\theta_W$ the Weinberg angle. From the amplitude in Eq.~(\ref{MHVV}) we obtain
\bea
{\cal M}_H(\lambdaA,\lambdaB) {\cal M}_H(\lambdaAp,\lambdaBp)^{\dag} &=&
g^2 \, M_V^2 \, \xi^2_V \, g_{\mu\nu}g_{\mup\nup}  \mathscr{P}^{\mu\mup}_{\lambdaA\lambdaAp}(k_1)
 \mathscr{P}^{\nu\nup}_{\lambdaB\lambdaBp}(k_2)\, .
\eea
where $\mathscr{P}^{\mu\nu}_{\lambda\lambda^{\prime}}(k)$ is given in \eq{proj} with $M=M_V$ or $M=M^*_V$ for the on-shell and off-shell boson, respectively.

Following the procedure explained in Section~\ref{sec:rho} for a CM energy $\sqrt s=m_H$, we obtain the coefficients $f_a$, $g_a$, and $h_{ab}$ ($a,b\in\{1, \dots,8\}$) reported below. There is no dependence on the scattering angle $\Theta$ because we are considering the decay of the Higgs boson at rest. 

The non-vanishing $f_{a}$ elements are
\bea
&&f_{3}=\frac{1}{6}\, \frac{-m_H^4 + 2 (1 + f^2) m_H^2 M_V^2 - (1 - f^2)^2 M_V^4}{m_H^4 - 2 (1 + f^2) m_H^2 M_V^2 +(1 + 10 f^2 + f^4) M_V^4}
\, ,
\nonumber\\
\nonumber\\
&&f_{8}=-\frac{1}{\sqrt{3}}f_3\, ,
\eea
and we find $g_{a}=f_a$ for $a\in\{1,\dots,8\}$. The non-vanishing $h_{ab}$ elements are 
\bea
&&h_{16}=h_{61}=h_{27}=h_{72}=\frac{f M_V^2 \Big[-m_H^2 + (1 + f^2) M_V^2\Big]}
    {m_H^4 - 2 (1 + f^2) m_H^2 M_V^2 + (1 + 10 f^2 + f^4) M_V^4}\, ,
    \nonumber\\
    \nonumber\\
&&h_{33}=\frac{1}{4}\, \frac{\Big[m_H^2 - (1 + f^2) M_V^2\Big]^2}{m_H^4 - 2 (1 + f^2) m_H^2 M_V^2 + (1 + 10 f^2 + f^4) M_V^4}\, ,
    \nonumber\\
    \nonumber\\
&&h_{38}=h_{83}=-\frac{1}{4\sqrt{3}}
    \nonumber\\
    \nonumber\\
&&h_{44}=h_{55}=\frac{2 f^2 M_V^4} {m_H^4 - 2 (1 + f^2) m_H^2 M_V^2 + (1 + 10 f^2 + f^4) M_V^4}\, ,
\nonumber\\
\nonumber\\
&& h_{88}=\frac{1}{12}\, \frac{m_H^4 - 2 (1 + f^2) m_H^2 M_V^2 + (1 - 14 f^2 + f^4) M_V^4}
    {m_H^4 - 2 (1 + f^2) m_H^2 M_V^2 + (1 + 10 f^2 + f^4) M_V^4}\, , \label{fghHiggs}
\eea
The unpolarized squared amplitude $|\xbar{{\cal M}}_H|^2$ of the process instead reads
\bea
|\xbar{{\cal M}}_H|^2&=&\frac{g^2\xi_V^2}{4 f^2 M_V^2}
\Big[ m_H^4 - 2 (1 + f^2) m_H^2 M_V^2 + (1 + 10 f^2 + f^4) M_V^4\Big]\, .
\eea

The main theoretical uncertainty affecting the correlation coefficients in \eq{fghHiggs} is due to higher order corrections to the tree-level values. To estimate the size of these contributions, we take as guidance the results in~\cite{Boselli:2015aha}---in which the NLO EW corrections have been computed. According to these results, we expect the error induced by these missing corrections yields at most a few percent of uncertainty on the main entanglement observables, in the relevant kinematic regions in which one of the two EW gauge boson are on-shell ~\cite{Boselli:2015aha}. This expectation is based on the fact that these corrections give a 1-2\% effect on the total width ~\cite{Boselli:2015aha}.

We then compute through \eq{eq:rhone} the polarization density matrix $\rho_H$ for the two vector bosons emitted in the decay of the Higgs boson
\be
\rho_H = 2 \begin{pmatrix} 
  0 & 0 & 0 & 0 & 0 & 0 & 0 & 0 & 0  \\
  0 & 0 & 0 & 0 & 0 & 0 & 0 & 0 & 0  \\
  0 & 0 &  h_{44} & 0 &  h_{16} & 0 & h_{44} & 0 & 0  \\
  0 & 0 & 0 & 0 & 0 & 0 & 0 & 0 & 0  \\
  0 & 0 &  h_{16} & 0 & 2\, h_{33} & 0 & h_{16} & 0 & 0  \\
  0 & 0 & 0 & 0 & 0 & 0 & 0 & 0 & 0  \\
  0 & 0 &  h_{44} & 0 &  h_{16} & 0 &  h_{44}& 0 & 0  \\
  0 & 0 & 0 & 0 & 0 & 0 & 0 & 0 & 0  \\
  0 & 0 & 0 & 0 & 0 & 0 & 0 & 0 & 0  \\
\end{pmatrix} \, ,
\label{rhoH}
\ee
with the condition $\Tr[\rho_H]=1$ following from the relation $4(h_{33}+h_{44})=1$.

\begin{figure}[h!]
\begin{center}
\includegraphics[width=3in]{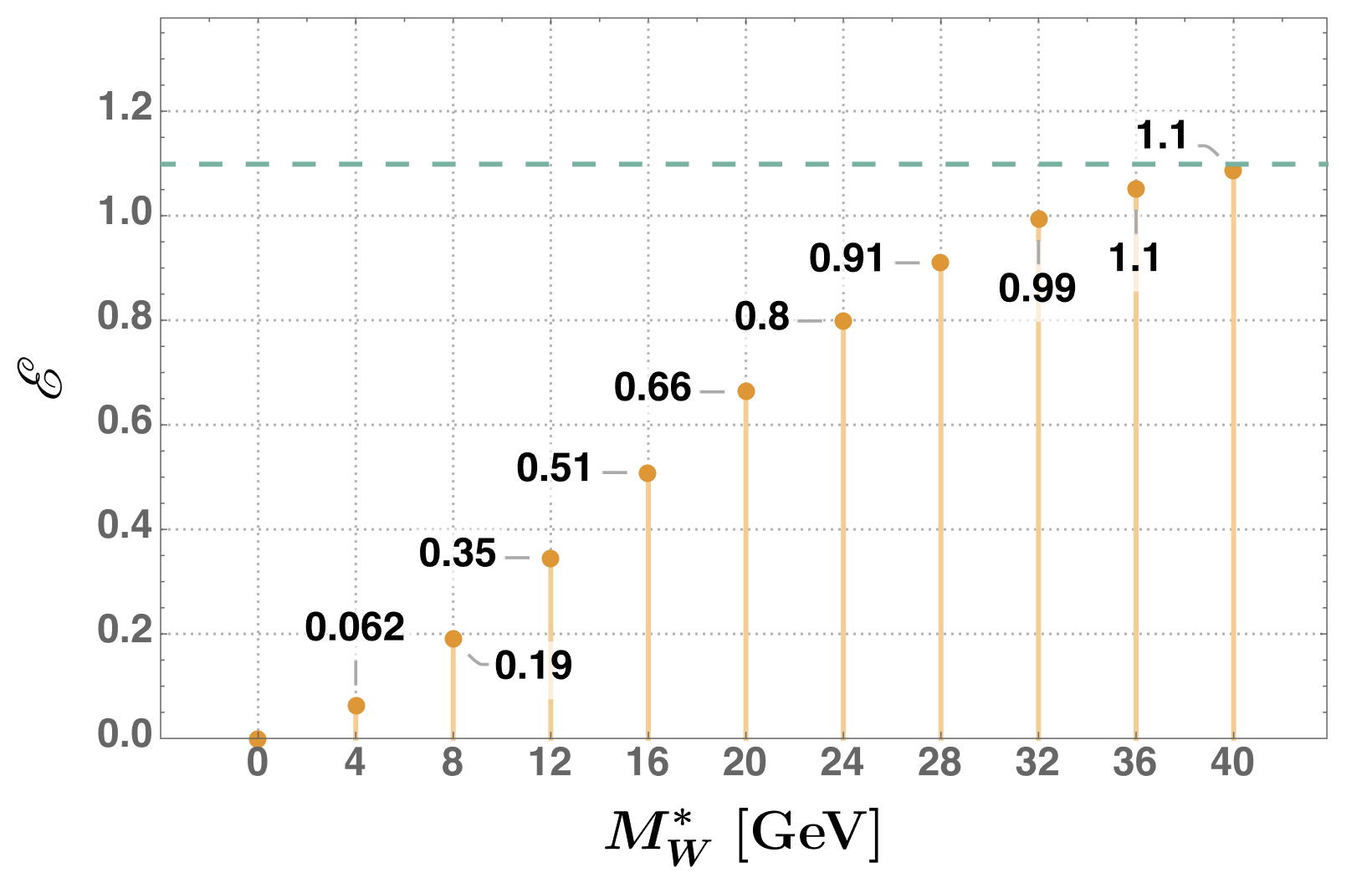}
\includegraphics[width=3in]{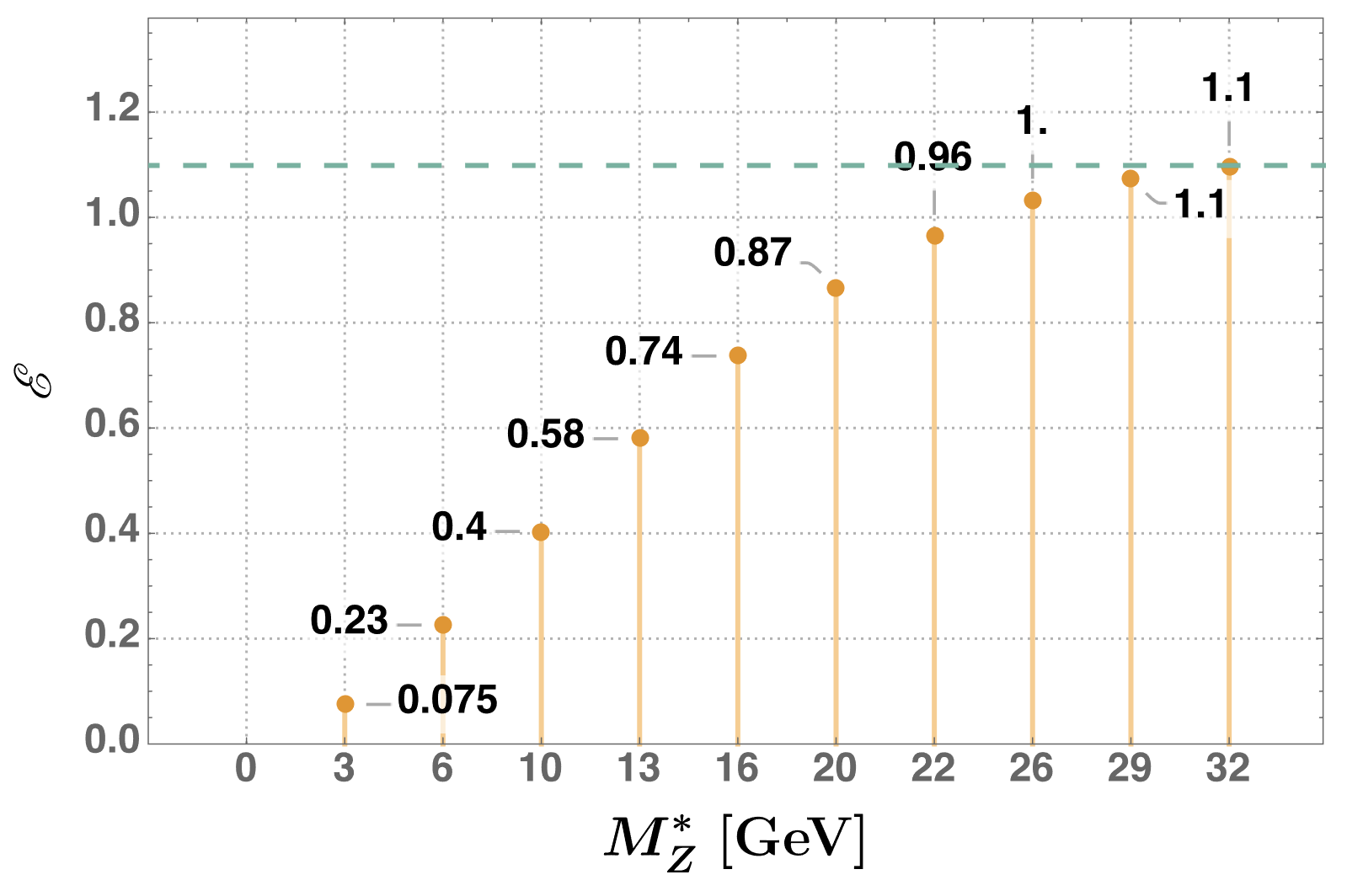}
\caption{\small The  entropy of entanglement (left plot for $H\to WW^*$ and 
 right plot for $H\to ZZ^*$)  as functions of the virtual  mass of one of the two weak gauge bosons. The dashed line marks the maximum value $\log 3$.
\label{fig:entropy} 
}
\end{center}
\end{figure}

We remark that although some $f_a$ and $g_a$ are non vanishing, the dependence of $\rho_H$ on these quantities cancels in the final expression. Furthermore, due to the following identity among the correlation coefficients 
\begin{equation}
  h_{44} = 2 \left(h_{16}^2 + 2 h_{44}^2\right)\,,
\end{equation}
the above polarization density matrix is idempotent
\bea
\rho_H^2=\rho_H\, ,
\label{rhorel}
\eea
signaling that the final $VV^*$ state is a pure state. The density matrix in \eq{rhoH} can then be written~\cite{Aguilar-Saavedra:2022wam}
\be
\rho_H = |\Psi_H \rangle \langle \Psi_H | \, ,
\ee
where (in the basis $|\lambda\, \lambdap\rangle = |\lambda\rangle\otimes|\lambdap\rangle$ with $\lambda,\lambdap\in\{+,0,-\}$)  
\be
|\Psi_H \rangle = \frac{1}{\sqrt{2 +  \varkappa^2}} \left[  |{\small +-}\rangle -  \varkappa \,|{\small 0\, 0}\rangle + |{\small -+}\rangle  \right] \label{pure}
\ee
with
\be
 \varkappa = 1+ \frac{m_H^2 - (1+f)^2 M_V^2}{2 f M^2_V} 
\ee
and $ \varkappa=1$ corresponding to the production of two gauge bosons at rest.

Because the di-boson system is described by a pure state, we can measure its entanglement through the  entropy of entanglement defined in \eq{E}. This quantity is plotted in Fig.~\ref{fig:entropy} as a function of the of the mass of virtual $W$ or $Z$ boson and reaches the theoretical maximum at the kinematic threshold, signaling a maximally entangled state. The dependence of the polarization entanglement on the  mass of the virtual state  is due  the contribution of the longitudinal polarization, the coefficient $\varkappa$ in \eq{pure}: it  starts out bigger  and decreases  to 1  at the threshold. The value of 1 yields a singlet state and the maximum in the entanglement of the state.

In Figs.~\ref{fig:HWW} and \ref{fig:HZZ} we show the results for the
main observables targeting quantum entanglement, ${\cal I}_{3}$ (left panel) and
$\cmb$ (right panel), in the $H\to WW^*$ and $H\to ZZ^*$ decays. The plots are for different values of the virtual gauge boson masses $M_{{W}^{*}}$ and $M_{{Z}^{*}}$, respectively. 

The maximization of the ${\cal I}_{3}$ observable, which depends in this case only on the $M_V^{*}$ mass, is obtained through the rotation 
\be
{\cal B} \to (U_V\otimes V_V)^{\dag} \cdot {\cal B} \cdot (U_V\otimes V_V)\, ,
\ee
by unitary matrices $U_V,V_V$ (with index $V\in\{W,Z\}$), as defined in Section~\ref{sec:obs}. The maximization must be performed point by point as the density matrix varies with $M_V^{*}$.

\begin{figure}[h!]
\begin{center}
\includegraphics[width=3in]{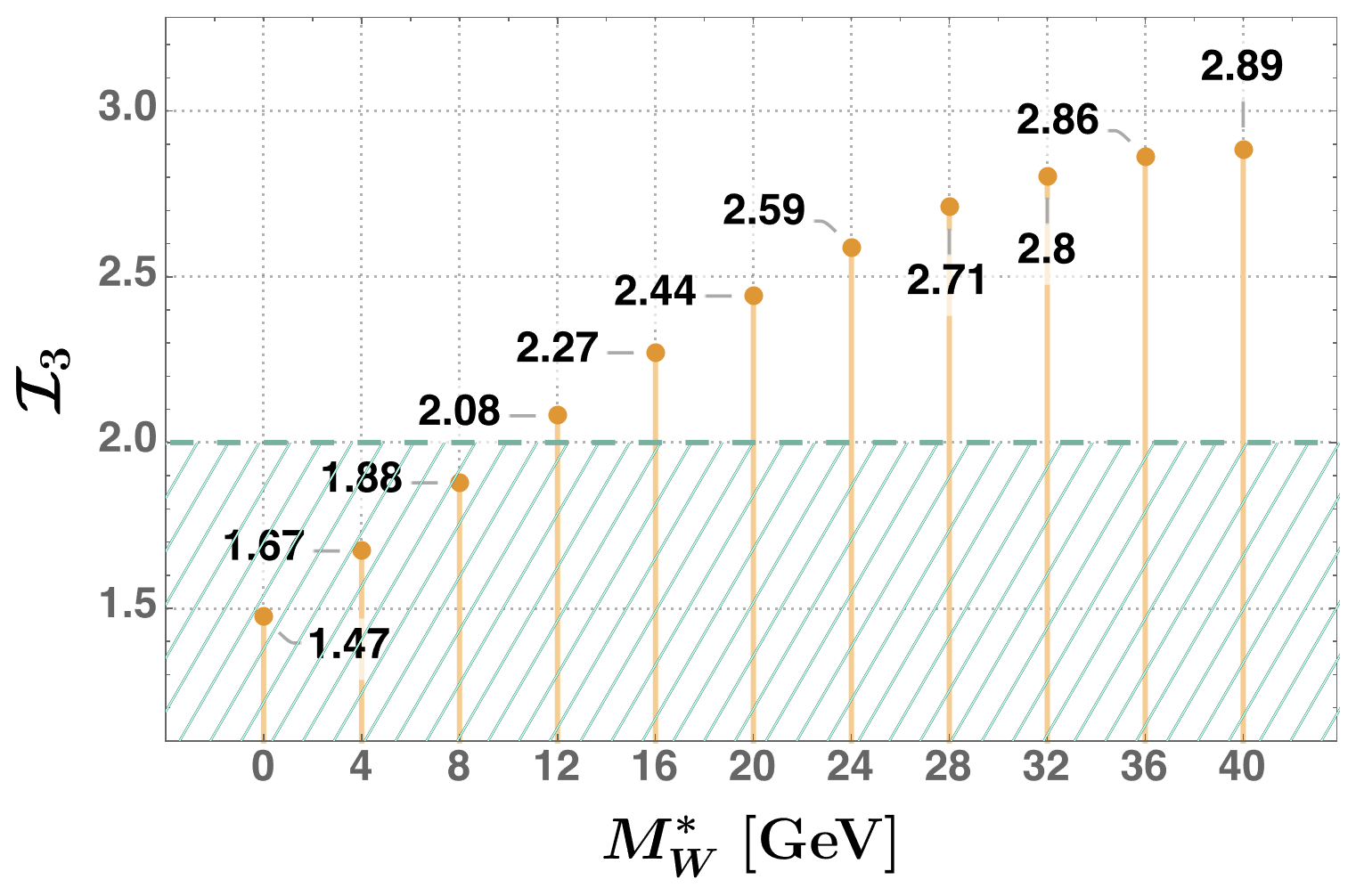}
\includegraphics[width=3in]{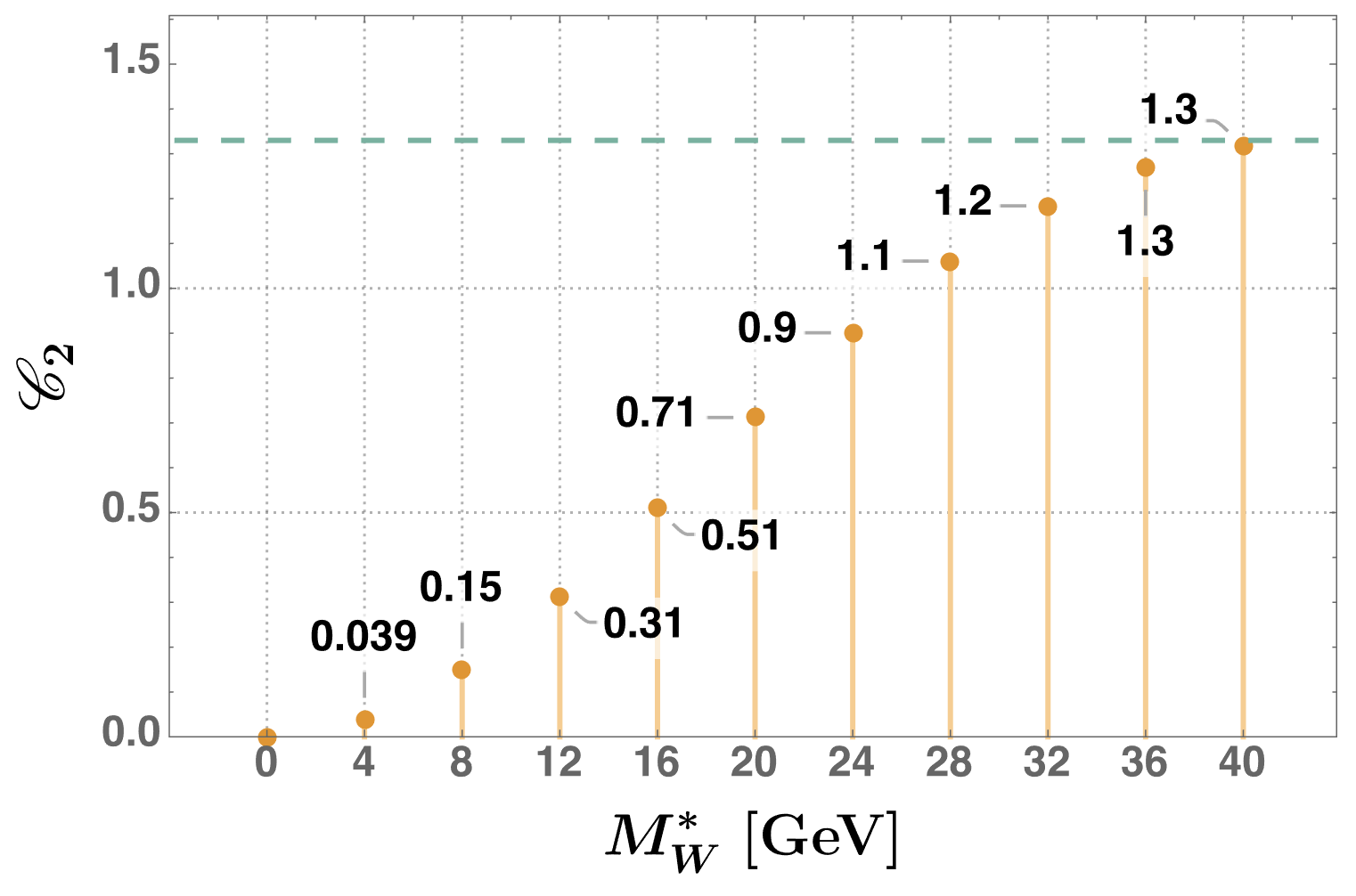}
\caption{\small The observables ${\cal I}_3$ (left plot) and 
$\cmb$ (right plot) for the pair production of $W$ bosons in Higgs boson decays as functions of the virtual $W^{*}$ mass in the range $0<M_{W^*}<40$~GeV. The dashed horizontal line in the left-hand side plot marks the Bell-inequality violation condition ${\cal I}_3>2$. The dashed line in the right-hand side plot illustrates for the maximum value 4/3 corresponding to a pure state.
\label{fig:HWW} 
}
\end{center}
\end{figure}

\begin{figure}[h!]
\begin{center}
\includegraphics[width=3in]{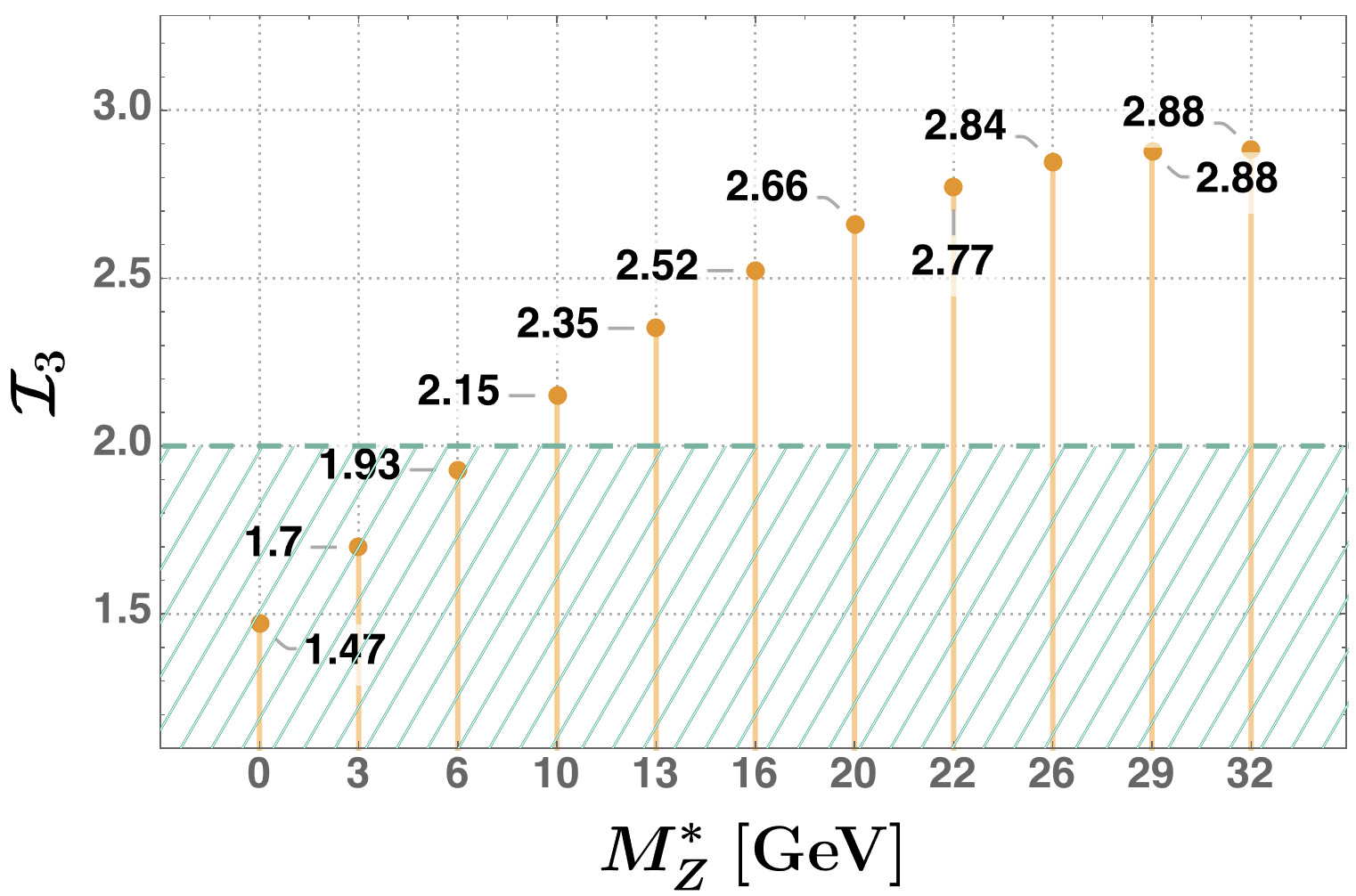}
\includegraphics[width=3in]{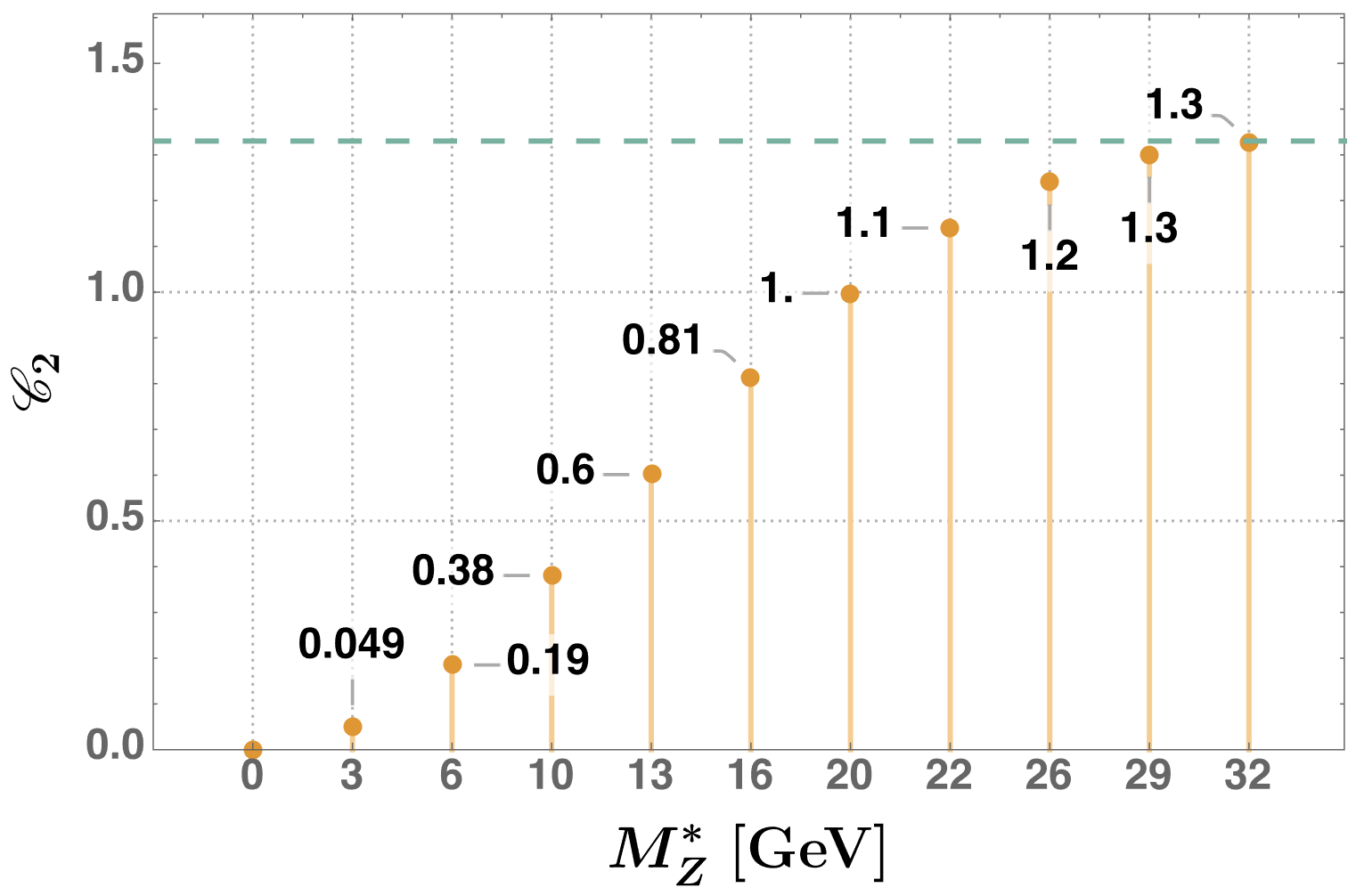}
\caption{\small The observables ${\cal I}_3$ (left plot) and 
  $\cmb$ (right plot) for the pair production of $Z$ bosons in Higgs boson decays as functions of the virtual $Z^{*}$  mass in the range $0<M_{Z^*}<32$~GeV. The dashed horizontal line in the left plot stands for the Bell-inequality violation condition ${\cal I}_3>2$. The dashed line in the right plot denotes the maximum value 4/3 corresponding to a pure state.
\label{fig:HZZ} 
}
\end{center}
\end{figure}

We provide in \eqs{UVWW}{UVZZ} the expressions for the unitary matrices $U$ and $V$ that maximizes the ${\cal I}_{3}$ observable in the last bins (in which $M_{W^*}=40$ GeV and $M_{Z^*}=32$ GeV) for the $H\to WW^*$ and $H\to ZZ^*$ decays, respectively. For the $WW$ channel, we find
 \be
U_W = \begin{pmatrix} 
  \dfrac{4}{11}+\dfrac{i}{14} & \dfrac{1}{6}+\dfrac{9i}{13} & \dfrac{3}{5} +\dfrac{i}{14}  \\\\
  -\dfrac{1}{9}-\dfrac{6i}{7} &0& \dfrac{1}{10}+\dfrac{i}{2}  \\\\
  \dfrac{4}{11}+\dfrac{i}{12} & -\dfrac{1}{7}-\dfrac{7i}{10} & \dfrac{3}{5}
  +\dfrac{i}{10}\\
\end{pmatrix} \, , ~~~
V_W = \begin{pmatrix} 
  -\dfrac{1}{7}-\dfrac{7i}{12} & -\dfrac{7}{10}-\dfrac{i}{10} & -\dfrac{1}{9} - \dfrac{6i}{17}  \\\\
  \dfrac{11}{21}+\dfrac{i}{17} &0& -\dfrac{6}{7}-\dfrac{i}{26}  \\\\
  -\dfrac{1}{8}-\dfrac{3i}{5} & \dfrac{7}{10}+\dfrac{i}{8} & -\dfrac{1}{10}-\dfrac{5i}{14}\\
\end{pmatrix} \, , ~~~
\label{UVWW}
\ee
while for the $ZZ$ channel
 \be
U_Z = \begin{pmatrix} 
 - \dfrac{1}{2}+\dfrac{3 i}{11} & \dfrac{7}{13}+\dfrac{5 i}{11} & \dfrac{4}{13}-\dfrac{3 i}{10}  \\\\
- \dfrac{1}{2}+\dfrac{3 i}{8} &   0 &  - \dfrac{15}{31}+\dfrac{5 i}{8} & \\\\
- \dfrac{1}{5}+\dfrac{10 i}{19} &   -\dfrac{5}{7} &  + \dfrac{1}{22}-\dfrac{3 i}{7} & \\
\end{pmatrix} \, ,~~~~
V_Z = \begin{pmatrix} 
  -\dfrac{1}{7}+\dfrac{5 i}{12} &   \dfrac{7}{11}+\dfrac{2 i}{7} &   \dfrac{1}{25}-\dfrac{5 i}{9}  \\\\
   \dfrac{2}{11}+\dfrac{10 i}{13} & 0  &  \dfrac{2}{7}+\dfrac{6 i}{11} \\\\
  \dfrac{1}{6}+\dfrac{2 i}{5} &   -\dfrac{11}{16}+\dfrac{i}{5} & -\dfrac{1}{3} -\dfrac{4 i}{9}\\
\end{pmatrix} \, .
\label{UVZZ}
\ee
The matrices in \eq{UVWW} and \eq{UVZZ} are given in terms of rational numbers which approximate the corresponding real values with a 1\% precision. The unitary condition is satisfied barring $O(10^{-2})$ factors. These matrices cannot be directly compared with the similar expressions given in~\cite{Aguilar-Saavedra:2022wam} because of the different assumptions in the utilized optimization procedure.

The $\cmb$ observable admits here a simple analytical expression
\bea
\cmb &=& \frac{32 f^2 M_V^4 \Big[m_H^4 - 
     2 (1 + f^2) m_H^2 M_V^2 + (1 + 4 f^2 + f^4) M_V^4\Big]}{\Big[m_H^4 - 
    2 (1 + f^2) m_H^2 M_V^2 + (1 + 10 f^2 + f^4) M_V^4\Big]^2}\,.
\label{cmbformula}
\eea
The plots on the right-hand side in Figs.~\ref{fig:HWW} and~\ref{fig:HZZ}  nicely show that the value of $\cmb$ decreases as the pure state in \eq{rhoH}  becomes less and less entangled, for decreasing values of $M^*_V$. 

\subsection{Events and sensitivity}

In order to evaluate the sensitivity of current experiments to the observables ${\cal I}_3$ and $\cmb$, we estimate the number of suitable events available. These are given in Table~\ref{tab:events_h_ww} for the run2 at the LHC.

\begin{table}[h!]
\bc
\vskip1cm
\begin{tabular}{ccc}
&\hskip0.5cm ${\color{oucrimsonred} \ell^+ \nu_\ell \, j_s j_c}$  \hskip0.5cm &  \hskip0.5cm${\color{oucrimsonred} \ell^- \ell^{+} \ell^- \ell^+}$ \hskip0.5cm \\[0.2cm]
\hline\\
 \underline{LHC run2} \hskip0.5cm (${\color{oucrimsonred} {\cal L}=140\ \text{fb}^{-1}}$) \hskip0.4cm  &\hskip0.4cm  $3718$   \hskip0.4cm &\hskip0.4cm  $28$ \hskip0.4cm \\[0.4cm]
    \underline{Hi-Lumi} \hskip0.5cm (${\color{oucrimsonred} {\cal L}=3\ \text{ab}^{-1}}$) \hskip0.4cm  &\hskip0.4cm  $8.0\times 10^{4}$   \hskip0.4cm &\hskip0.4cm  $589$ \hskip0.4cm \\[0.4cm]
   \hline%
\end{tabular}
\caption{\small \label{tab:events_h_ww}Number of expected events for the Higgs boson decays into $WW^*$ and $ZZ^*$  assuming a luminosity ${\cal L}=140\ \text{fb}^{-1}$ for the run2 at the LHC.  The cut in invariant mass is at 30 and 40 GeV respectively for the $WW$ and $ZZ$ channels. A benchmark efficiency of 70\% is assumed in the identification of each charged lepton.}
\ec
\end{table}

The cross sections for $p \,p \to H\to W^+\ell^-\bar \nu_\ell$ and  $p \,p \to H\to  Z\ell^+\ell^-$ utilized in the estimates are computed with {\tt MADGRAPH5}~\cite{Alwall:2014hca} at the LO and then corrected by the $\kappa$-factor given at the N3LO+N3LL~\cite{Bonvini:2016frm}.  

Even the definition of the rest frame of one decaying $W$-boson introduces an essential uncertainty because of the ambiguity in the reconstruction of the longitudinal momentum of the neutrino and the possibility of misidentifications (and other errors) in the identification of the missing momentum. There is no such a problem in the case of the $Z$-boson decay which may though suffer of other generic inefficiencies. In the case of the Higgs boson decay into two $W$-bosons, the problem is exacerbated: the full reconstruction is not  possible even in principle because  there are more variables than constraints (since one of the masses of the gauge bosons is necessarily off-shell and the missing transverse momentum includes both neutrinos). 
   
The problem of actually estimating the size of these uncertainties (for a given choice of an algorithm for the neutrino momenta reconstruction) is the central problem of any physical analysis from the actual or simulated data  of the process and cannot be resolved here. 

We   take into account the problem of these irreducible uncertainties  in the evaluation of the operators in  the decays of the $WW^*$  by introducing a systematic error that mimics  the significant uncertainty in the reconstruction of the neutrino momenta, which has to come from a dedicated algorithm (see, for instance, \cite{Choi:2009hn,Betchart:2013nba,Grossi:2020orx,Leigh:2022lpn}). Since the uncertainty would be dominated by the error in the reconstruction of the two neutrino momenta, it is better  to consider  the semi-leptonic decay $H\to jj\ell \nu_\ell$ and use the momentum from the $s$-quark jet ($s$-jet)--identified via the $c$-tagging of the companion jet--to measure the polarization of one of the two $W$-bosons. It has been shown that the efficiency of the jet tagging and the decreased uncertainty in the single neutrino momentum may improve the polarization reconstruction~\cite{maurin}. In this case, we take a conservative benchmark value  of 0.3  for the systematic error in the single neutrino momentum and an overall efficiency of 40\% in the $c$-jet tagging  and the identification of the momentum associated to the $s$-jet that carries (by the same degree  as the charged lepton in leptonic decays) the polarization of the $W$. 

In addition we include an  efficiency factor  of 70\% in the identification of each charged lepton~\cite{Jain:2021bis}. 

\begin{figure}[h!]
\begin{center}
\includegraphics[width=2.9in]{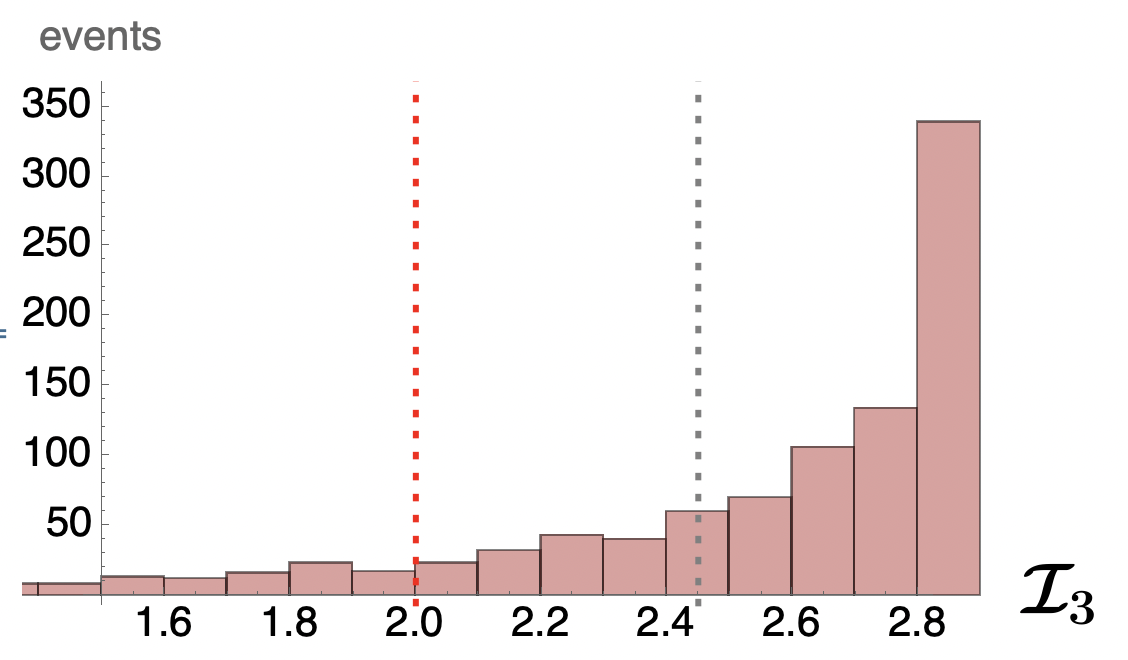}
\includegraphics[width=3.04in]{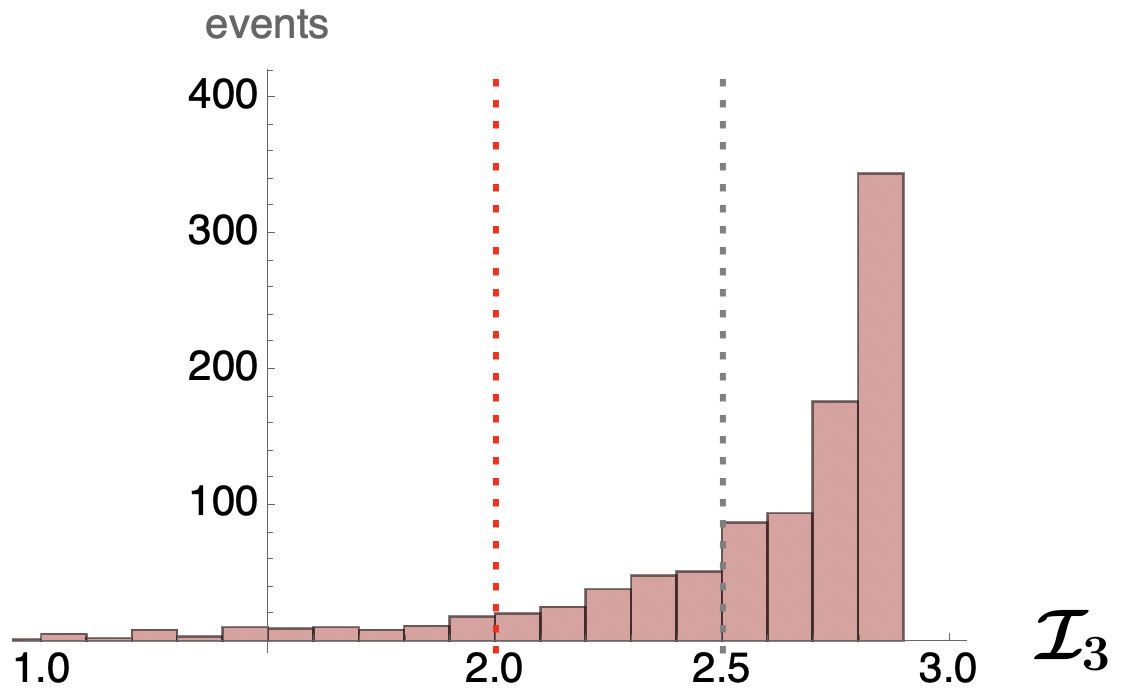}
\caption{\small  Distribution of the  events at the LHC run 2 for the $H\to W^+\ell^-\bar \nu_\ell$ and $H\to Z\ell^+\ell^-$ processes.The  set of events for $WW^*$ has mean value ${\cal I}_3=2.4$, that for $ZZ^*$ has mean value ${\cal I}_3=2.5$. The threshold value of 2 for Bell inequality violation is shown as a dashed red line.
\label{fig:eventsH1} 
}
\end{center}
\end{figure}

\begin{figure}[h!]
\begin{center}
\includegraphics[width=2.9in]{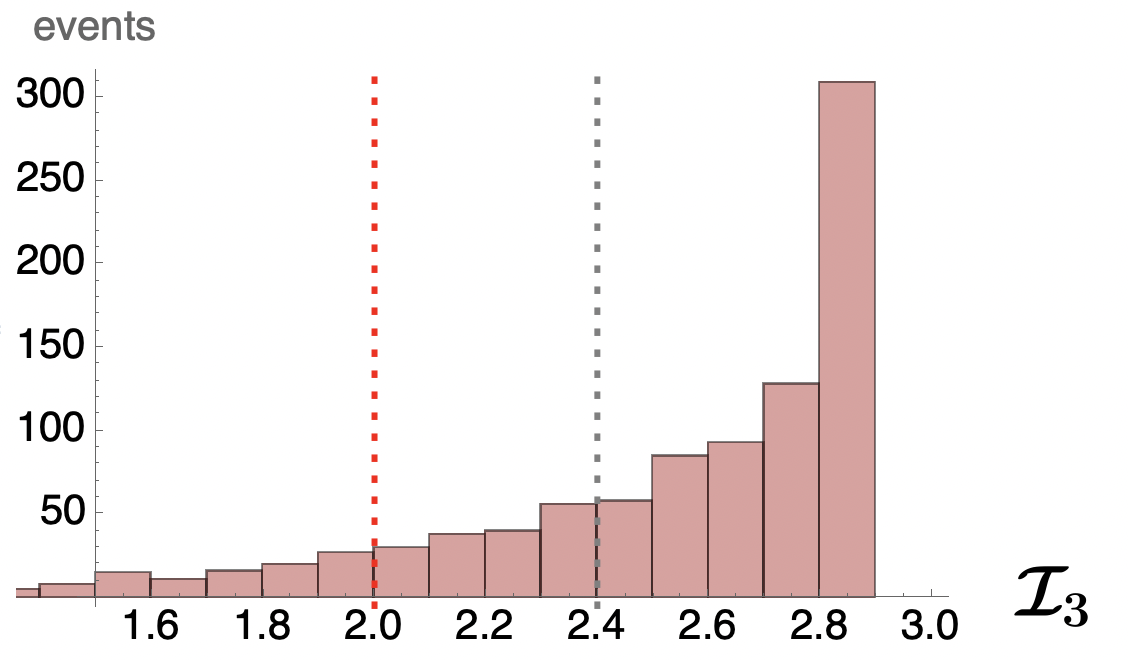}
\includegraphics[width=3.04in]{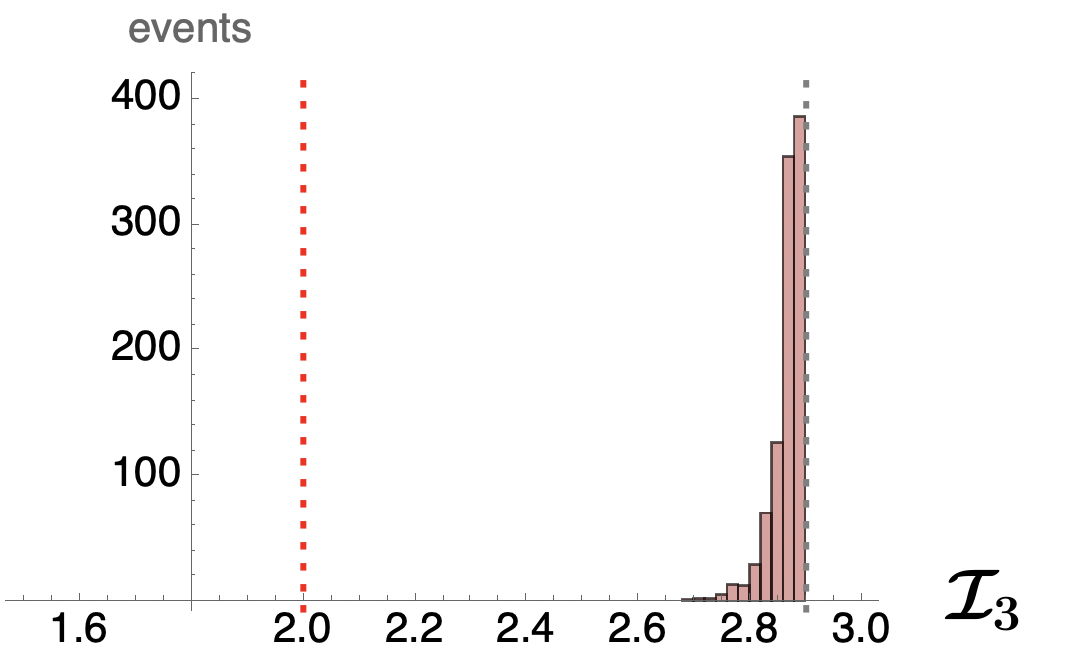}
\caption{\small Distribution of the  events at the LHC Hi-Lumi for the $H\to W^+\ell^-\bar \nu_\ell$ and $H\to Z\ell^+\ell^-$ processes.  The  set of events for $WW^*$ has mean value ${\cal I}_3=2.5$, that for $ZZ^*$ has mean value ${\cal I}_3=2.9$. The threshold value of 2 for Bell inequality violation is shown as a dashed red line.
\label{fig:eventsH2} 
}
\end{center}
\end{figure}

 The irreducible background for the $H\to W^+\ell^-\bar \nu_\ell$ signal comes from the continuum electroweak production of $W^+W^-$ pairs. It can be reduced  by considering  the characteristic  distribution of the kinematical variables to a manageable size~\cite{ATLAS:2022ooq,maurin}.  In addition, one has to remove the reducible background events from $t\bar t$ and $W t$ production.
  The irreducible background for the $H\to Z\ell^+\ell^-$ signal is rather small and dominated by the electroweak process $p p \to  ZZ/Z\gamma \to  4\ell$, which is about 4 times smaller at the Higgs peak~\cite{CMS:2021ugl}.
  We   neglect all backgrounds in our assessment of  the significance even though they will have to be included in the actual analysis from the data and will affect the uncertainty. 
  
  To show   the impact of  possible irreducible backgrounds we show in Fig.~\ref{fig:bckg} the values for ${\cal I}_3$ and $\cmb$ found as a generic factorizable density matrix is added with weight $(1-\alpha)$ to that of the $H\to WW^*$ process, which is accordingly multiplied by a factor $\alpha$ (with $\alpha$ between 0 and 1). For the case shown in Fig.~\ref{fig:bckg},  for values of $\alpha$ smaller than 0.7, the uncertainty substantially reduce  the possibility of assessing the Bell inequality violation.

\begin{figure}[h!]
\begin{center}
\includegraphics[width=2.9in]{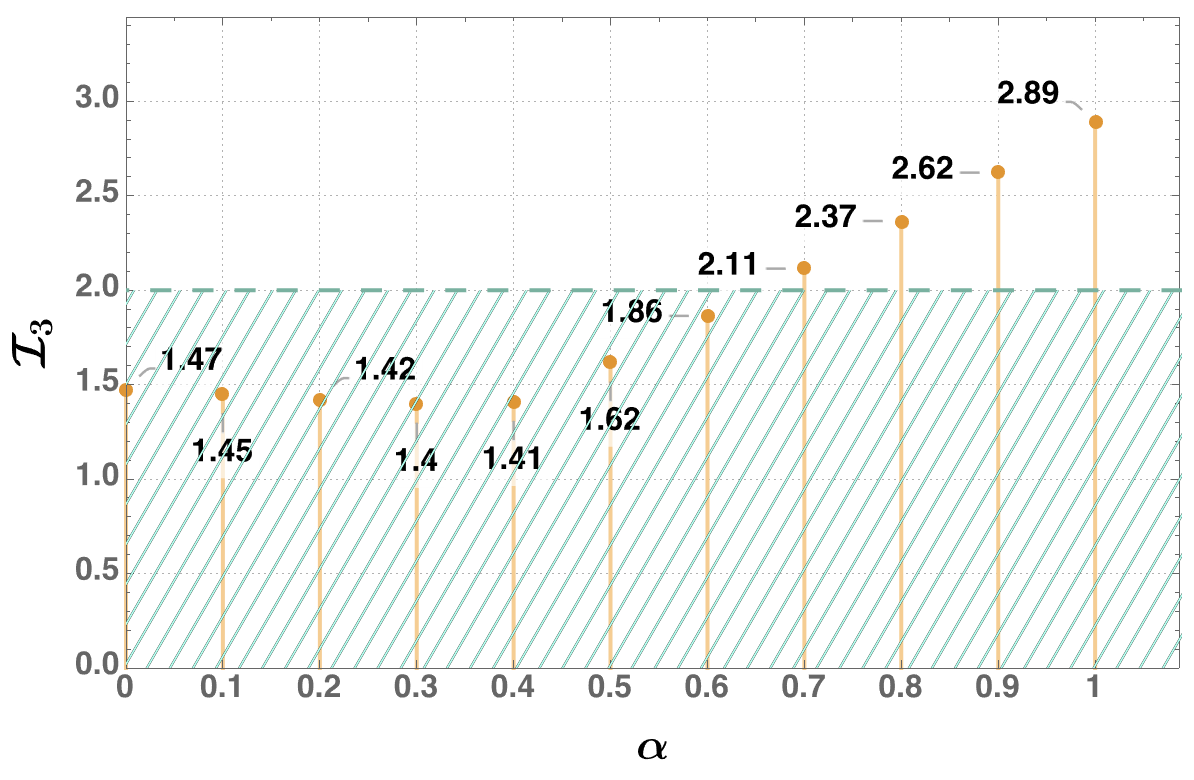}
\includegraphics[width=2.9in]{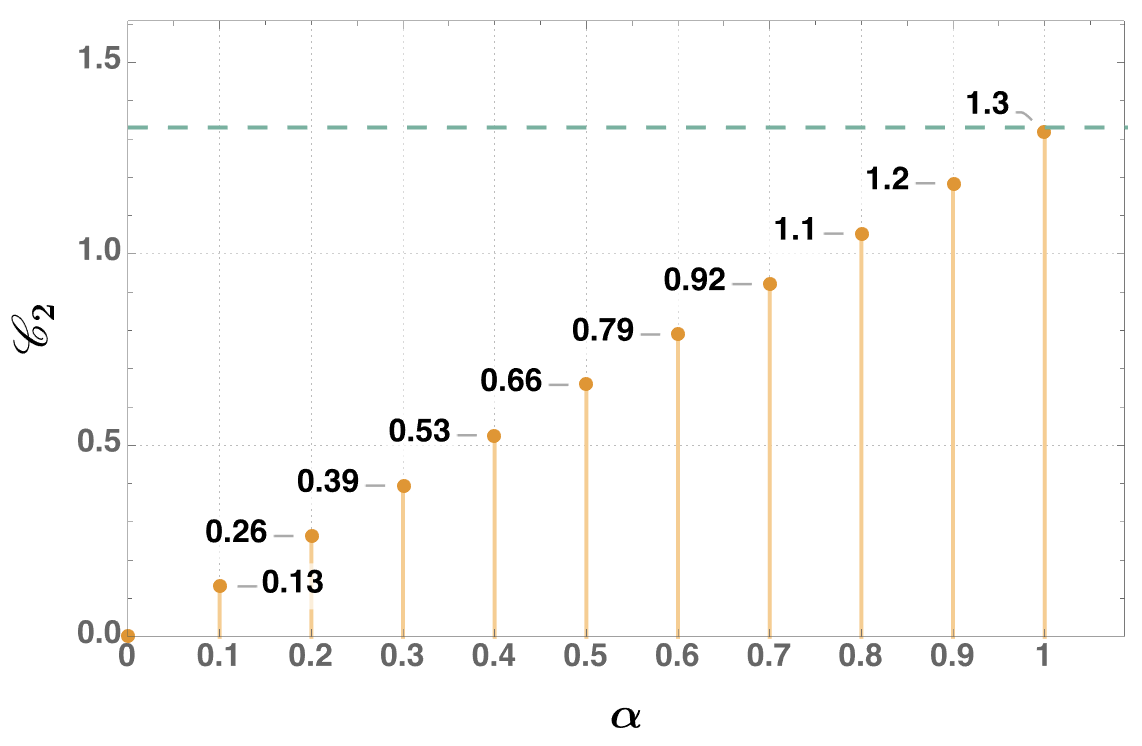}
\caption{\small Effect of the background events on  ${\cal I}_3$ and $\cmb$ the signal $H\to W W^*$. For $\alpha=1$ the background-free result is reproduced. For values $\alpha<0.7$ the mixing of the background to the signal decreases the entanglement beyond the possibility of assessing Bell inequality violation.
\label{fig:bckg} 
}
\end{center}
\end{figure}

We run 1000 pseudo experiments as we vary the invariant mass of the off-shell gauge boson around the mean value with a dispersion given by the (statistical and systematic) uncertainty as discussed above,  and compute the observable ${\cal I}_3$. Fig.~\ref{fig:eventsH1} and Fig.~\ref{fig:eventsH2} show the distributions which are obtained for, respectively, LHC run 2 and Hi-Lumi. The distributions are skewed because the observable is computed near its maximum value and the random variation can only reduce this value.

Fig.~\ref{fig:eventsH1} shows that, at the LHC run 2,  the significance for rejecting the null hypothesis ${\cal I}_{3}\leq 2$ is 1 for the $WW^*$ case and 1.3 for the $ZZ^*$ case. 
Fig.~\ref{fig:eventsH2} shows that, at the LHC Hi-Lumi,  the significance for rejecting the null hypothesis ${\cal I}_{3}\leq 2$ remains 1 for the $WW^*$ case, since the uncertainty is dominated by the statistical error,  while it  reaches 5.6 for the $ZZ^*$ case. These significances are likely to decrease in a more complete analysis based on a full simulation because of the reconstruction from the final lepton angular distributions and the systematic uncertainties of the unfolding, which is particular severe  for the $W^+W^-$ case due to the presence of neutrinos and background events.

Our results confirm the numerical simulations presented in~\cite{Barr:2021zcp} for the $H\to WW^*$ process and in~\cite{Aguilar-Saavedra:2022wam} for the $H\to ZZ^*$ case. These works estimate the uncertainties from a parton-level  reconstruction of the final lepton angular distributions. Yet, a fully realistic estimate of the uncertainty is still missing as  uncertainties due to detector unfolding and  background  have not been modeled.

\section{Di-boson production via quark fusion 
\label{sec:DY}}  

Final $WW$, $ZZ$, and $WZ$ states can be produced via electroweak processes in a continuous range of di-boson invariant masses. We show in the following how the polarization density matrix of the di-boson system can be computed starting from the density matrices obtained for the involved parton contributions, presented in Fig.~\ref{fig:DYVV} for the processes at hand. For the sake of simplicity, when possible we leave implicit the dependence of the correlation coefficients $h_{ab}(\mVV,\Theta)$, $g_a(\mVV,\Theta)$ and $f_a(\mVV,\Theta)$ on the scattering angle $\Theta$ in the CM frame and on the invariant mass of the bosons $\mVV$. 

\begin{figure}[h!]
  \begin{center}
  \includegraphics[width=5.0in]{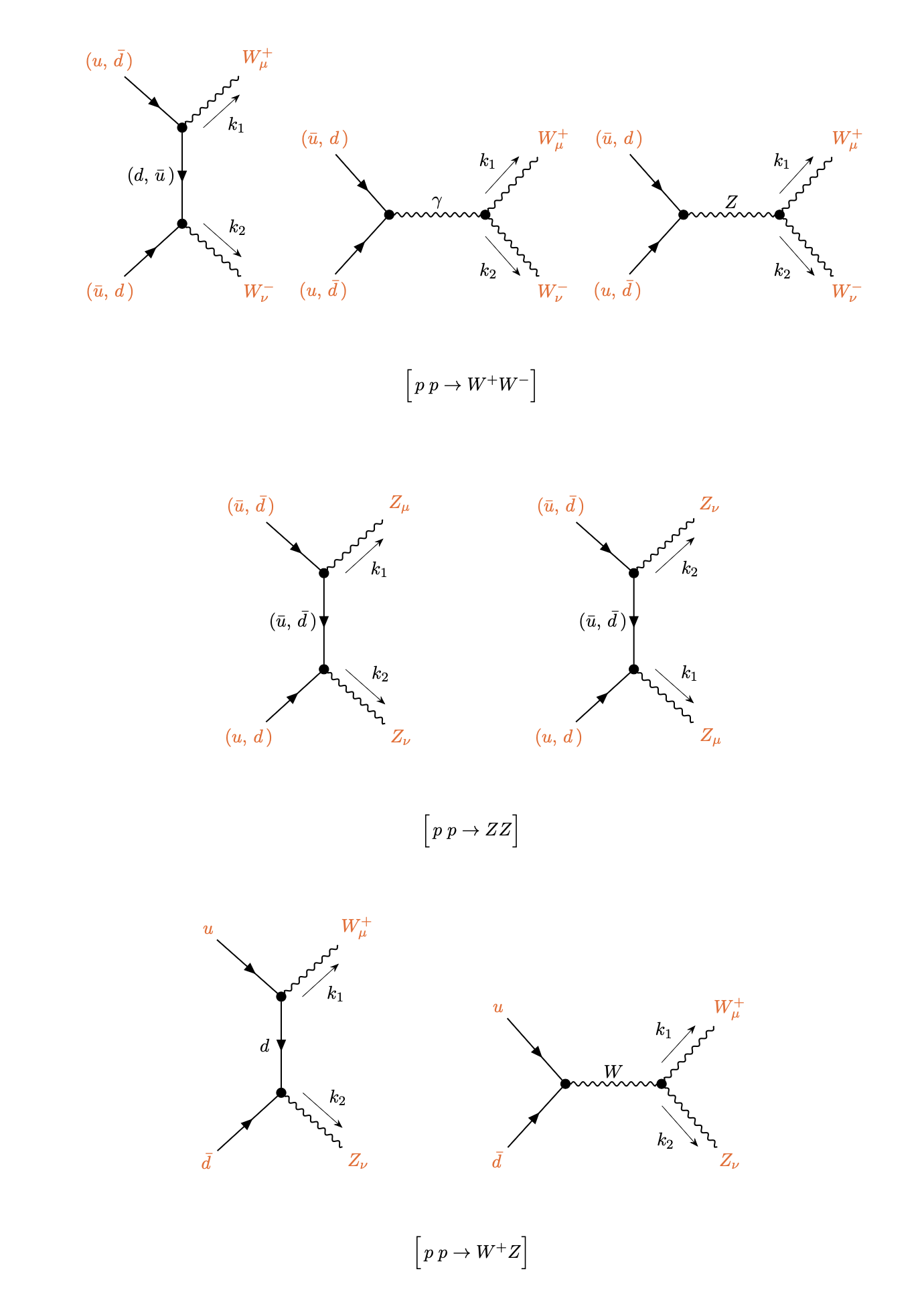}
  \caption{\small Feynman diagrams for the processes $p\; p\to W^+W^-$ (first row),  $p\; p\to Z Z$ (second row) and $p\; p\to W^+Z$ (third row) at the parton level for the first quark generation. We neglect diagrams mediated by the Higgs boson considering the limit of massless quarks. The arrows on the fermion lines indicate the momentum flow. 
  \label{fig:DYVV} 
  }
  \end{center}
  \end{figure}

The polarization density matrix $\rho$ for two bosons produced in proton collisions is given by the convex combination of the density matrices of the involved parton contributions. Given initial state quarks $q_1$ and $\bar q_2$, we compute through Eq.~(\ref{rho}) the polarization density matrix $\rho^{q_1 \bar q_2}$ of the parton contribution, from the scattering amplitude of the process $q_1\, \bar q_2 \to V_1 V_2$, $V_1, V_2\in\{W,Z\}$. Then,

\begin{equation}
  \rho = \sum_{\{q_1 \bar q_2\}} w^{q_1\bar q_2} \, \rho^{q_1 \bar q_2}
\end{equation}
where the sum runs over all the allowed initial states, including both the configurations where the anti-quark originates from either proton.\footnote{The kinematics of di-boson production is such that for each pair of these `specular' configurations it holds
\begin{equation}
  \rho^{\bar q_2  q_1} (\Theta) = \rho^{q_1 \bar q_2}(\Theta+\pi)\, ,   
\end{equation}
where the ordering of the quark fields symbolically tracks the proton of origin. 
}     
The coefficients $w^{q_1 \bar q_2}$, satisfying $\sum_{\{q_1, \bar q_2\}} w^{q_1 \bar q_2} = 1$, are given by 
\begin{equation}
  w^{q_1 \bar q_2} 
  = 
  \dfrac{L^{q_1\bar q_1} \, |\xbar{\mathcal{M}}^{\;q_1 \bar q_2}_{V_1V_2}|^2 }
  {\sum_{\{q_1 \bar q_2\}} L^{q_1\bar q_1} \, |\xbar{\mathcal{M}}^{\;q_1 \bar q_2}_{V_1V_2}|^2}
\end{equation}
and depend on the unpolarized squared amplitude of the parton process, $|\xbar{\mathcal{M}}^{\,q_1, \bar q_2}_{V_1V_2}|^2$, as well as on the parton luminosity of the initial $q_1\bar{q_2}$ state  
\be
L^{q_1\bar q_1} (\tau)= \frac{4 \tau}{\sqrt{s}} \int\limits_{\tau}^{1/\tau} \frac{\di z}{z} \, q_{q_1} (\tau \,z) \, q_{\bar q_2} \left( \frac{\tau}{z}\right)\, .
\ee
In the formula above $q_j(x)$ is the parton distribution function (PDF) of the parton $j$ and $\tau=\mVV/\sqrt s$. We utilized the numerical values provided by the recent {(\tt PDF4LHC21)} release~\cite{PDF4LHCWorkingGroup:2022cjn} for $\sqrt{s}=13$ TeV and factorization scale $\mVV$ (see Fig.~\ref{fig:pdf}). 

\begin{figure}[h!]
  \begin{center}
  \includegraphics[width=3.5in]{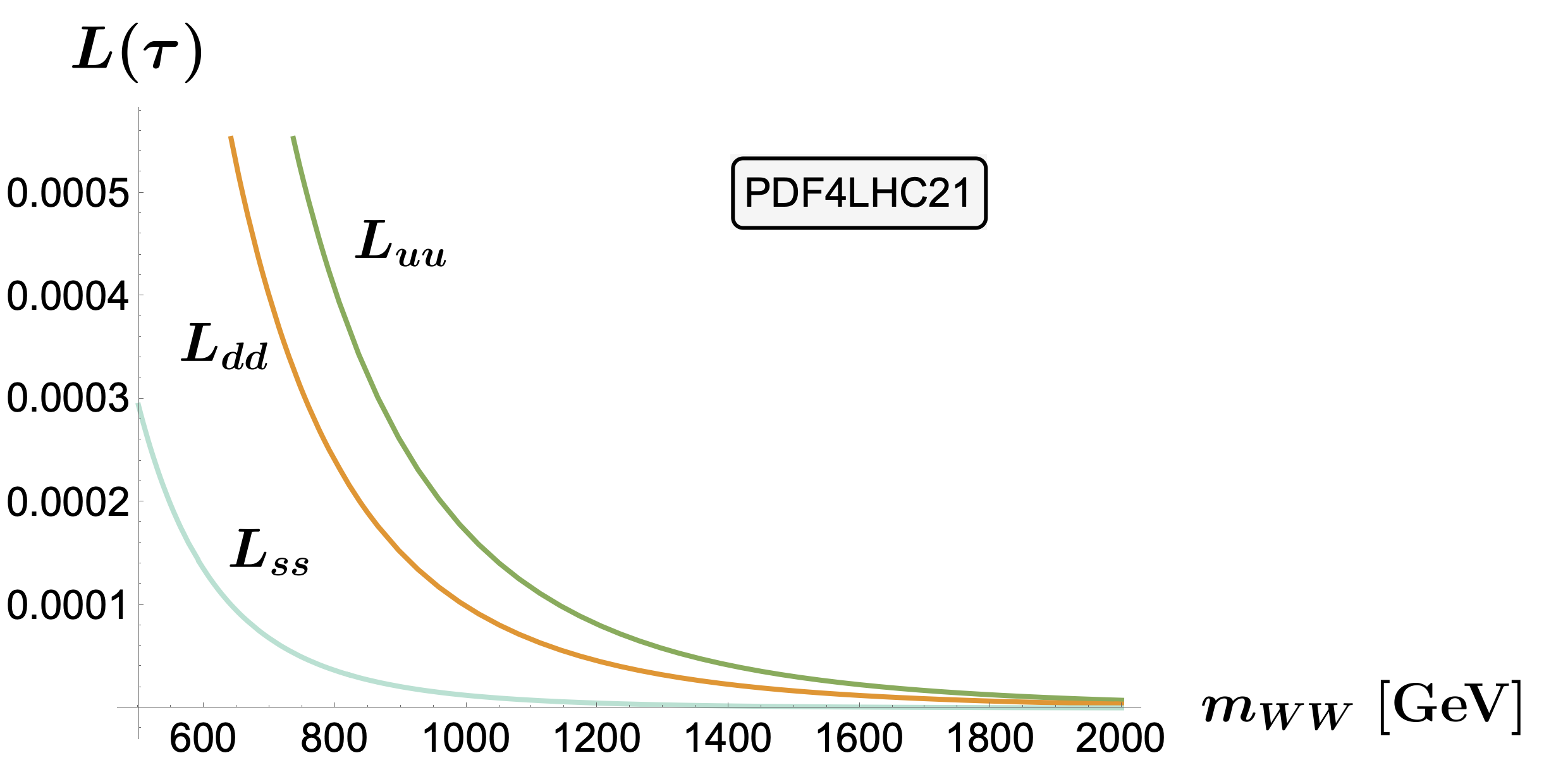}
  \caption{\small  Parton luminosity functions as functions of the invariant mass. 
  \label{fig:pdf} 
  }
  \end{center}
  \end{figure}

As an example, for the $W$ boson pair production we find 
\be
h_{ab} [\mVV,\Theta]= \frac{\sum_{q=u,d,s} L^{q\bar q} (\tau)
\Big(  \tilde{h}^{q \bar q}_{ab}[\mVV,\Theta] +
\tilde{h}^{q \bar q}_{ab}[\mVV,\Theta+\pi]\Big)}
{\sum_{q=u,d,s} L^{q\bar q} (\tau)\Big(A^{q \bar q}[\mVV,\Theta] +A^{q\bar q}[\mVV,\Theta+\pi]
  \Big)}\,
\label{habDY}
\ee
where we introduced the abbreviations $A^{q \bar q} = |\xbar{\mathcal{M}}^{\;q \bar q}_{WW}|^2$ and $\tilde h_{ab} = A^{q \bar q} h_{ab} $. Similar results hold for the remaining correlation coefficients $f_a$ and $g_a$, $a\in\{1, \dots, 8\}$. The explicit expressions of $\tilde h_{ab}$, $\tilde f_{a} = A^{q \bar q} f_{a}$, $\tilde g_{a} = A^{q \bar q} g_{a}$ for all the analyzed processes are collected in~\ref{sec:a2}. 

The main source of theoretical uncertainty in the determination of the correlation coefficients comes from higher order QCD corrections. Taking as a guidance the results in~\cite{Denner:2020bcz}, we assume that the error induced by these missing corrections yields approximately a 10\% uncertainty on the main entanglement observables in the relevant kinematic regions.
The theoretical uncertainties coming from the PDFs and the top-quark mass are negligible: by comparing results obtained with two different set of PDFs, we estimate the related uncertainty to be of the order of per mille.
This is due to the fact that only ratios of PDFs enter in \eq{habDY} for the $h_{ab}$ coefficients, and analogously  for the  $g_a$ and $f_a$ ones, and therefore  most of the PDF uncertainty   cancels out.
  
This is of the same order as the uncertainty due to the  top-quark mass, obtained by varying the parameter around its experimental value at most by two standard deviations. 
  
In the following, we present our results for the entanglement observables for the $WW$, $WZ$ and $ZZ$ cases separately.

Our results differ from those presented in~\cite{Ashby-Pickering:2022umy}, obtained through a parton-level numerical simulation. In particular, we find substantially lower values for the observable ${\cal I}_3$ in the $W^+Z$ process, larger for the $W^+W^-$ and $ZZ$, and a general reduction of all $\mathscr{C}_2$ values. Since the results in~\cite{Ashby-Pickering:2022umy} come without an estimate of the uncertainty, the comparison is not straightforward; dedicated work is needed to fully understand the origin of these discrepancies.\footnote{A revised version of \cite{Ashby-Pickering:2022umy} has since appeared and it now agrees with our estimates.}
  
\subsection{$p\, p \to W^+W^-$\label{sec:WW}}

The tree-level Feynman diagrams contributing to the process
\be
\bar{q}(p_1) q(p_2) \to W^+(k_1,\lambda_1) W^-(k_2,\lambda_2)\, ,
\label{qqWW}
\ee
at the parton level are shown in the top part of Fig.~\ref{fig:DYVV}.
The polarization vectors of $W^+$ and $W^-$ are $\varepsilon^{\mu}(k_1,\lambda_1)$ and $\varepsilon^{\nu}(k_2,\lambda_2)$, respectively.

The polarized amplitude for the process in \eq{qqWW}, for $u$ and $\bar u$ initial states, is given by
\bea
{\cal M}^{u\bar u}_{WW}(\lambdaA,\lambdaB)&=&-ie^2 \Big[\bar{v}(p_1) \Gamma^{\W\W}_{\mu\nu} u(p_2)\Big]
\varepsilon^{\mu}(k_1,\lambda_1)^{\star}\varepsilon^{\nu}(k_2,\lambda_2)^{\star}
\, ,
\label{MDYWW}
\eea
where the effective vertex $\Gamma^{WW}_{\alpha\beta}$ is
\bea
\Gamma^{\W\W}_{\mu\nu}&=&\frac{1}{s}
\left(\gamma^{\alpha} \bar{g}_V^q-\gamma^{\alpha}\gamma_5 \bar{g}_A^q\right)
V_{\alpha\nu\mu}(q,-k_2,-k_1)+\frac{1}{4 t \ssW^2}\gamma_{\nu}\left(\slashed{p}_2-\slashed{k}_1\right)\gamma_{\mu}(1-\gamma_5)\,,
\eea
with $\ssW= \sW$ and $e$ being the unit of electric charge. The effective couplings $\bar{g}^q_{V,A}$ are given by
\be
\bar{g}_V^q=Q^q+\frac{g_V^q\chi}{\ssW^2}\, ,~~
\bar{g}_A^q=\frac{g_A^q\chi}{\ssW^2}\, ,
~~ \chi=\frac{s}{2(s-M_Z^2)}\, ,
\label{effgva}
\ee
where  $g_{V}^q=T_3^q-2Q^q\ssW^2$, $g_{A}^q=T_3^q$ and $T_3^q$ and $Q^q$ are the isospin and electric charge (in unit of $e$) of the quark $q$. The $\chi$ term in \eq{effgva}, which weights the contribution of the virtual $Z$ 
channel, is real since we neglect the $Z$ width contribution. The function $V_{\alpha\nu\mu}(k_1,k_2,k_3)$ is the Feynman rule of the trilinear vertex $V_{\alpha}(k_1)~W^+_{\nu}(k_2)~W^-_{\mu}(k_3)$, $V\in\{\gamma ,Z\}$, given by 
\be
V_{\alpha\nu\mu}(k_1,k_2,k_3)=(k_1-k_2)_{\mu}g_{\alpha\nu}
+(k_2-k_3)_{\alpha} g_{\mu \nu}+(k_3-k_1)_{\nu} g_{\alpha\mu}\, ,
\label{V3}
\ee
for incoming momenta ($k_1+k_2+k_3=0)$. The Mandelstam variables are defined as
\be
s=(p_1+p_2)^2, \quad t=(p_2-k_1)^2, \quad u=(p_1-k_2)^2\,\,.
\ee

From the amplitude in Eq.~(\ref{MDYWW}), summing over the spin of quarks we obtain
\be
{\cal M}^{u\bar u}_{WW}(\lambdaA,\lambdaB) \Big[ {\cal M}^{u \bar u }_{WW}(\lambdaAp,\lambdaBp)\Big]^{\dag}\,=\,
\Tr\Big[\bar{\Gamma}^{\W\W}_{\mu\nu}\,\slashed{p}_1\,\Gamma^{\W\W}_{\mup\nup}\,\slashed{p}_2\Big]
 \mathscr{P}^{\mu\mup}_{\lambdaA\lambdaAp}(k_1)
 \mathscr{P}^{\nu\nup}_{\lambdaB\lambdaBp}(k_2)\, ,
 \label{M2WWpol}
\ee
where $\bar{\Gamma}_{\mu\nu}= \gamma_0 (\Gamma_{\mu\nu})^{\dag} \gamma_0$ and the projector $\mathscr{P}^{\mu\nu}_{\lambda\lambda^{\prime}}(k)$ is given in \eq{proj} with $M=M_W$.

The unpolarized square amplitude for the process $u\,\bar{u}\to W^+W^-$ is given by
\bea
|\xbar{{\cal M}}^{\; u\bar u}_{WW}|^{2}&=&
\frac{4\fWW}{(1 - \betaW^2)^2 \DW}
\Big\{4 + 
       16 \Ct\betaW + 
       \betaW^2 \big[9 + 11 \Ct^2 + 4 \betaW \Ct (1 - \Ct^2)
\nonumber\\ 
       &-&
          4 \betaW^3 \Ct (3 + \Ct^2) + \betaW^4 (1 - 5 \Ct^2)
          -2 \betaW^2 (5 + 3 \Ct^2 + 2 \Ct^4)\big]
\nonumber\\ 
&+&2 \betaW(1 + \betaW^2 + 2 \betaW \Ct) \Big[-8 \Ct +
             \betaW \big[-19 + 3 \Ct^2 + 2 \betaW^2 (9-\Ct^2) + 
             3 \betaW^4 (\Ct^2-1)
\nonumber\\ 
             &+& 2 \betaW \Ct (\Ct^2-1) + 
                2 \betaW^3 \Ct (3 + \Ct^2)\big]\Big] \big(\gAb + \gVb\big)\ssW^2
\nonumber\\ 
          &+&2 \betaW^2 (1 + \betaW^2 + 2 \betaW \Ct)^2 \big[19 - 3 \Ct^2 + 
             2 \betaW^2 (\Ct^2-9) + 3 \betaW^4 (1 - \Ct^2)\big] \big(\gAbb + 
\gVbb\big) \ssW^4\Big]\Big\}~~~~~~~~
  \label{M2WW}
\eea 
with
\be
 \fWW = \frac{4\pi^2\alpha^2 N_c}{\ssW^4\DW}\, ,
 \quad \text{and} \quad
 \DW =  1+\betaW ^2+2 \betaW \Ct\, ,
\label{fWW}
\ee          
where $N_c=3$, $\Ct=\cos{\Theta}$, $\betaW=\sqrt{1-4M_W^2/\mWW^2}$, $\mWW$ is the invariant mass of the $W$ pair and we chose $\Theta$ as the angle between the anti-quark and $W^+$ momenta in the CM frame. Our convention for the polarization density matrix is that the $W^+$ momentum defines the $\hat{k}$ unit vector of the basis in \eq{basis}.
   
The result for the $d\bar{d}\to W^+W^-$ process follows from \eqs{M2WWpol}{M2WW} through the substitutions 
\be
\gVb \to -\gVbD,\quad \gAb \to -\gAbD,\quad\betaW\to -\betaW
\, ,
\label{transfUD}
\ee
with the angle $\Theta$ being defined as before  by the anti-quark and $W^+$ momenta. The contribution of strange quark initial states equals that of $d$ quarks in the considered massless limit.

The \eq{M2WWpol} (together with the corresponding ones for $d\bar d$ and $s\bar s$ processes) makes it possible to compute the unnormalized correlation coefficients $\tilde{f}_a$, $\tilde{g}_a$, and $\tilde h_{ab}$ of the density matrix for the process at hand (given in Appendix~\ref{sec:a2}) and consequently, the value of the operators ${\cal I}_3$  and  $\cmb$. As explained in Section~\ref{sec:obs}, for the observable ${\cal I}_3$ we find at each point in the kinematic space the unitary matrices $U$ and $V$ that maximize the violation of Bell inequalities.

The results obtained for the two observables of interest are shown in Fig.~\ref{fig:WW}, as functions of the kinematic variables. We observe that the violation of the Bell inequalities takes place only in a limited range of the kinematic variables. The bin in which ${\cal I}_3>2$ is indicated by the hatched area in first plot of Fig.~\ref{fig:WW}. The matrices maximizing the Bell observable are given by
\be
U_W = \begin{pmatrix} 
  \dfrac{1}{50}-\dfrac{5 i}{9} & -\dfrac{1}{6}+\dfrac{3 i}{7} & -\dfrac{1}{13} +\dfrac{9 i}{13}  \\\\
  \dfrac{1}{4}-\dfrac{4 i}{7} &  \dfrac{2}{9}-\dfrac{5 i}{7}& \dfrac{1}{5}+\dfrac{i}{12}  \\\\
  \dfrac{2}{5}-\dfrac{2 i}{5} & -\dfrac{1}{9}+\dfrac{4 i}{9} & \dfrac{1}{3}-\dfrac{3 i}{5}\\
\end{pmatrix} \, , ~~~
V_W = \begin{pmatrix} 
  -\dfrac{1}{16}-\dfrac{4 i}{7} & -\dfrac{2}{11}+\dfrac{3 i}{7} & -\dfrac{1}{8} +\dfrac{2 i}{3}  \\\\
 - \dfrac{2}{13}+\dfrac{3 i}{5} & - \dfrac{3}{11}+\dfrac{5 i}{7} & -\dfrac{1}{5}-\dfrac{i}{13}  \\\\
  \dfrac{1}{3}-\dfrac{4 i}{9} & -\dfrac{1}{8}+\dfrac{3 i}{7} & \dfrac{3}{8}-\dfrac{3 i}{5}\\
\end{pmatrix} \, , ~~~
\label{UV}
\ee
with a precision of 1\% with respect to the numerical solutions we found. Accordingly, unitarity is satisfied barring $O(10^{-2})$ terms. These expressions might be useful in a future simulation of the process.

\begin{figure}[h!]
  \begin{center}
  \includegraphics[width=3.2in]{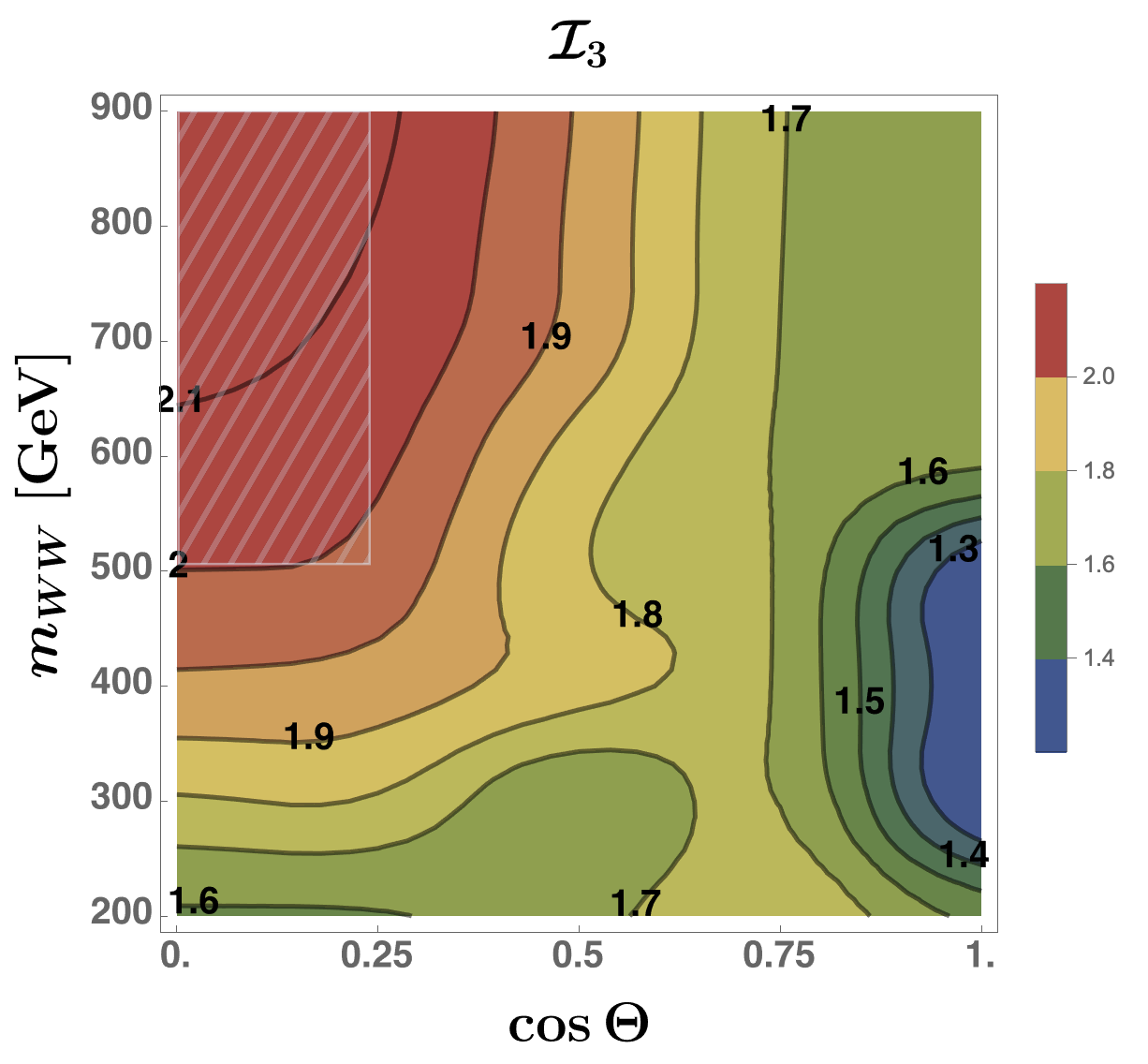}
  \includegraphics[width=3.16in]{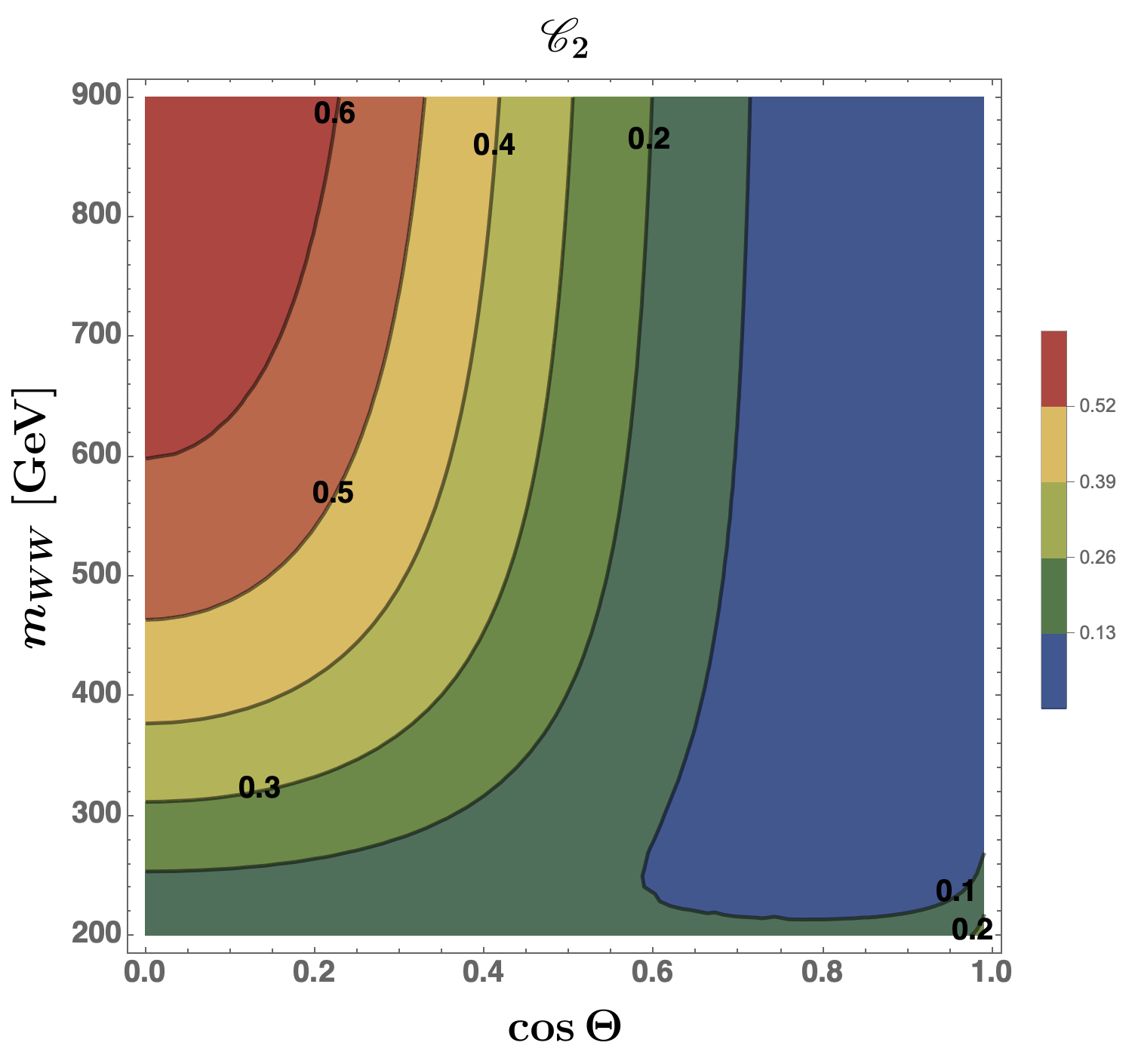}
  \caption{\small The observables ${\cal I}_3$ (left plot) and  $\cmb$ (right plot) for the process $p\;p\to W^+ W^-$ as functions of the invariant mass and scattering angle. The hatched area in the plot on the left represents the bin used as reference for our estimate of the significance.
  \label{fig:WW} 
  }
  \end{center}
  \end{figure}

The observable  $\cmb$   follows roughly the pattern of ${\cal I}_3$ and reaches the largest values in the upper-left quadrant, thus witnessing the presence of states more entangled than in the rest of the kinematic space. 
This feature can be made manifest by considering the density matrix of the process. For instance, at $m_{WW}=900$ GeV and $\cos \Theta=0$, the  polarization density matrix for the $W^+W^-$ states can be approximated
up to terms $O(10^{-3})$ by the following combination of pure state density matrices
\be
\rho= \alpha\, |\Psi_{+-}\rangle \langle \Psi_{+-}| + \beta\, |\Psi_{+-\,0}\rangle \langle \Psi_{+-\,0}| +
\gamma\, |00\rangle\langle 00| + \delta\, |\Psi_{0\,-}\rangle \langle \Psi_{0\,-}|\ ,
\label{density_WW}
\ee
with decreasing weights: $\alpha\simeq 0.72$, $\beta\simeq 0.18$, $\gamma\simeq 0.07$ and $\delta\simeq 0.02$;
the normalization condition $\alpha+\beta+\gamma+\delta=1$ is satisfied within the adopted approximation. The involved pure states are
\begin{eqnarray}
\nonumber
&&|\Psi_{+-}\rangle = \frac{1}{\sqrt{2}} \big( |++\rangle - |--\rangle \big)\ ,\\
&&|\Psi_{0\,-}\rangle = \frac{1}{\sqrt{2}} \big( |0\, -\rangle + |-0\rangle \big)\ ,\\
\nonumber
&&|\Psi_{+-\,0}\rangle = \frac{1}{\sqrt{3}} \big( |++\rangle - |--\rangle +|0\, 0\rangle \big)\ ,\\
\nonumber
\label{WW-states}
\end{eqnarray}
where $|a\, b\rangle= |a\rangle\otimes |b\rangle$ with $a,b\in \{+,\,0,\,-\}$ 
are the polarization states of the two $W$~gauge bosons at rest in the single spin-1 basis. Though the dominant contribution in (\ref{density_WW}) comes from the entangled pure state $|\Psi_{+-}\rangle$---a result that justifies the high value of $\cmb$---the actual density matrix $\rho$ describes a mixture, even more so if the discarded $O(10^{-3})$ terms were included. This feature explains why the corresponding value of $\cmb$, in this corner of the kinematic space, is large but far from maximal.

\subsubsection{Events and sensitivity}

Having identified the best region to test the data, we estimate the corresponding number of events expected at the LHC. This is given in Table~\ref{tab:events_WW}, where the cross sections needed for the estimates were computed with {\tt MADGRAPH5}~\cite{Alwall:2014hca} at the LO and then correcting by the $\kappa$-factor given at the  NNLO~\cite{Grazzini:2019jkl}. This is a good approximation, 
since there is little variation in the $k$-factors in the range of $WW$, $ZZ$, $WZ$ invariant masses between 200 and 800 GeV ~\cite{Grazzini:2019jkl}---which is the one we consider. We reduce the number of events thus found by the efficiency  in the identification of the final leptons---which we take conservatively to be 70\% for each lepton~\cite{Jain:2021bis}. We consider semi-leptonic decays of the $W$ and proceed as explained in Section~\ref{sec:hvv}.

\begin{table}[h!]
\bc
\vskip1cm
\begin{tabular}{ccc}
&\hskip0.5cm (run2) ${\color{oucrimsonred} {\cal L}=140\ \text{fb}^{-1}}$  \hskip0.5cm &  \hskip0.5cm (Hi-Lumi) ${\color{oucrimsonred} {\cal L}= 3\ \text{ab}^{-1}}$  \hskip0.5cm \\[0.2cm]
\hline\\
 \underline{events} \hskip0.4cm  &\hskip0.4cm $288$  \hskip0.4cm &\hskip0.4cm $6145$ \hskip0.4cm \\[0.4cm]
   \hline%
\end{tabular}
\caption{\small \label{tab:events_WW} Number of expected events in the kinematic region $m_{WW} > 500$ GeV and $\cos \Theta < 0.25$ at the LHC with   $\sqrt{s}=13$ TeV and luminosity ${\cal L}$ =140 fb$^{-1}$ (run2) and luminosity ${\cal L}$ = 3 ab$^{-1}$ (Hi-lumi). A benchmark efficiency of 0.70 is assumed for the identification of each lepton.}
\ec
\end{table}

Though there are irreducible background events from the $H\to W^+ W^-$ decay, they are few in the $\cos \Theta\leq 0.25$ bin where the observable is to be estimated. Events of the reducible background, coming from $t\bar t$ and $Wt$ production, must be selected out. 

\begin{figure}[h!]
\begin{center}
\includegraphics[width=2.9in]{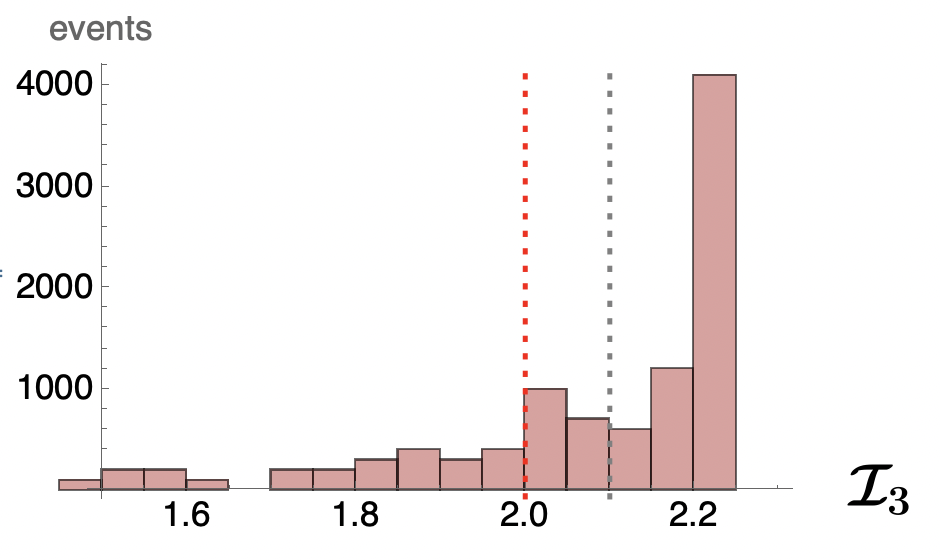}
\includegraphics[width=2.9in]{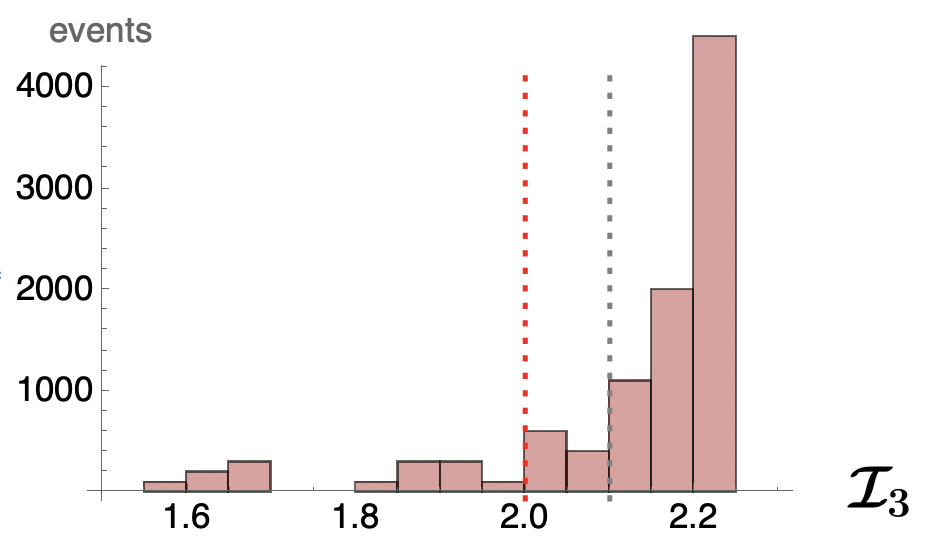}
\caption{\small Distribution of the  events of the  $W^+W^-$ process at the LHC run 2 (left) and Hi-lumi (right). The events have mean value ${\cal I}_3=2.1$ in both instances. The threshold value of 2 for Bell inequality violation is shown as a dashed red line.
\label{fig:eventsWW} 
}
\end{center}
\end{figure}

We run $10^{4}$ pseudo experiments as we vary the invariant mass and the scattering angle around the mean value with a dispersion given by the (statistical and systematic) uncertainty as discussed in the previous Section,  and compute the observable ${\cal I}_3$. Fig.~\ref{fig:eventsWW}  shows the distribution which is obtained for LHC run2 and Hi-Lumi. The distributions are skewed because the observable is computed near its maximum value and the random variation can only reduce this value.

\begin{figure}[h!]
\begin{center}
\includegraphics[width=2.9in]{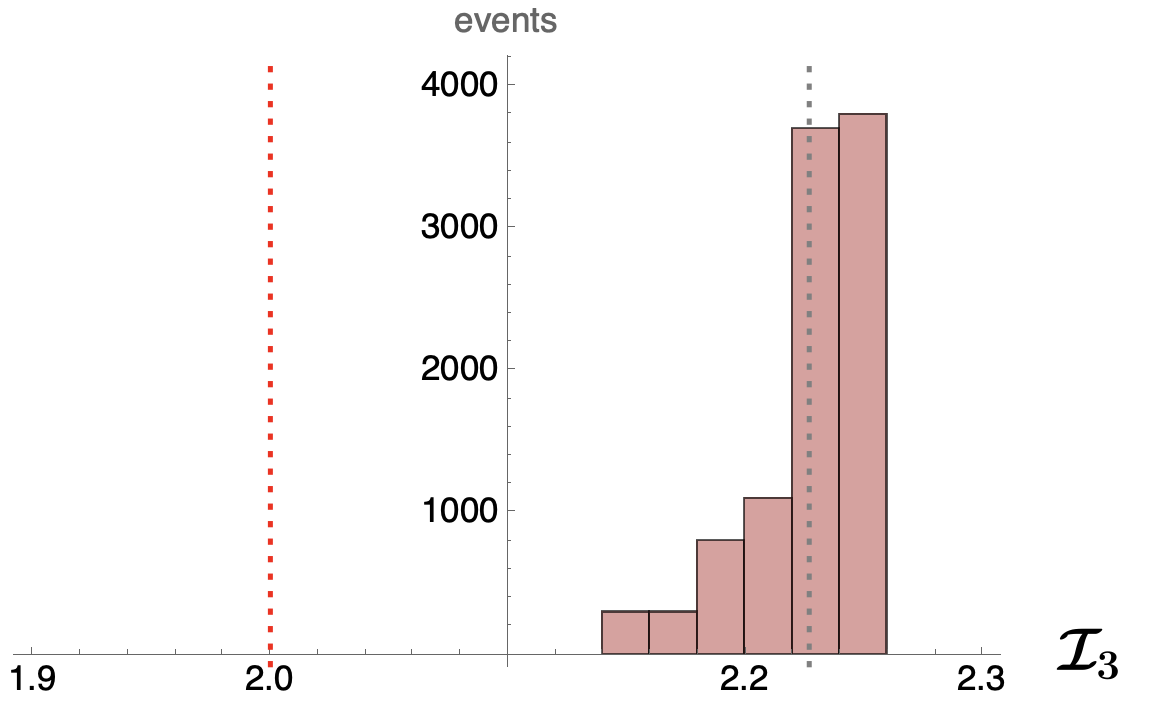}
\caption{\small Distribution of the  events of the  $W^+W^-$ process at the LHC run 2 for a systematic uncertainty of 0.1. The events have mean value ${\cal I}_3=2.23$. The threshold value of 2 for Bell inequality violation is shown as a dashed red line.
\label{fig:eventsWWsys} 
}
\end{center}
\end{figure}

We find that run2 yields a significance of 0.8 for rejecting the null hypothesis $ {\cal I}_3\leq 2$ (see Fig.~\ref{fig:eventsWW}) which remains about the same at Hi-Lumi because the uncertainty  is dominated by the systematic error. Fig.~\ref{fig:eventsWWsys} shows how the distribution of the pseudo-experiment changes as the systematic error is decreased to 0.1. In the latter case, the significance grows and reaches the value 6. Not surprisingly, the better the reconstruction of the neutrino momenta, the higher the significance of the violation of Bell inequality.

The significances we quote are bound to decrease  in a full simulation because of other systematic uncertainties and the smearing of the events in the detector.

\subsection{$p\,p \to ZZ$}

The tree-level Feynman diagrams contributing to the process
\bea
\bar{q}(p_1) q(p_2) &\to& Z(k_1,\lambda_1) Z(k_2,\lambda_2)\, ,
\label{qqZZ}
\eea
at the parton level are shown in the middle row of Fig.~\ref{fig:DYVV}. We indicate the polarization vectors of the two $Z$ bosons with 
$\varepsilon^{\mu}(k_1,\lambda_1)$ and $\varepsilon^{\nu}(k_2,\lambda_2)$.

The polarized amplitude for the process in \eq{qqZZ} is given by
\bea
{\cal M}^{q \bar q}_{ZZ}(\lambdaA,\lambdaB)&=&-\frac{ie^2}{4\ccW^2\ssW^2} \Big[\bar{v}(p_1) \Gamma^{\Z\Z}_{\mu\nu} u(p_2)\Big]
\varepsilon^{\mu}(k_1,\lambda_1)^{\star}\varepsilon^{\nu}(k_2,\lambda_2)^{\star}\, ,
\label{MDYZZ}
\eea
where
\be
\Gamma^{\Z\Z}_{\mu\nu}\,=\,V^q_{\mu}
\frac{(\slashed{k}_1-\slashed{p}_1)}{u}\,V^q_{\nu}
+
V^q_{\nu}
\frac{(\slashed{k}_1-\slashed{p}_2)}{t}V^q_{\mu}\, .
\label{GammaZZ}
\ee
The Mandelstam variables $u$ and $t$ are defined as
\be
u=(k_1-p_1)^2\, ,~~t=(k_1-p_2)^2\, ,
\ee
and
\be
V^q_{\mu}\,=\,g_V^q\gamma_{\mu} -g_A^q \gamma_{\mu}\gamma_5
\label{VZ}
\ee
with the $g_{V,A}^q$ couplings defined as in \eq{effgva}.

Summing over the quark polarizations and colors we then obtain
\be
{\cal M}^{q \bar q}_{ZZ}(\lambdaA,\lambdaB) \Big[{\cal M}^{q \bar q }_{ZZ}(\lambdaAp,\lambdaBp)\Big]^{\dag}\,=\,
\Tr\Big[\bar{\Gamma}^{\Z\Z}_{\mu\nu}\,\slashed{p}_1\,\Gamma^{\Z\Z}_{\mup\nup}\,\slashed{p}_2\Big]
 \mathscr{P}^{\mu\mup}_{\lambdaA\lambdaAp}(k_1)
 \mathscr{P}^{\nu\nup}_{\lambdaB\lambdaBp}(k_2)\, ,
 \label{M2ZZpol}
\ee
where $\mathscr{P}^{\mu\nu}_{\lambda\lambda^{\prime}}(k)$ is given in \eq{proj} with $M=M_Z$.

The corresponding unpolarized square amplitude is then obtained by summing over the polarizations of the two $Z$ bosons
\bea
|\xbar{{\cal M}}^{\;q \bar q}_{ZZ}|^{2}=
\frac{8\fZZ (\gAAAA + 6 \gAA \gVV + \gVVVV)
}{\DZ}\Big\{2 -
       \betaZ^2 \big[\betaZ^4 + (9 - 10 \betaZ^2 + \betaZ^4) \Ct^2 + 
         4 \betaZ^2 \Ct^4-3\big]\Big\}\, ,
\label{M2ZZ}
\eea
where
\be
\fZZ =  \frac{8\alpha^2\pi^2N_c}
      {\DZ\ccW^4 \ssW^4}\, ,
      \quad \text{and} \quad
      \DZ =  1 +\betaZ^4 + 2\betaZ^2 (1 - 2 \Ct^2)\, ,
\label{fZZ}
\ee
with $\betaZ=\sqrt{1-4M_Z^2/\mZZ^2}$. The angle $\Theta$ is here defined as the angle between the anti-quark momentum and $k_1$ in the CM frame. The orientation of the latter coincides with that of the $\hat{k}$ unit vector of the basis in \eq{basis}.

 \begin{figure}[h!]
\begin{center}
\includegraphics[width=3.1in]{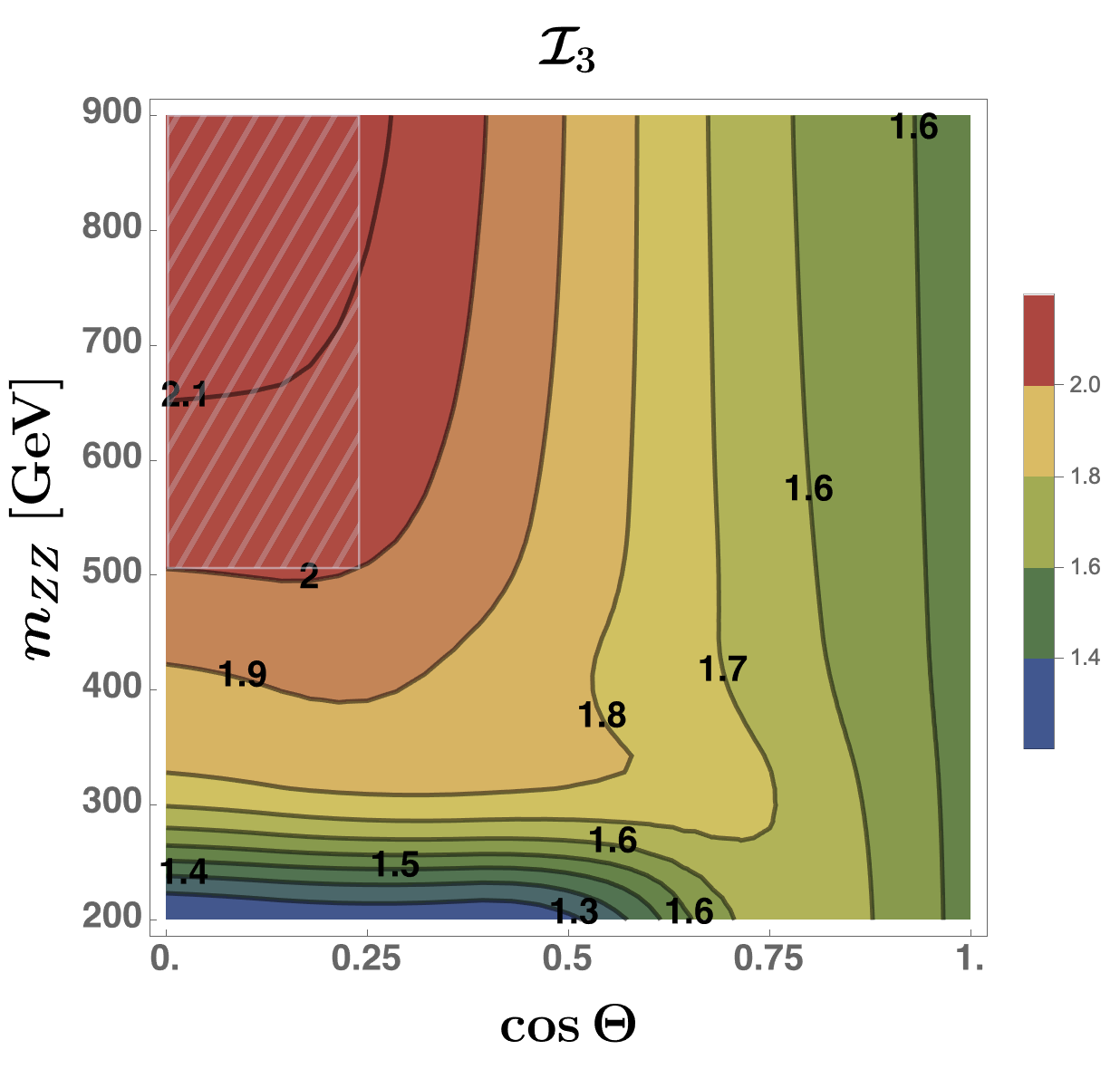}
\includegraphics[width=3.2in]{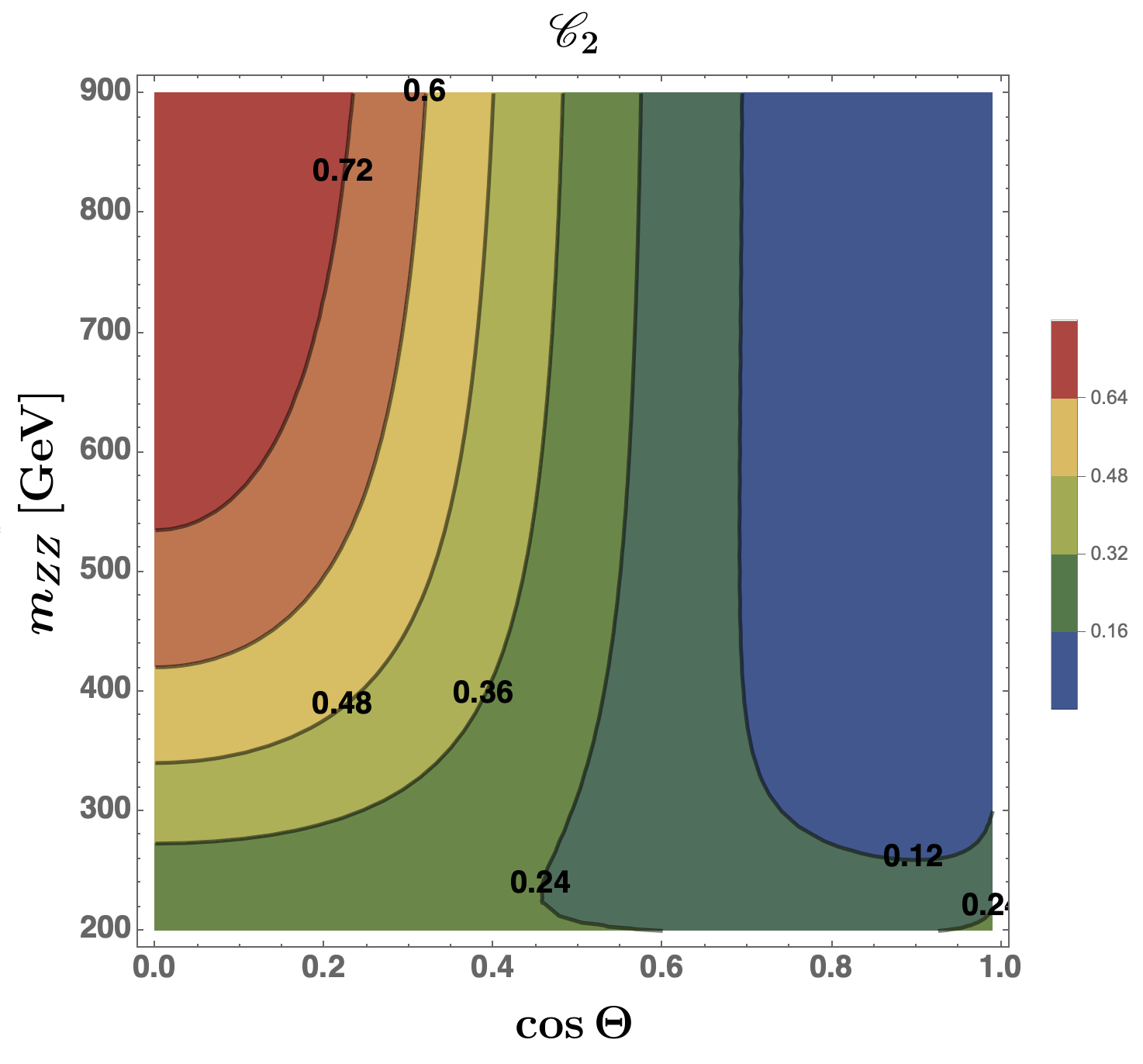}
\caption{\small The observables ${\cal I}_3$ (left plot) and  $\cmb$ (right plot) for the process $p\, p\to Z Z$ as functions of the invariant mass and scattering angle in the CM frame. The hatched area in the plot on the left indicates the bin in which the observable is to be evaluated.
\label{fig:ZZ} 
}
\end{center}
\end{figure}

The \eq{M2ZZpol} makes it possible through \eq{fgh} to compute the unnormalized correlation coefficients $\tilde{f}_a$, $\tilde{g}_a$, and $\tilde h_{ab}$ (given in Appendix~\ref{sec:a2}) of the density matrix for the process at hand and consequently, the value of the operators ${\cal I}_3$  and  $\cmb$.

In Fig.~\ref{fig:ZZ} we present our results for the entanglement observables. The violation of the Bell inequalities takes place only in a limited range of the kinematic variables. The bin in which ${\cal I}_3>2$ is shown as a hatched area in the left panel.

The observable $\cmb$ follows the pattern of ${\cal I}_3$---as it does in the case of the $W^+W^-$ final states---and reaches the largest values in the upper-left quadrant. In this region it witnesses the presence of states more entangled than in the rest of the kinematic space. 

\subsubsection{Events and sensitivity}

The number of expected events at the LHC is given in  Table~\ref{tab:events_ZZ}. As before, the relevant cross sections were computed with {\tt MADGRAPH5}~\cite{Alwall:2014hca} at the LO and then corrected by the $\kappa$-factor given at the NNLO~\cite{Grazzini:2019jkl}. We reduce the number of events thus found by the efficiency in the identification of the final leptons---which we take conservatively to be 70\% for each of the identified leptons~\cite{Jain:2021bis}. We consider semi-leptonic decays of the $W$ and proceed  as explained in Section~\ref{sec:hvv}.

Though there are irreducible background events from the $H\to ZZ$ decay, they are negligible in the kinematic bin where the observables are to be estimated. 
\begin{table}[h!]
\bc
\vskip1cm
\begin{tabular}{ccc}
&\hskip0.5cm (run2) ${\color{oucrimsonred} {\cal L}=140\ \text{fb}^{-1}}$  \hskip0.5cm &  \hskip0.5cm (Hi-Lumi) ${\color{oucrimsonred} {\cal L}= 3\ \text{ab}^{-1}}$  \hskip0.5cm \\[0.2cm]
\hline\\
 \underline{events} \hskip0.4cm  &\hskip0.4cm $4$  \hskip0.4cm &\hskip0.4cm $77$ \hskip0.4cm \\[0.4cm]
   \hline%
\end{tabular}
\caption{\small \label{tab:events_ZZ}Number of expected events in the kinematic region $m_{ZZ} > 500$ GeV and $\cos \Theta < 0.25$  at the LHC with   $\sqrt{s}=13$ TeV and luminosities ${\cal L}$ =140 fb$^{-1}$, run2, and ${\cal L}$ = 3 ab$^{-1}$ for Hi-lumi. A benchmark efficiency of 70\% is assumed in the identification of each charged lepton. }
\ec
\end{table}

\begin{figure}[h!]
\begin{center}
\includegraphics[width=2.9in]{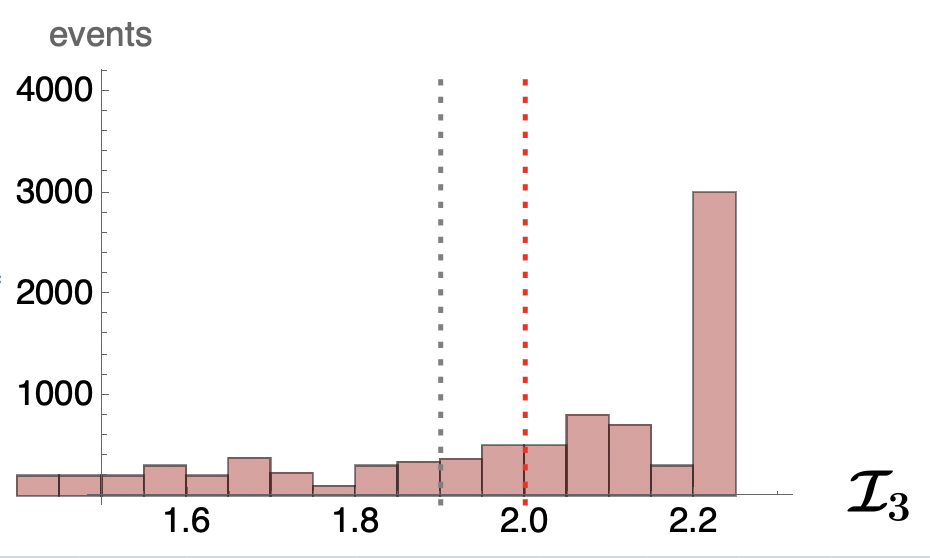}
\includegraphics[width=2.9in]{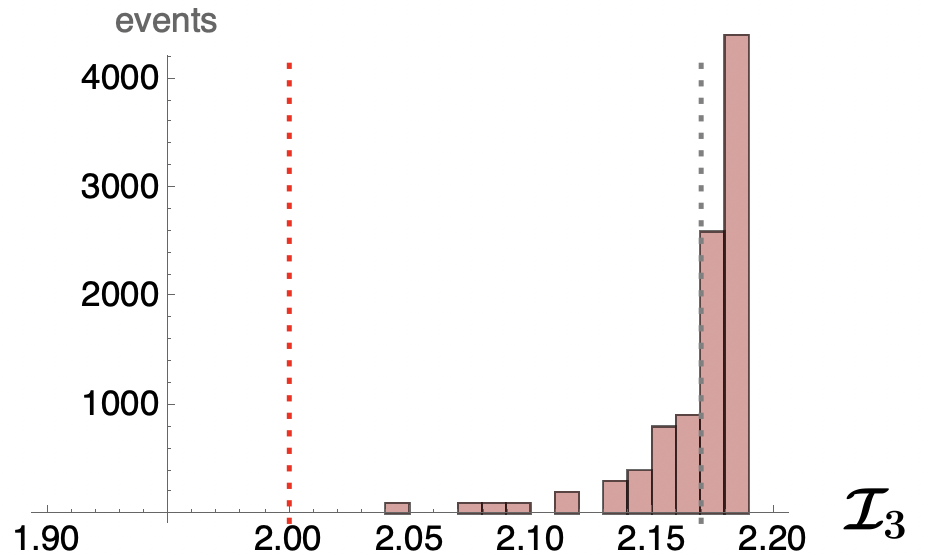}

\caption{\small Distribution of the  events of the  $ZZ$ process at the LHC run 2 (left) and Hi-lumi (right). The events have mean value ${\cal I}_3=1.9$  and $2.2$, respectively. The threshold value of 2 for Bell inequality violation is shown as a dashed red line.
\label{fig:eventsZZ} 
}
\end{center}
\end{figure}

We run $10^{4}$ pseudo experiments as we vary the invariant mass and the scattering angle  around the mean value with a dispersion given by the (statistical and systematic) uncertainty as discussed in the previous Section, and compute the observable ${\cal I}_3$. Fig.~\ref{fig:eventsZZ}  shows the distribution which is obtained for LHC run2 and Hi-Lumi. The distributions are skewed because the observable is computed near its maximum value and the random variation can only reduce this value.

We find that run2 yields an average value of  $ {\cal I}_3\leq 2$ that is below the threshold for Bell violation. At Hi-Lumi the significance  for rejecting the null hypothesis $ {\cal I}_3\leq 2$ (see Fig.~\ref{fig:eventsWW}) is  more than 2. 

The significance we quote is bound to decrease  in a full simulation because of the reconstruction from the final lepton angular distributions and the systematic uncertainties of the unfolding.

\subsection{$p\,p \to WZ$}
Let us consider the tree-level Feynman diagrams contributing to the process 
\be
\bar{d}(p_1) u(p_2) \to W^+(k_1,\lambda_1) Z(k_2,\lambda_2)\, ,
\label{duWZ}
\ee
at the partonic level, shown in the last row of Fig.~\ref{fig:DYVV}. We indicate the polarization vectors of the $W^+$ and $Z$ with $\varepsilon^{\mu}(k_1,\lambda_1)$ and $\varepsilon^{\nu}(k_2,\lambda_2)$, respectively. The polarized amplitude of the process is
\bea
{\cal M}^{u\bar d}_{WZ}(\lambdaA,\lambdaB)&=&-\frac{ie^2}{\sqrt{2}\ssW^2} \Big[\bar{v}(p_1) \Gamma^{\W\Z}_{\mu\nu} u(p_2)\Big]
\varepsilon^{\mu}(k_1,\lambda_1)^{\star}\varepsilon^{\nu}(k_2,\lambda_2)^{\star}\, ,
\label{MDYWZ}
\eea
where
\be
\Gamma^{\W\Z}_{\mu\nu}\,=\,\frac{\gamma^{\alpha}(1-\gamma_5)}{s-M_W^2}
V_{\nu\alpha\mu}(-k_2,q,-k_1)\ccW+\gamma_{\mu}(1-\gamma_5)
  \frac{\slashed{k}_1-\slashed{p}_1}{2t\,\ccW}V^u_{\nu}+
   V^d_{\nu}\frac{\slashed{k}_1-\slashed{p}_2}{2u\,\ccW}\gamma_{\mu}(1-\gamma_5)\, ,
\ee
with Mandelstam variables $s=(p_1+p_2)^2$, $t=(k_1-p_1)^2$, $u=(k_1-p_2)^2$, the vertex
$V^q_{\mu}$ ($q\in\{u,d\})$ being defined in \eq{VZ} and the vertex function $V_{\nu\alpha\mu}$ defined in \eq{V3}. We neglected  up-strange quark transitions by setting $\cos{\theta_c}=1$, with $\theta_c$ the Cabibbo angle. 
  
Summing over the internal degrees of freedom of the initial state quarks gives
\be
{\cal M}^{u \bar d}_{WZ}(\lambdaA,\lambdaB) \Big[ {\cal M}^{u \bar d}_{WZ}(\lambdaAp,\lambdaBp)\Big]^{\dag}\,=\,
\Tr\Big[\bar{\Gamma}^{\W\Z}_{\mu\nu}\,\slashed{p}_1\,\Gamma^{\W\Z}_{\mup\nup}\,\slashed{p}_2\Big]
 \mathscr{P}^{\mu\mup}_{\lambdaA\lambdaAp}(k_1,M_W)
 \mathscr{P}^{\nu\nup}_{\lambdaB\lambdaBp}(k_2,M_Z)\, .
 \label{M2WZpol}
\ee
where $\mathscr{P}^{\mu\nu}_{\lambda\lambda^{\prime}}(k,M_V)$ is defined as in \eq{proj} with $V\in\{W,Z\}$.

 \begin{figure}[h!]
\begin{center}
\includegraphics[width=3.2in]{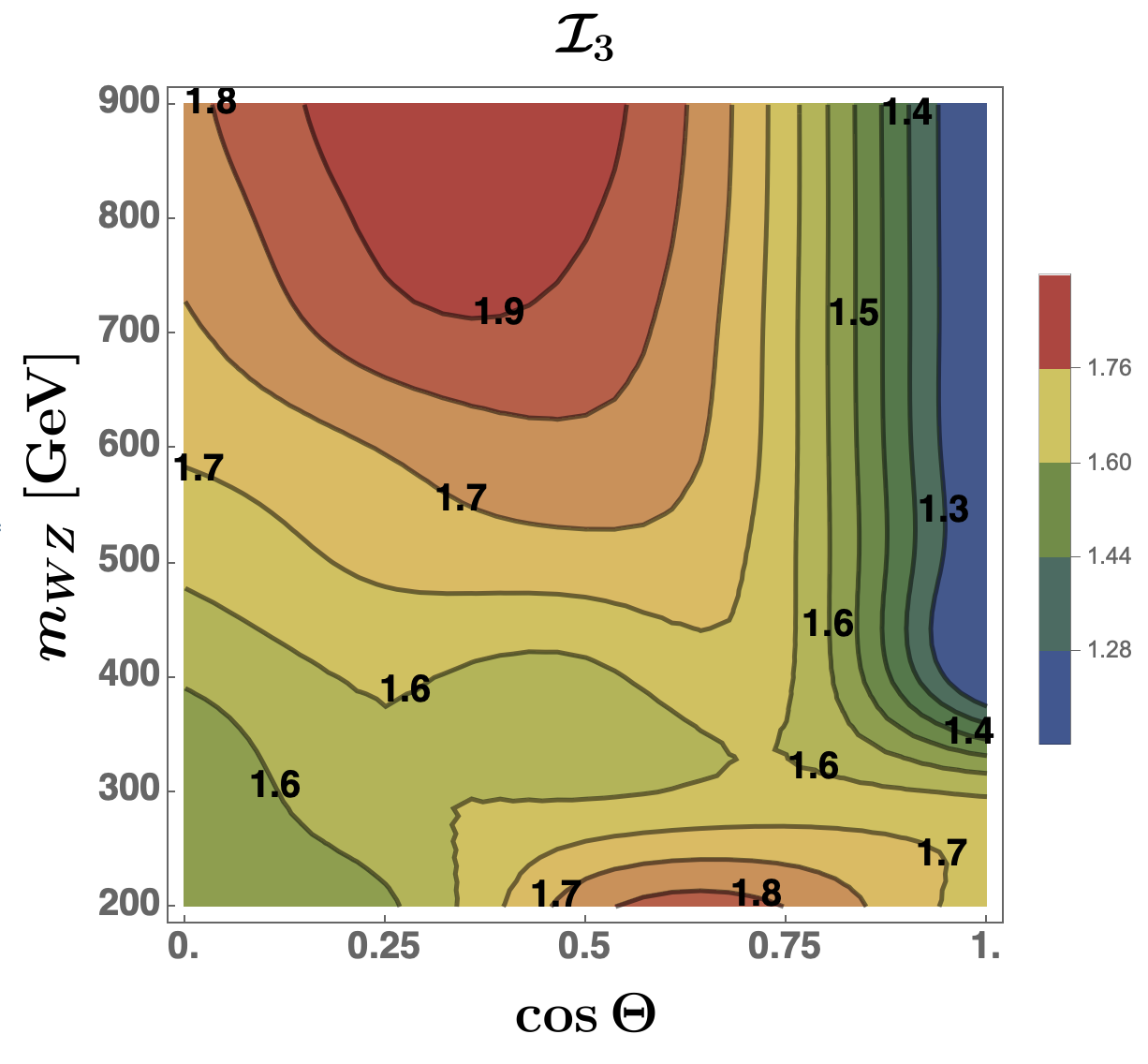}
\includegraphics[width=3.2in]{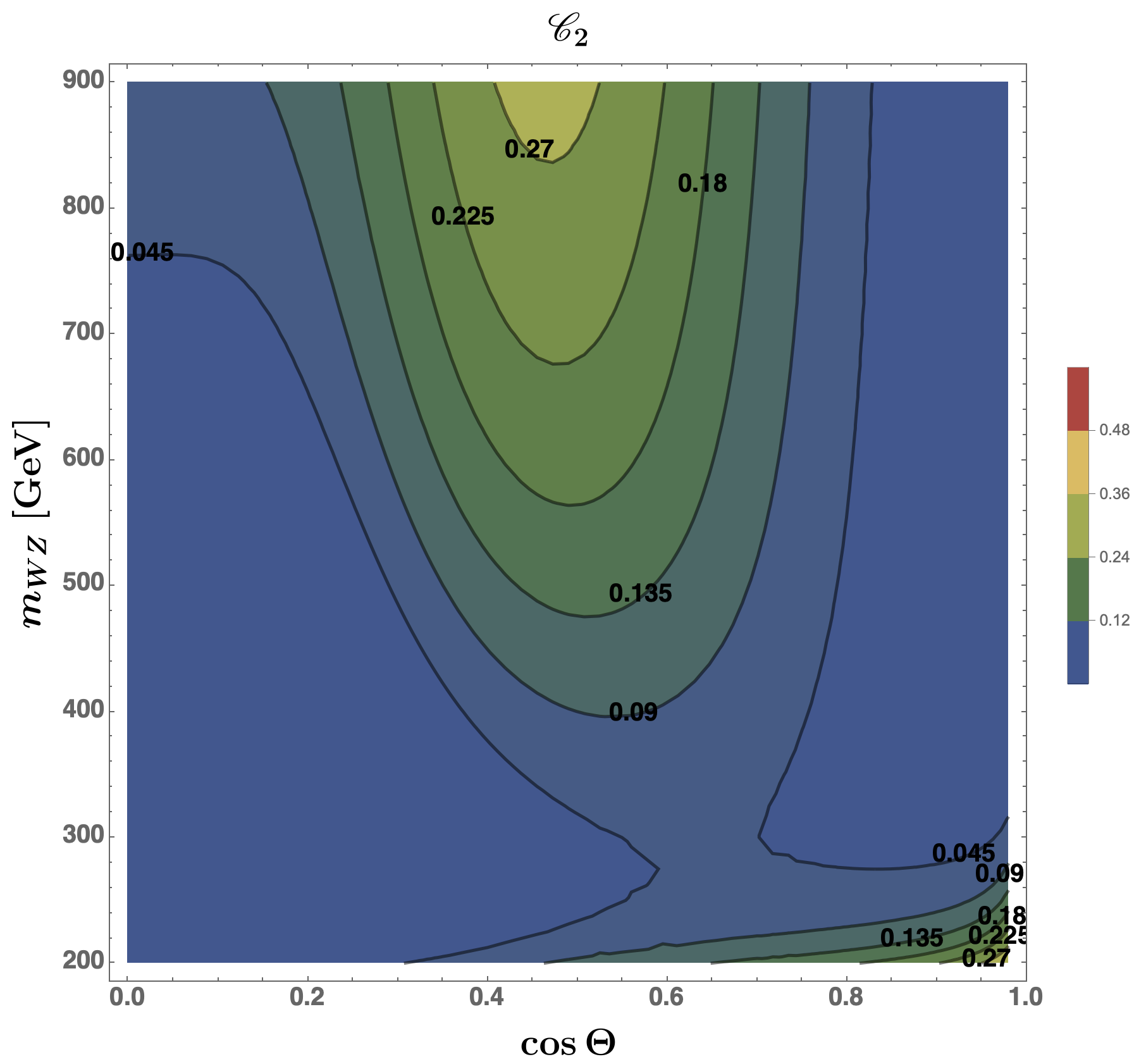}
\caption{\small The observables ${\cal I}_3$ (left plot) and  $\cmb$ (right plot) for the process $p\;p\to W^+ Z$  as a function of the invariant mass and scattering angle in the CM frame. 
\label{fig:WZ} 
}
\end{center}
\end{figure}
The following expressions approximate the quantity above in the limit  $M_W\simeq M_Z= M_V$ 
\bea
|\xbar{{\cal M}}^{\; u \bar d}_{WZ}|^{2}&=&
\frac{4\fWZ}{\DV}
\Big\{2 (3 + \betaV^2)^2 \Big[
     \betaV^2 \Big( \betaV^4 + (9 - 10 \betaV^2 + \betaV^4) \Ct^2 + 
     4 \betaV^2 \Ct^4-3\Big)-2\Big]
\nonumber\\
&-& 4 (3 + \betaV^2) \Big[(1 + \betaV^2)^2 (-6+\betaV^2 + \betaV^4) + 
24 \betaV (-1 - 2 \betaV^2 + \betaV^6) \Ct
\nonumber\\
&+& \betaV^2 (27 - 21 \betaV^2 - 7 \betaV^4 + \betaV^6) \Ct^2 - 
     120 \betaV^3 (-1 + \betaV^2) \Ct^3 + 4 \betaV^4 (3 + \betaV^2) \Ct^4 + 
     48 \betaV^5 \Ct^5\Big]\ccW^2 
\nonumber\\
     &-&     \Big[(1 + \betaV^2)^2 (36 + 177 \betaV^2 - 
        170 \betaV^4 + 25 \betaV^6) + 
        96 \betaV (1 + \betaV^2) (3 + \betaV^2) (1 + \betaV^2 - \betaV^4) \Ct
\nonumber\\
&+& \betaV^2 (243 + 756 \betaV^2 + 498 \betaV^4 - 748 \betaV^6 - 29 \betaV^8)
\Ct^2 + 480 \betaV^3 (-3 + 2 \betaV^2 + \betaV^4) \Ct^3
\nonumber\\
&+& 8 \betaV^4 (-333 + 336 \betaV^2 + 35 \betaV^4) \Ct^4 - 
192 \betaV^5 (3 + \betaV^2) \Ct^5 - 1296 \betaV^6 \Ct^6\Big]\ccW^4\Big\}\,,
\label{M2WZ}
\eea
where
\be
\fWZ= \frac{8\alpha^2\pi^2N_c}
      {9(3+\betaV^2)^2
        \DV\ccW^2 \ssW^4},
 \qquad 
 \DV= 1 +\betaV^4 + 2\betaV^2 (1 - 2 \Ct^2)\, ,
 \label{fWZ}
\ee
with $\betaV=\sqrt{1-4M_V^2/\mVV}$. The angle $\Theta$ is here implied by the momenta of anti-down quark and $W$ in the CM frame.
As before, our convention for the polarization density matrix for the $WZ$ production is that the momentum of $W$ is along $\hat{k}$, \textit{cf.} \eq{basis}. Analogous results hold for the process $p\;p\to W^- Z$ initiated by the $\bar u d$ quarks. 

We compute the unnormalized correlation coefficients $\tilde{f}_a$, $\tilde{g}_a$, and $\tilde h_{ab}$ (given explicitly in Appendix~\ref{sec:a2}) of the density matrix by using \eqs{M2WZpol}{fgh}. Fig.~\ref{fig:WZ} shows the values obtained for the observables ${\cal I}_3$ (left panel) and  $\cmb$ (right panel) for the process $p\;p\to W Z$. By inspection, the observable ${\cal I}_3$ is less than 2 regardless of the value of the kinematic variables. The final states are less entangled than in the case of the weak gauge boson pairs and the observable $\cmb$ presents low values everywhere.

\subsection{Lepton colliders}

We consider now the charged di-boson production at $e^+ e^-$ and muon colliders, proceeding from the process
 \bea
 \ell^+(p_1) \ell^-(p_2) &\to& W^+(k_1,\lambda_1)\, W^-(k_2,\lambda_2)\, ,
 \eea
where $\ell\in\{e,\mu\}$. We neglect the contribution of an intermediate Higgs boson regarding the leptons as massless.

\begin{figure}[h!]
\begin{center}
\includegraphics[width=3.3in]{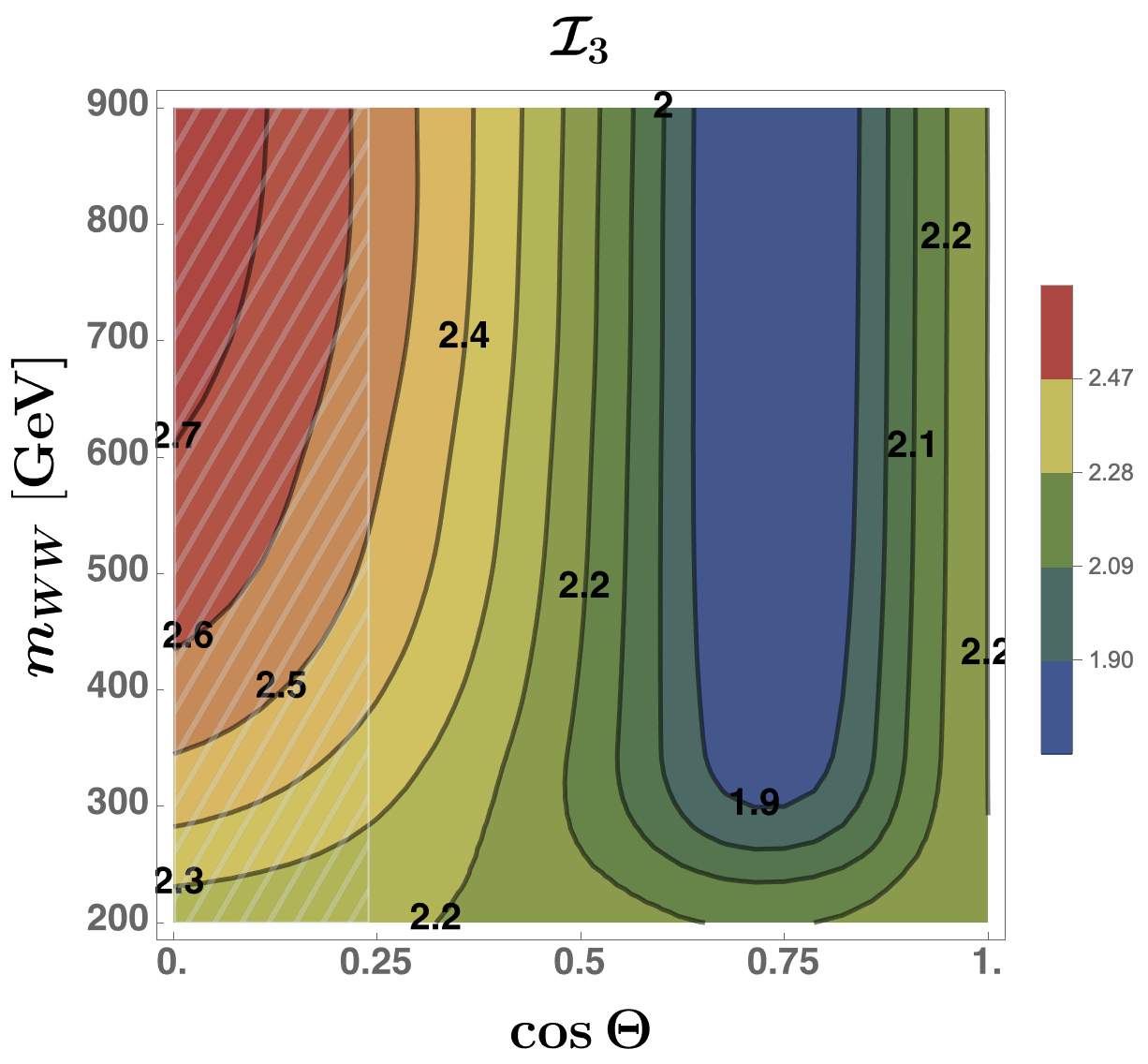}
\includegraphics[width=3.3in]{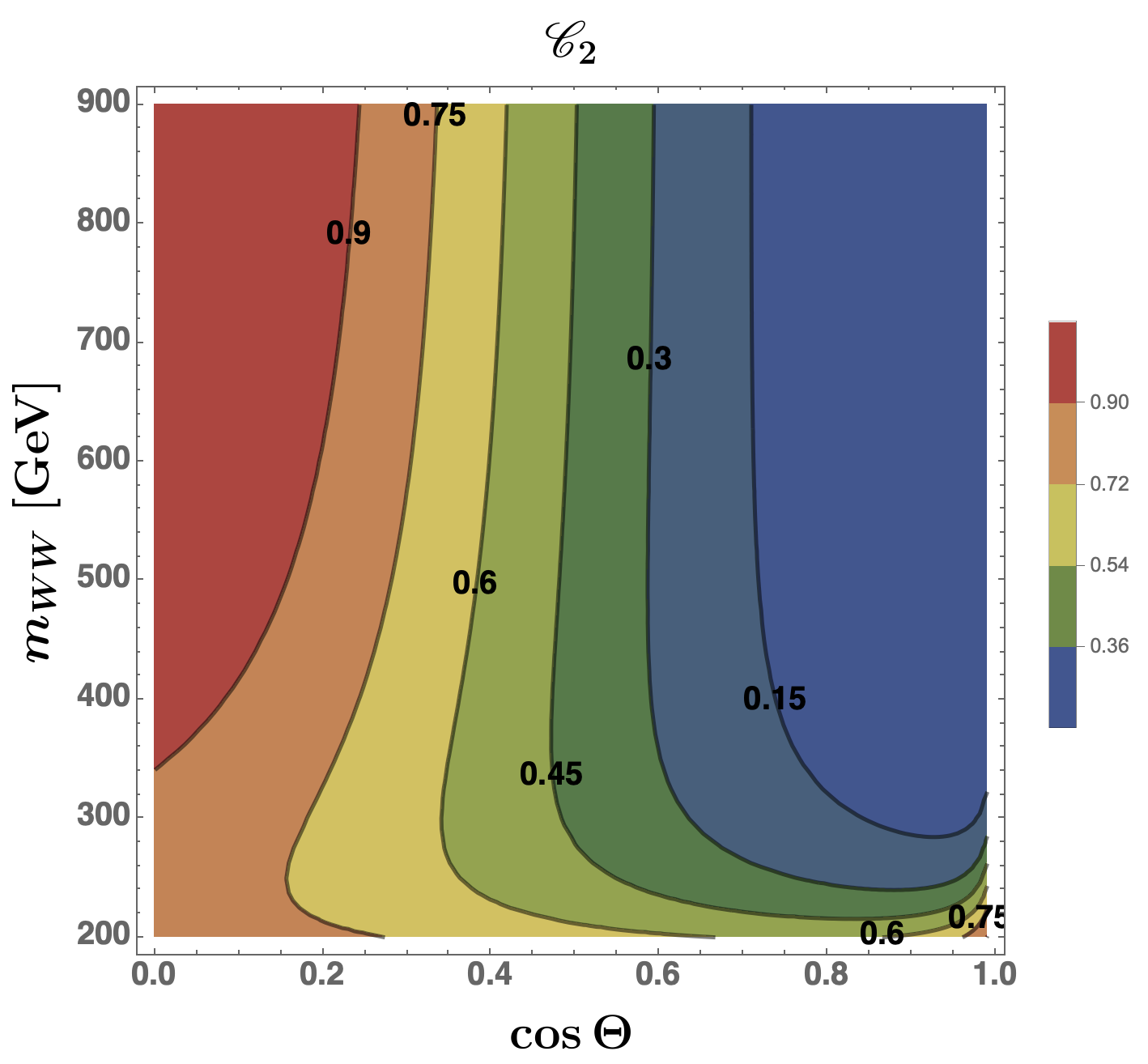}
\caption{\small The observables ${\cal I}_{3}$ and $\cmb$ for the process $\ell^+\ell^-\to W^+ W^-$  as functions of the invariant mass and scattering angle in the CM frame. The hatched area in the plot on the left represents the bin in which the observable ${\cal I}_{3}$ is to be evaluated.
\label{fig:WWmuon} 
}
\end{center}
\end{figure}
 
\begin{figure}[h!]
\begin{center}
\includegraphics[width=3.2in]{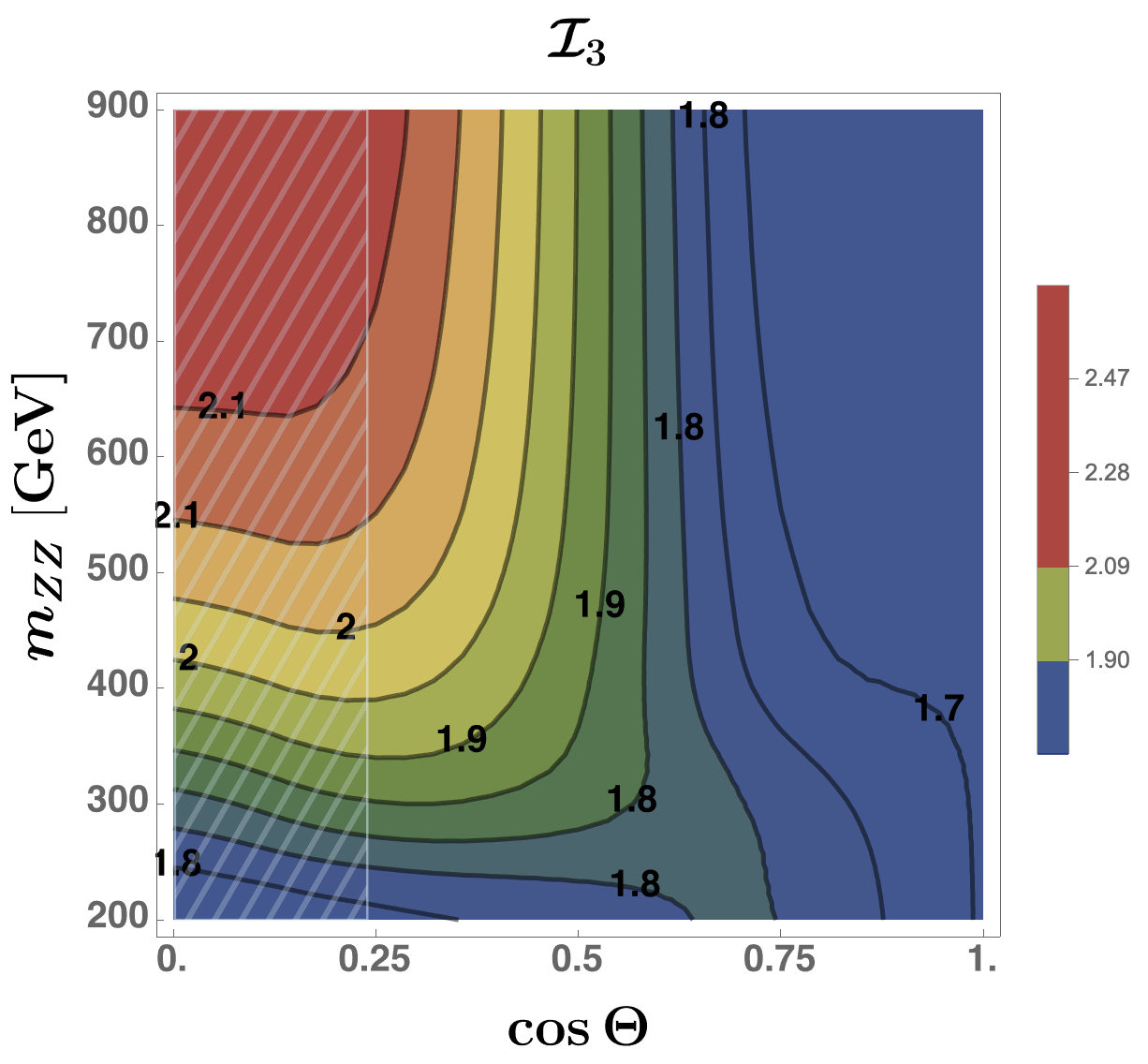}
\includegraphics[width=3.3in]{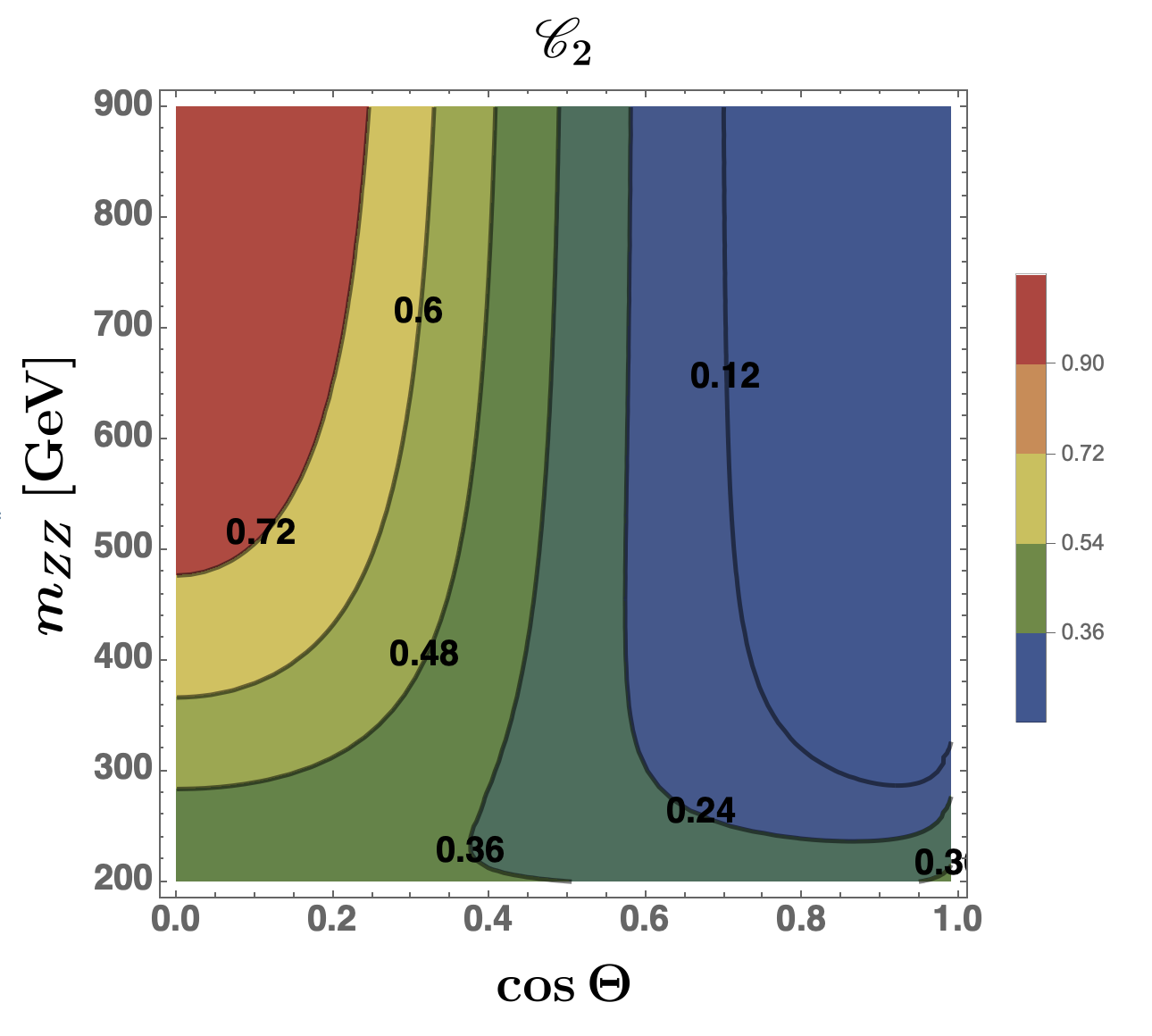}
\caption{\small The observables ${\cal I}_{3}$ and $\cmb$ for the process $\ell^+\ell^-\to Z Z$  as functions of the invariant mass and scattering angle in the CM frame. The hatched area in the plot on the left represents the bin in which the observable ${\cal I}_{3}$ is to be evaluated.
\label{fig:ZZmuon} 
}
\end{center}
\end{figure}
 
The analytical results for the amplitude and the polarization density matrix coefficients can be obtained from those given in Section~\ref{sec:WW} and Appendix~\ref{sec:a2} through the replacements $\bar{g}_{V,A}^d\to\bar{g}_{V,A}^{\ell}$. Because the initial state is unique, the total density matrix comprises only one contribution. For the correlation coefficients $h_{ab}$, $f_{a}$, $g_{a}$  we then find 
\bea
h_{ab} [\mWW,\Theta]&=& \frac{
\htee_{ab}[\mWW,\Theta]}{\Aee[\mWW,\Theta]}
\, ,
\nonumber\\
f_{a} [\mWW,\Theta]&=& \frac{\ftee_{a}[\mWW,\Theta]}
{\Aee[\mWW,\Theta]}  
\, ,
\nonumber\\         
g_{a} [\mWW,\Theta]&=& \frac{\gtee_{a}[\mWW,\Theta]}{\Aee[\mWW,\Theta]}
\, ,
\label{rholepton}
\eea
where the scattering angle $\Theta$ is defined as the angle between the anti-lepton and $W^+$ momenta.

The results for the entanglement observables are shown in Fig.~\ref{fig:WWmuon}. The violation of the Bell inequalities takes place  in a  range of the kinematic variables broader than in the LHC case and it is larger. The theoretical uncertainty of the result is negligible.
The same results  for the $ZZ$ di-bosons are shown in Fig.~\ref{fig:ZZmuon}.  The violation of the Bell inequalities in this case takes place  in a  range of the kinematic variables more or less equivalent to that at the LHC 

In this instance, as for  the $WZ$ process, the process is generated by only one kind of diagram (see bottom diagrams of Fig.~\ref{fig:DYVV}) and the PDF dependence exactly cancels out in the $h_{ab}$  coefficients in \eq{habDY}, as well as in the  $f_a,g_a$ ones. This PDF factorization in the density matrix for the $WZ$ production at the LHC takes always place at the lepton colliders, where  no dependence on the PDF appears.

\subsubsection{Events and sensitivity}

The bin in which ${\cal I}_3>2$ for lepton colliders is shown as a hatched area in  Fig.~\ref{fig:WWmuon}.  
Having identified the best region to confront the data, we can estimate the number of events expected at a muon collider working at an energy of $\sqrt{s}=1$ TeV and at the future circular collider (FCC) working at an energy of $\sqrt{s}=368$ GeV. These numbers are given in Table~\ref{tab:events_muon}, where the relevant cross sections were computed with {\tt MADGRAPH5}~\cite{Alwall:2014hca} at the LO. We reduce the number of events thus found by the efficiency  in the identification of the final leptons---which we take conservatively to be 70\% per lepton as we did for the LHC. We consider semi-leptonic decays of the $W$ and proceed as explained in Section~\ref{sec:hvv}.

It is  premature to discuss any background---except for stressing that at $\sqrt{s}=1$ TeV the leptons initiated production is 10 times that of vector boson fusion (see, for example,~\cite{Costantini:2020stv}).

\begin{table}[h!]
\bc
\begin{tabular}{ccc}
&\hskip0.5cm ${\color{oucrimsonred} \ell^+ \nu_\ell \, j j}$  \hskip0.5cm &  \hskip0.5cm${\color{oucrimsonred} \ell^- \ell^{+} \ell^- \ell^+}$ \hskip0.5cm \\[0.2cm]\hline\\
 \underline{muon} \hskip0.5cm (${\color{oucrimsonred} {\cal L}=1\ \text{ab}^{-1}}$) \hskip0.4cm  &\hskip0.4cm  $5.7\times 10^3$   \hskip0.4cm &\hskip0.4cm  $44$ \hskip0.4cm \\[0.4cm]
    \underline{FCC} \hskip0.5cm (${\color{oucrimsonred} {\cal L}=1.5\ \text{ab}^{-1}}$) \hskip0.4cm  &\hskip0.4cm  $9.2\times 10^{4}$   \hskip0.4cm &\hskip0.4cm  $748$ \hskip0.4cm \\[0.4cm]
   \hline%
\end{tabular}
\caption{\small \label{tab:events_muon} Number of expected events in the kinematic region $m_{WW} > 200$ GeV and $\cos \Theta < 0.25$ for a muon collider with   $\sqrt{s}=1$ TeV and luminosity ${\cal L}$ =1 ab$^{-1}$ and FCC with   $\sqrt{s}=364$ GeV and luminosity ${\cal L}$ = 1.5 ab$^{-1}$.  A benchmark efficiency of 70\% is assumed in the identification of each charged lepton.}
\ec
\end{table}

\begin{figure}[h!]
\begin{center}
\includegraphics[width=2.9in]{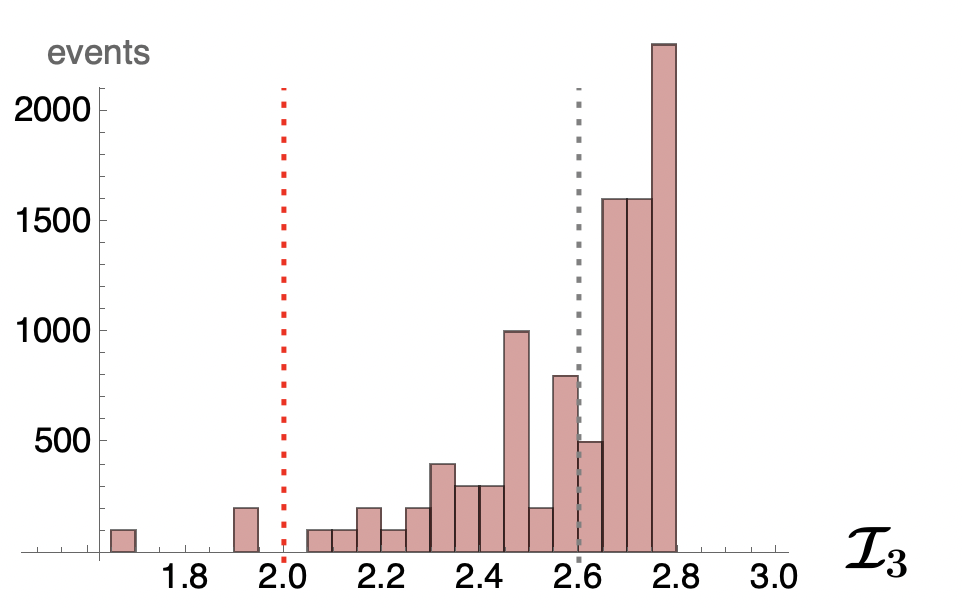}
\includegraphics[width=2.9in]{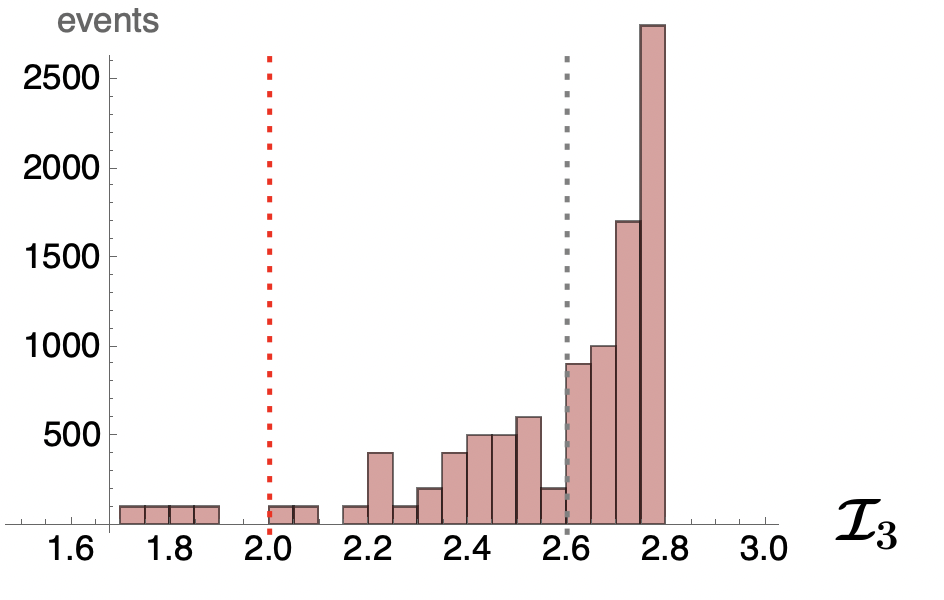}
\caption{\small Distribution of the  events  (muon collider, left, FCC, right) in the  $W^{+}W^{-}$ process. The events have mean value ${\cal I}_3=2.6$. The threshold value of 2 for Bell inequality violation is shown as a dashed red line.
\label{fig:eventsWWmuon} 
}
\end{center}
\end{figure}

\begin{figure}[h!]
\begin{center}
\includegraphics[width=2.8in]{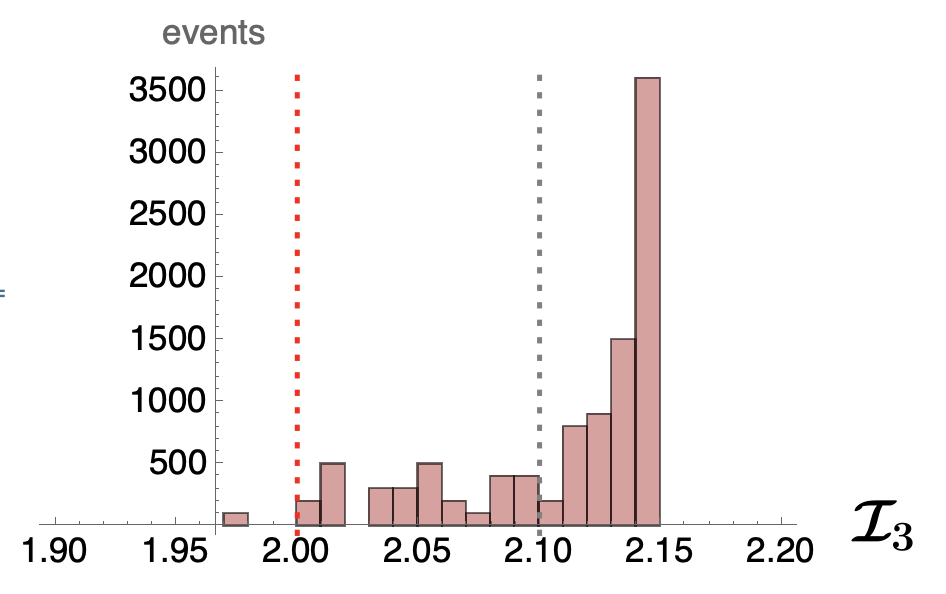}
\includegraphics[width=2.85in]{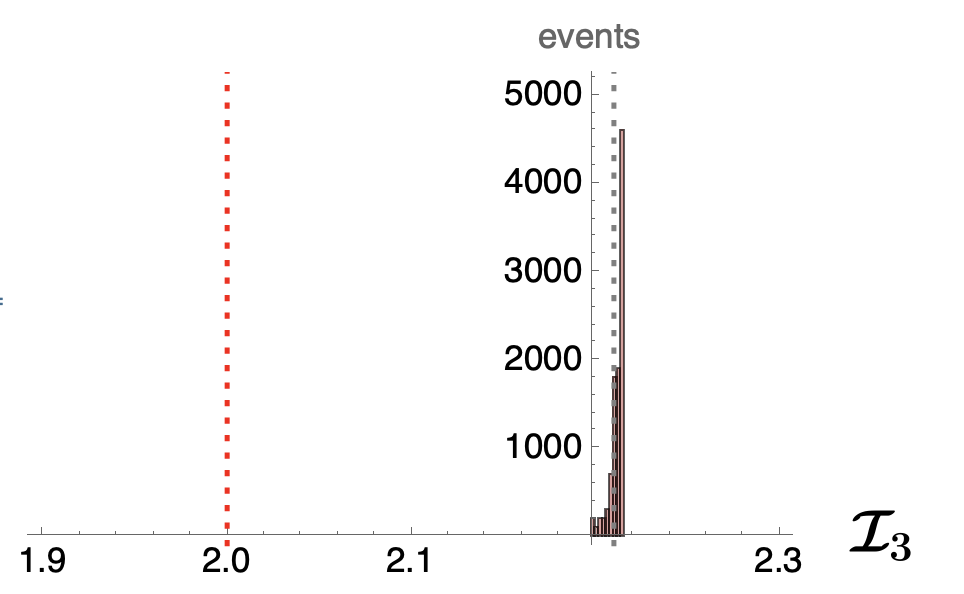}
\caption{\small Distribution of the  events  (muon collider, left, FCC, right) in the  $ZZ$ process. The events have mean value ${\cal I}_3=2.17$. The threshold value of 2 for Bell inequality violation is shown as a dashed red line.
\label{fig:eventsZZmuon} 
}
\end{center}
\end{figure}

In the case of $WW$  di-bosons, both the future  muon collider  and the FCC can provide a significance equal to 2 for rejecting the null hypothesis ${\cal I}_{3}\leq 2$ (see Fig.~\ref{fig:eventsWWmuon}). In the case of $ZZ$  di-bosons, the future  muon collider  can provide a significance equal to 2 for rejecting the null hypothesis ${\cal I}_{3}\leq 2$,  the FCC---which is expected to produce many more events---a significance of more than  4   (see Fig.~\ref{fig:eventsZZmuon}).

A more realistic estimate of these  numbers can only be provided by a full numerical simulation. In particular, the systematic uncertainties of the unfolding because of the presence of the neutrinos, and the  background events may further decrease these significances.

\section{Summary\label{sec:sum}}

We have computed the value of two observables---$\cmb$ and ${\cal I}_3$, linked respectively to quantum entanglement and violation of Bell inequalities---in  processes yielding two weak interaction gauge bosons in the final state. These particles, being spin-1 and massive, are qutrit states and, as such, more complicated to treat than the more ordinary qubit states implemented with fermions and photons.

We find that the most promising processes for testing Bell inequalities and the presence of entanglement are, by far, those in which the final gauge bosons result from the decay of a Higgs boson.  In this case the null hypothesis that the Bell inequalities be satisfied can be excluded using the data of Hi-Lumi at the LHC  with a significance of  6 for a $ZZ$ final state. The  systematic uncertainty in the reconstruction of the neutrino momenta in the case $WW$ final states   makes it very hard to reach a satisfactory significance in this channel.  We hope that these provisional results encourage the experimental collaborations to  estimate the actual significance in a full simulation. In our opinion this is where Bell inequalities violation stands the best chance of being observed at energies around the weak scale.

The same observables can also be measured in the di-boson production initiated by electroweak quark fusion, reminiscent of the Drell-Yan processes. In this case, the invariant mass required to achieve significant values of the observables is rather large and only few events are expected at the LHC. These processes will become more competitive at future lepton colliders, with both the FCC and the muon collider  reaching a  significance of about  2  in testing the violation of Bell inequalities with $WW$ di-bosons. A better result is expected in the case of $ZZ$ di-bosons, in particular at the FCC.

\section*{Acknowledgements}
LM was supported by the European Regional Development Fund through the CoE program grant TK133 and by the Estonian Research Council grant PRG356. The authors thank Kristjan Kannike for useful comments.

\newpage
\appendix
\section{Spin and Gell-Mann matrices\label{sec:a1}}

The spin-1 representation of the three $SU(2)$ generators $S_i$, $i\in\{1,2,3\}$, used throughout the text is 
\be
S_1 =  \frac{1}{\sqrt{2}} \begin{pmatrix} 0& 1 & 0 \\1& 0 & 1 \\ 0 & 1 &0 \end{pmatrix}\,, 
\qquad  
S_2 =  \frac{1}{\sqrt{2}} \begin{pmatrix} 0& -i & 0 \\i& 0 & -i \\ 0 & i &0 \end{pmatrix}\,, 
\qquad S_3 = \begin{pmatrix} 1& 0 & 0 \\0& 0 & 0 \\ 0 & 0 &-1 \end{pmatrix} \, .
\ee
These can be expressed as a function of the Gell-Mann matrices $T^a$ as
\be
\label{eq:s3}
S_1=\frac{1}{\sqrt{2}} \Big( T^1 + T^6 \Big)\, , \quad S_2=\frac{1}{\sqrt{2}} \Big( T^2 + T^7 \Big)\,,\quad 
S_3=\frac{1}{2} \, T^3 + \frac{\sqrt{3}}{2} \,T^8  \, , 
\ee
and the matrices $S_{ij}$ in \eq{Sij} are:
\bea
S_{31} =S_{13}&=& \frac{1}{\sqrt{2}} \, \Big( T^1 - T^6 \Big)\, ,  \nn \\
S_{12} =S_{21}&=& T^5 \, , \nn \\
S_{23} =S_{32}&=& \frac{1}{\sqrt{2}} \, \Big( T^2 - T^7 \Big)  \nn \\
S_{11} &=&   \frac{1}{2 \sqrt{3}} \, T^8 +T^4 - \frac{1}{2} \,T^3 \, , \nn \\
S_{22} &=&  \frac{1}{2 \sqrt{3}} \, T^8 -  T^4 - \frac{1}{2} \, T^3  \, , \nn \\
S_{33} &=&     T^3 -  \frac{1}{\sqrt{3}} \,T^8 \, , 
\eea
with $\mathbb{1}$ being the $3\times3$ unit matrix. The Gell-Mann matrices $T^a$ are: 
\bea
T^1 = \begin{pmatrix} 0& 1 & 0 \\1& 0 & 0 \\ 0 & 0 &0 \end{pmatrix}\,, & \quad  T^2 = \begin{pmatrix} 0& -i & 0 \\i& 0 & 0 \\ 0 & 0 &0 \end{pmatrix}\,, & \quad T^3 = \begin{pmatrix} 1& 0 & 0 \\0& -1 & 0 \\ 0 & 0 &0 \end{pmatrix}\,, \nn \\
T^4 = \begin{pmatrix} 0& 0 & 1 \\0& 0 & 0 \\ 1 & 0 &0 \end{pmatrix}\,, &\quad T^5 =\begin{pmatrix} 0& 0 & -i \\0& 0 & 0 \\ i & 0 &0 \end{pmatrix}\,, & \quad T^6 =\begin{pmatrix} 0& 0 & 0 \\0& 0 & 1 \\ 0 & 1&0 \end{pmatrix}\,, \nn \\
T^7 =\begin{pmatrix} 0& 0 & 0 \\0& 0 & -i \\ 0 & i &0 \end{pmatrix}\,, &\quad T^8 = \dfrac{1}{\sqrt{3}} \begin{pmatrix} 1& 0 & 0 \\0& 1 & 0 \\ 0 & 0 &-2 \end{pmatrix} \, .
\eea

\section{The functions $\mathfrak{q}^n_\pm$  and $\mathfrak{p}^n_\pm$  and the matrix   $\mathfrak{a}^{n}_{m}$\label{sec:Aqp}}

In this Appendix we follow~\cite{Ashby-Pickering:2022umy}.
The  $\mathfrak{q}^n_\pm$ functions introduced in Section~\ref{sec:qp} are given by the following expressions
\bea
 \mathfrak{q}^1_\pm &=&  \dfrac{1}{\sqrt{2}} \, \sin \theta^\pm \Big( \cos \theta^\pm \pm 1\Big) \cos \phi^\pm \,,\nn\\
 \mathfrak{q}^2_\pm&=&   \dfrac{1}{\sqrt{2}} \, \sin \theta^\pm \Big( \cos \theta^\pm \pm 1 \Big) \sin \phi^\pm  \,,\nn\\
 \mathfrak{q}^3_\pm&=& \dfrac{1}{8} \,  \Big( 1 \pm 4 \,   \cos \theta^\pm  + 3 \cos 2 \theta^\pm  \Big) \,,\nn\\
 \mathfrak{q}^4_\pm&=&  \dfrac{1}{2} \, \sin^{2} \theta^\pm \cos 2 \,\phi^\pm\,, \nn\\
 \mathfrak{q}^5_\pm&=&  \dfrac{1}{2} \, \sin^{2} \theta^\pm \sin 2 \,\phi^\pm  \,,\nn\\
 \mathfrak{q}^6_\pm&=&  \dfrac{1}{\sqrt{2}} \, \sin \theta^\pm \Big(- \cos \theta^\pm \pm 1\Big) \cos \phi^\pm \,,\nn\\
 \mathfrak{q}^7_\pm&=&  \dfrac{1}{\sqrt{2}} \, \sin \theta^\pm \Big( - \cos \theta^\pm \pm 1 \Big) \sin \phi^\pm \,, \nn\\
 \mathfrak{q}^8_\pm&=&  \dfrac{1}{8\sqrt{3}} \,  \Big(- 1 \pm  12   \cos \theta^\pm  -3 \cos 2 \theta^\pm  \Big)  \, , \label{Q}
\eea
in terms of the spherical coordinates of the two decaying particle rest frames. 

The $\mathfrak{p}^n_\pm$ functions utilized in Section~\ref{sec:qp}  are given  by the following expressions:
\bea
 \mathfrak{p}^1_\pm &=&  \sqrt{2} \, \sin \theta^\pm \Big( 5\, \cos \theta^\pm \pm 1\Big) \cos \phi^\pm \,,\nn\\
 \mathfrak{p}^2_\pm&=&  \sqrt{2} \, \sin \theta^\pm \Big( 5\, \cos \theta^\pm \pm 1 \Big) \sin \phi^\pm  \,,\nn\\
 \mathfrak{p}^3_\pm&=& \dfrac{1}{4} \,  \Big( 5 \pm 4 \,   \cos \theta^\pm  + 15\,  \cos 2 \theta^\pm  \Big) \,,\nn\\
 \mathfrak{p}^4_\pm&=&  5 \, \sin^{2} \theta^\pm \cos 2 \,\phi^\pm \,,\nn\\
 \mathfrak{p}^5_\pm&=& 5 \, \sin^{2} \theta^\pm \sin 2 \,\phi^\pm  \,,\nn\\
 \mathfrak{p}^6_\pm&=& \sqrt{2} \, \sin \theta^\pm \Big(- 5\, \cos \theta^\pm \pm 1\Big) \cos \phi^\pm \,,\nn\\
 \mathfrak{p}^7_\pm&=&  \sqrt{2}\, \sin \theta^\pm \Big( -5 \cos \theta^\pm \pm 1 \Big) \sin \phi^\pm  \,,\nn\\
 \mathfrak{p}^8_\pm&=&  \dfrac{1}{4\sqrt{3}} \,  \Big(- 5 \pm  12   \cos \theta^\pm  -15 \cos 2 \theta^\pm  \Big)  \, . \label{P}
\eea

The matrix $\mathfrak{a}^{n}_{m}$ used in Section~\ref{sec:qp} is the following
\be
\mathfrak{a}^{n}_{m}= \dfrac{1}{g_L^2-g_R^2} \begin{pmatrix} g_{R}^2&0&0&0&0&g_L^2&0&0\\
0& g_{R}^2&0&0&0&0&g_L^2&0\\
0&0&  g_{R}^2-\frac{1}{2}\, g_L^2&0&0&0&0&\frac{\sqrt{3}}{2}\, g_L^2\\
0&0&  0&g_{R}^2- g_L^2&0&0&0&0\\
0&0&  0&0& g_{R}^2- g_L^2&0&0&0\\
g_L^2&0&  0&0&0& g_{R}^2&0&0\\
0& g_L^2&0&  0&0&0& g_{R}^2&0\\
0&0&\frac{\sqrt{3}}{2}\, g_L^2&0&0&0&0&\frac{1}{2}\, g_L^2-g_R^2
\end{pmatrix} \label{Anm}\, .
\ee
The coefficients in \eq{Anm} are $g_{L} =-1/2 + \sin^2 \theta_{W}\simeq -0.2766$  and $g_{R}= \sin^2 \theta_{W}\simeq 0.2234$.


\section{Analytic expressions for the density matrices\label{sec:a2}}
{\small 
\subsection{Polarization density matrix for $q\,\bar{q}\to W^+W^-$}

We write below the expressions for the coefficients
 $\Aqq [\Theta,\mVV]$, $\ftqq_a[\Theta,\mVV],\, \gtqq_a[\Theta,\mVV]$, and $\htqq_{ab}[\Theta,\mVV]$, with $q=u,d$, appearing in the polarization density matrix for $q\,\bar{q}\to W^+W^-$. The angle $\Theta$ is the scattering angle in the CM frame from the anti-quark and $W^+$ momenta.  Our convention for the polarization matrix is that the  momentum of $W^+$ is chosen parallel to the $\hat{k}$ unit vector of  the spin right-handed basis in \eq{basis} and
\be
A^{u \bar u}=|\xbar{{\cal M}}_{\W\W}^{\;u\bar u}|^2
\ee
where the expression for the unpolarized square amplitude $|\xbar{{\cal M}}_{\W\W}^{\;u \bar u}|^2$ is given in \eq{M2WW}. Throughout the following expressions we use $\Ct\equiv \cos{\Theta}$, $\St\equiv \sin{\Theta}$ and $\DW$, $\fWW$ which are given in \eq{fWW}.

The non-vanishing elements $\htuu_{ab}$ ($\htuu_{ba}=\htuu_{ab}$)
are given by
\vspace{0.5cm}
\bea
\htuu_{11}[\Theta,\mWW]&=&\frac{\fWW}{(1-\betaW^2)}
\Bigg\{
\big(\betaW
   +1\big)^2
   (\Ct+1)^2
   \nonumber\\
   &-&2\betaW  (\Ct+1)^2 
   \Big[\betaW  \big(\betaW  \Ct+\betaW
   +4\big)+\Ct+1\Big]
   \left(\bar{g}_A^{u}+\bar{g}^{u}_V\right)\ssW^2
\nonumber\\
&+&8
   \betaW ^2\DW
   \Big[\big(1+\Ct^2\big)\left(
   \bar{g}_A^{\U 2}+\bar{g}^{\U 2}_V\right)+4 \Ct
   \bar{g}_A^{\U} \bar{g}^{\U}_V\Big]\ssW^4
   \Bigg\}\, ,
   \nonumber\\
   \nonumber\\
  \htuu_{15}[\Theta,\mWW]&=&
\frac{\fWW\sqrt{2}\,\St}
{\sqrt{1 - \betaW^2}}
\Bigg\{(1 + \Ct)(1+\betaW) 
     -\betaW(1 + \Ct) \big(1 + 4\betaW + \betaW^2 + 2 \Ct\big) \left(\gAb + \gVb\right) \ssW^2
   \nonumber\\
     &+& 
      4 \betaW^2 \DW
   \Big[2 \gAb \gVb + 
           \Ct \left(\gAbb + \gVbb\right)\Big]\ssW^4\Bigg\}
   \nonumber\\
   \nonumber\\
\htuu_{16}[\Theta,\mWW]&=&
\fWW\St^2
  \Bigg\{1 -
         2 \betaW \left(2 \betaW + \Ct\right)\left(\gAb + \gVb\right) \ssW^2
        + 
        4 \betaW^2\DW 
        \left(\gAbb + \gVbb\right) \ssW^4\Bigg\}
   \nonumber\\
   \nonumber\\
  \htuu_{22}[\Theta,\mWW]&=&\frac{\fWW}{1-\betaW^2}
     \Bigg\{(1 + \Ct)^2 \Big[1 + \betaW^4 -
       2 \betaW^3 (\Ct-3)+ 8 (\Ct-1) \Ct + 
  \betaW^2 (2 + 4 \Ct) + 2 \betaW (7 \Ct-5)\Big] \nonumber\\
&+& 
2 \betaW \Big[5 - \betaW \left(4 + 3\betaW-\Ct\betaW\right) + 7 \Ct\Big]
(1 + \Ct)^2 \DW
\left(\gAb + \gVb\right) \ssW^2
\nonumber\\
&+& 
8 \betaW^2 \DW^2
\Big[(1 + \Ct^2) \gAbb + 
     4 \Ct \gAb \gVb + (1 + \Ct^2) \gVbb\Big] \ssW^4\Bigg\}
\nonumber\\
\nonumber\\
  \htuu_{23}[\Theta,\mWW]&=&\frac{\fWW\sqrt{2}\St}
         {(1-\betaW^2)^{3/2}\DW}
         \Bigg\{
(1 + \Ct) \big[\betaW (2 + \betaW) + 2 \Ct-1\big] \big[  
   \betaW^2 (1 + \betaW + \Ct)-1 -3\betaW-3 \Ct   \big]
\nonumber\\
&-&
\betaW (1 + \Ct) \DW
\Big[1 - 12 \Ct  
  -12\betaW + \betaW^2 \big(4\betaW + \betaW^2 + 4 \Ct-2\big)\Big]
\big(\gAb + \gVb\big) \ssW^2
\nonumber\\
&+&
4 \betaW^2 (\betaW^2-3) \DW^2
\Big[2 \gAb \gVb + 
    \Ct (\gAbb + \gVbb)\Big] \ssW^4
\Bigg\}
\eea
\bea
  \htuu_{24}[\Theta,\mWW]&=&
-\frac{\fWW\sqrt{2}\,\St}
{\sqrt{1 - \betaW^2}\DW}
\Bigg\{(1 + \Ct) \big[\betaW^2 (3 + \betaW)
+(3\betaW+2\Ct)(2\Ct-1)-1\big]
\nonumber\\
&-& \betaW (1 + \Ct) ( 4\betaW + \betaW^2 + 6 \Ct-3)
\DW(\gAb + \gVb) \ssW^2
\nonumber\\
&+&
4 \betaW^2 \DW^2 \Big[2 \gAb \gVb + 
  \Ct \left(\gAbb + \gVbb\right)\Big]\ssW^4\Bigg\}
\nonumber\\
\nonumber\\
\htuu_{27}[\Theta,\mWW]&=&-\frac{\fWW\,\St^2}
{(1 - \betaW^2)\DW}
\Bigg\{1 - 12 \betaW^2 + 3 \betaW^4 - 18 \betaW \Ct + 2 \betaW^3 \Ct - 8 \Ct^2
\nonumber\\
&+&
2 \betaW \DW
\Big[2 \betaW (5 - \betaW^2) + (9 - \betaW^2) \Ct\Big] (\gAb + \gVb)\ssW^2
\nonumber\\
&+&
4 \betaW^2 (\betaW^2-5) \DW^2 (\gAbb + \gVbb)\ssW^4
\Bigg\}
\nonumber\\
\nonumber\\
\htuu_{28}[\Theta,\mWW]&=&-\frac{\sqrt{2}\fWW\,\St}
      {\sqrt{3}(1 - \betaW^2)^{3/2}\DW}\Bigg\{
      -1 +\betaW^5 (3 - \Ct) + 3 \betaW^4 (1 - \Ct)^2
+\betaW\big(7 + 3\Ct - 16 \Ct^2\big)
\nonumber\\
&+& 2 \betaW^2 \Ct\big(\Ct^2-2\Ct-7\big) + 
\Ct (4 + 3 \Ct - 6 \Ct^2) - 2 \betaW^3 \big(5 +5\Ct- 4 \Ct^2\big)
\nonumber\\
&+&
\betaW \DW\Big[-7 + \betaW^3 (4 - 12 \Ct) + 
\betaW^4 (\Ct-3) + 4 \betaW (1 + 5 \Ct)
\nonumber\\
&+& \Ct (16 \Ct-3) + 
     2 \betaW^2 (5 + 5\Ct - 4 \Ct^2)\Big] (\gAb + \gVb)\ssW^2
\nonumber\\
&-&
4 \betaW^2 \DW^2
\Big[2 (1 + \betaW^2) \gAb \gVb + (5 - 
        3 \betaW^2) \Ct (\gAbb + \gVbb)\Big]\ssW^4
\Bigg\}
\nonumber\\
\nonumber\\
\htuu_{33}[\Theta,\mWW]&=&-\frac{\fWW}{(1 - \betaW^2)^2\DW}\Bigg\{
-2(1 + \Ct) \Big[2\betaW + \betaW^2(\betaW-1)(\betaW + 3 \betaW^2 + 
  \betaW^3-3)
\nonumber\\
&+&
     4  \Ct\big[1 + \betaW(\betaW-1)  (2 + \betaW) \big( 
     2\betaW + \betaW^2-1\big)\big]
\nonumber\\
&+&\Ct^2 \big(3 
        -20\betaW  -2\betaW^2 + 12\betaW^3 + 3\betaW^4\big)  +(3 - 
\betaW^2)^2 \Ct^3\Big]
\nonumber\\
&+&2 \betaW (1 + \Ct)\DW \Big[2 + 
  5\betaW + \betaW^2 (2 + \betaW) (\betaW^2-2) 
\nonumber\\
&+&\Ct\big(4- 13\betaW - 2\betaW^2 +10\betaW^3 +2\betaW^4 -\betaW^5\big) - 
2 (5 - 3 \betaW^2) \Ct^2\Big](\gAb + \gVb) \ssW^2
\nonumber\\
&+&2 \betaW^2 \DW^2 \Big[
16 (1 - \betaW^2) \Ct \gAb \gVb
+  \big[2 \betaW^2 - \betaW^4-5
+ (13 - 10 \betaW^2 + \betaW^4) \Ct^2\big] (\gVbb+\gAbb) \Big] \ssW^4
\Bigg\}
\nonumber\\
\nonumber\\
\htuu_{34}[\Theta,\mWW]&=&-\frac{\fWW\St^2}{(1 - \betaW^2)}\Bigg\{
1 + \betaW^4 - 2 \Ct - 6 \Ct^2 + 2 \betaW^3 (1 + \Ct) - 2 \betaW (1 + 5 \Ct) +
2 \betaW^2 (\Ct + \Ct^2-3)
\nonumber\\
     &+&2 \betaW \DW \Big[1 + 5 \Ct - 
        \betaW \big(\betaW + \betaW \Ct-4)\Big] (\gAb + \gVb) \ssW^2 - 
     8 \betaW^2 \DW^2(\gAbb + 
        \gVbb)\ssW^4
        \Bigg\}
\eea
\bea
\htuu_{37}[\Theta,\mWW]&=&-\frac{\sqrt{2}\fWW\St}{(1 - \betaW^2)^{3/2}\DW}
\Bigg\{
 2 \betaW^5 + \betaW^4 \Ct (5 + \Ct) + \betaW (2 + 4 \Ct - 14 \Ct^2) + 
 \Ct (3 + \Ct - 6 \Ct^2)
 \nonumber\\
 &+& \betaW^3 (4 \Ct + 6 \Ct^2-8) + 
 2 \betaW^2 (2 - 8 \Ct + \Ct^2 + \Ct^3)
 -2 \betaW\DW\Big[2 + \betaW^4 + \betaW (4 - 8 \Ct)
\nonumber\\
   &+& 
4 \betaW^3 \Ct + (2 - 7 \Ct) \Ct + \betaW^2 \big[\Ct (2 + 3 \Ct) -3\big]\Big](\gAb +    \gVb) \ssW^2
 \nonumber\\
&+& 8 \betaW^2 \DW^2 \Big[2 \gAb \gVb + (\betaW^2-2) \Ct (\gAbb + \gVbb)\Big]\ssW^4
 \Bigg\}
\nonumber\\
\nonumber\\
\htuu_{38}[\Theta,\mWW]&=&\frac{\fWW}{\sqrt{3}(1 - \betaW^2)^2\DW}
\Bigg\{
2 + 4 \Ct - 9 \Ct^2 - 6 \Ct^3 + 9 \Ct^4 + 2 \betaW^6 (\Ct + \Ct^2-2)
\nonumber\\
&+& 
 4 \betaW^5 (1 - 3 \Ct + \Ct^2 + \Ct^3) - 
 4 \betaW^3 (2 - 5 \Ct - 2 \Ct^2 + 5 \Ct^3) + 
 4 \betaW (1 - 4 \Ct - 3 \Ct^2 + 6 \Ct^3)
\nonumber\\
 &+& 
 \betaW^4 (10 + 8 \Ct - 21 \Ct^2 + 2 \Ct^3 + \Ct^4) - 
 2 \betaW^2 (6 + 7 \Ct - 14 \Ct^2 - 2 \Ct^3 + 3 \Ct^4)
\nonumber\\
&+&2 \betaW (1 - \Ct) \DW \Big[-2 + 
   7\betaW + \betaW^2 (2 - 3 \betaW) (2 - \betaW^2) + \big(6  + 
   15\betaW -6\betaW^2 -14 \betaW^3
\nonumber\\
   &+& 4\betaW^4 + 3 \betaW^5\big) \Ct + 
2 (6 - 5 \betaW^2 + \betaW^4) \Ct^2\Big] (\gAb + \gVb)\ssW^2
\nonumber\\
&+&2 \betaW^2 \DW^2 \Big[\big[6 \betaW^2 - 
    3 \betaW^4 -7 +(\betaW^2-3) (3 \betaW^2-5) \Ct^2\big](\gVbb+ \gAbb)
\nonumber\\
 &+& 16 (\betaW^2-1) \Ct \gAb \gVb
   \Big] \ssW^4
\Bigg\}
\nonumber\\
\nonumber\\
\htuu_{44}[\Theta,\mWW]&=&\frac{\fWW \St^2}{\DW}
\Bigg\{
-1 + \betaW^2 + 2 \betaW \Ct + 2 \Ct^2 - 
2 \betaW (\betaW + \Ct) \DW(\gAb + \gVb)\ssW^2
\nonumber\\
&+& 2 \betaW^2 \DW^2(\gAbb + \gVbb)\ssW^4
\Bigg\}
\nonumber\\
\nonumber\\
\htuu_{47}[\Theta,\mWW]&=&\frac{\sqrt{2}\fWW \St}{\sqrt{1-\betaW^2}\DW}
\Bigg\{
(\Ct-1) \Big[-1 + 3 \betaW^2 - \betaW^3 + 2 \Ct + 4 \Ct^2 +
\betaW (3 + 6 \Ct)\Big]
\nonumber\\
&+&\betaW \Big[3 + (4 - \betaW) \betaW + 6 \Ct\Big] (1 - \Ct) \DW
(\gAb + \gVb)\ssW^2
\nonumber\\
&+&4 \betaW^2 \DW^2 \Big[-2 \gAb \gVb + 
\Ct (\gAbb + \gVbb)\Big]\ssW^4
\Bigg\}
\nonumber\\
\nonumber\\
\htuu_{48}[\Theta,\mWW]&=&\frac{\fWW \St^2}{\sqrt{3}(1-\betaW^2)\DW}
\Bigg\{1 + \betaW^4 - 2 \betaW^3 (3 - \Ct) + 6 \Ct - 6 \Ct^2 + 
2 \betaW (3 - 5 \Ct)
\nonumber\\
&-& \betaW^2 (6 + 6 \Ct - 2 \Ct^2)
-2 \betaW \Big[3 -4\betaW - 3\betaW^2 + \Ct\betaW^2 - 5 \Ct\Big]\DW(\gAb + \gVb)\ssW^2
\nonumber\\
&-&8 \betaW^2 \DW^2 (\gAbb + \gVbb)\ssW^4
\Bigg\}
\nonumber\\
\nonumber\\
\htuu_{55}[\Theta,\mWW]&=&2\fWW \St^2
\Big\{1 - 2 \betaW (\betaW + \Ct) (\gAb + \gVb)\ssW^2 +
  2\betaW^2 \DW (\gAbb + \gVbb)\ssW^4
\Big\}
\eea
\bea
\htuu_{56}[\Theta,\mWW]&=&\frac{\sqrt{2}\fWW \St}{\sqrt{1-\betaW^2}}
\Bigg\{(1-\betaW)(1 - \Ct)
+\betaW(1 -4 \betaW+ \betaW^2 - 2 \Ct) (1 - \Ct) (\gAb + 
\gVb)\ssW^2
\nonumber\\
&+& 
          4 \betaW^2 \DW\Big[2 \gAb \gVb -
          \Ct (\gAbb + \gVbb)\Big]\ssW^4
\Bigg\}
\nonumber\\
\nonumber\\
\htuu_{66}[\Theta,\mWW]&=&\frac{\fWW}{1-\betaW^2}
\Bigg\{(1-\Ct)^2(1-\betaW)^2
\nonumber\\
&-&2 \betaW (1 - \Ct)^2 \Big[ 4\betaW -(1+\betaW^2 )(1 - \Ct)\Big] (\gAb + 
   \gVb) \ssW^2
\nonumber\\
&+&8 \betaW^2 \DW \Big[(1 + \Ct^2) (\gVbb+\gAbb) - 
   4 \Ct \gAb \gVb\Big] \ssW^4
\Bigg\}
\nonumber\\
\nonumber\\
\htuu_{77}[\Theta,\mWW]&=&\frac{\fWW}{(1-\betaW^2)\DW}
\Bigg\{
(1 -\Ct)^2 \Big[1 + \betaW^4 + \betaW^2 (2 - 4 \Ct) + 8 \Ct (1 + \Ct) - 
  2 \betaW^3 (3 + \Ct)
\nonumber\\
&+& 2 \betaW (5 + 7 \Ct)\Big]
-2 \betaW (1 - \Ct)^2 \DW \Big[5 + 7 \Ct 
  +4\betaW - \betaW^2 (3 + \Ct)\Big] (\gAb + \gVb)\ssW^2
\nonumber\\
&+&8 \betaW^2 \DW^2\Big[(1 + \Ct^2) (\gVbb+\gAbb) - 4 \Ct \gAb \gVb \Big]\ssW^4
\Bigg\}
\nonumber\\
\nonumber\\
\htuu_{78}[\Theta,\mWW]&=&-
\frac{\sqrt{2}\fWW\St}{\sqrt{3}(1-\betaW^2)^{3/2}\DW}
\Bigg\{2 - \Ct + 2 \betaW^5 \Ct - 9 \Ct^2 + 6 \Ct^3 - 2 \betaW^3 (1 - 4 \Ct + \Ct^2)
\nonumber\\
&+& 3 \betaW^4 (2 - \Ct + \Ct^2) + 2 \betaW (1 - 9 \Ct + 5 \Ct^2) - 
2 \betaW^2 (6 - 4 \Ct - \Ct^2 + \Ct^3)
\nonumber\\
&-&2 \betaW \DW \Big[1 + 4 \betaW^3 + 4 \betaW (\Ct-2) + 
  \betaW^4 \Ct + \Ct (5 \Ct-9)
\nonumber\\
 &-& \betaW^2 (1 -4\Ct + \Ct^2)\Big] (\gAb + 
\gVb)\ssW^2
+8 \betaW^2 \DW^2 \Big[2 (\betaW^2-2) \gAb \gVb + 
   \Ct (\gAbb + \gVbb)\Big]\ssW^4
\Bigg\}
\nonumber\\
\nonumber\\
\htuu_{88}[\Theta,\mWW]&=&\frac{\fWW}{3(1-\betaW^2)^2\DW}
\Bigg\{
4 - 12 \Ct + 3 \Ct^2 + 18 \Ct^3 - 9 \Ct^4 - \betaW^6 (2 + 6 \Ct - 4 \Ct^2)
\nonumber\\
&+&  4 \betaW^3 (6 + \Ct - 6 \Ct^2 - \Ct^3) -
4 \betaW^5 (3 + 3 \Ct^2 - 2 \Ct^3) - 4 \betaW (3 - \Ct - 9 \Ct^2 + 3 \Ct^3)
\nonumber\\
&+&\betaW^4 (8 - 24 \Ct + 3 \Ct^2 - 6 \Ct^3 - \Ct^4) -
2 \betaW^2 (3 - 21 \Ct + 5 \Ct^2 + 6 \Ct^3 - 3 \Ct^4)
\nonumber\\
&-&2 \betaW \DW \Big[\betaW + 3 \betaW (8 - 3 \Ct) \Ct +
2 \betaW^2 \St^2 (6 + \Ct)+2 \betaW^3 (3 -12\Ct + \Ct^2)
\nonumber\\
&-&3 \betaW^5 \St^2 -
   2 (3 - \Ct) (1 - 3 \Ct^2) - \betaW^4 (6 + 6 \Ct^2 - 4 \Ct^3)\Big] (\gAb + 
   \gVb)\ssW^2
\nonumber\\
&+&2 \betaW^2 \DW^2 \Big[\big[1+ 2 \betaW^2(3  + \Ct^2)
    - 9 \Ct^2 - 3 \betaW^4 \St^2 \big] (\gVbb+\gAbb)
  \nonumber\\
&+& 48 (1 - \betaW^2) \Ct \gAb \gVb \Big] \ssW^4
\Bigg\}
\eea

\newpage
The non-vanishing elements $\ftuu_{a}$ are given by
\bea
\ftuu_{2}[\Theta,\mWW]&=&
\frac{2\sqrt{2}\fWW\St}{3(1-\betaW^2)^{3/2}\DW}
\Bigg\{
1 - \Ct - 
\betaW \Big(4 +3 \betaW^3 (1 - \Ct) - 6 \Ct + 2 \betaW^4 \Ct
\nonumber\\
  &-& 4 \Ct^2 - 
   4 \betaW^2 (1 + \Ct + \Ct^2) - 2 \beta (3 + \Ct + 2 \Ct^2 + 2 \Ct^3)\Big) 
+2 \betaW\DW \Big[2 + 4 \betaW^3
\nonumber\\
  &-& 3 \Ct + \betaW^4 \Ct - 
   2 \Ct^2 - 4 \betaW (2 + \Ct) - 2 \betaW^2 (1 + \Ct + \Ct^2)\Big] (\gAb + 
\gVb)\ssW^2
\nonumber\\
&+&8 \betaW^2 \DW^2 \Big[2 (2 - \betaW^2) \gAb \gVb +
   \Ct (\gAbb + \gVbb)\Big]\ssW^4
\Bigg\}
\nonumber\\
\nonumber\\
\ftuu_{3}[\Theta,\mWW]&=&
\frac{\fWW}{3(1-\betaW^2)^2\DW}
\Bigg\{
1 + \betaW^6 (1 - \Ct)^2 + 2 \Ct - 3 \Ct^2
+ 8 \betaW^3 (1 - \Ct)\St^2
\nonumber\\
&-& 4 \betaW^5 (1 - 3 \Ct + \Ct^2 + \Ct^3) -
4 \betaW (1 + 5 \Ct - 3 \Ct^2 - 3 \Ct^3)
\nonumber\\
&-&  \betaW^2 (21 - 2 \Ct + \Ct^2 - 8 \Ct^3 - 12 \Ct^4) + 
\betaW^4 (11 - 2 \Ct + 3 \Ct^2 - 8 \Ct^3 - 4 \Ct^4)
\nonumber\\
&+&4 \betaW (1 - \Ct)\DW \Big[1 + 3 \Ct (2 + \Ct) -
2 \betaW^3 (3 + \Ct) + 2 \betaW (5 + 3 \Ct)
\nonumber\\
&-& 2 \betaW^2 \St^2 + 
\betaW^4 (1 - 2\Ct - \Ct^2)\Big] (\gAb + \gVb)\ssW^2
\nonumber\\
&-&8 \betaW^2 \DW^2 \Big[\big[5 - 3 \Ct^2 + \betaW^2 (\Ct^2-3)\big]( \gVbb +\gAbb) + 
   4 (1 - \betaW^2) \Ct \gAb \gVb\Big]\ssW^4
\big\}
\nonumber\\
\nonumber\\
\ftuu_{4}[\Theta,\mWW]&=&
\frac{2\fWW\St^2}{3(1-\betaW^2)\DW}
\Bigg\{
1 - \betaW \big[4 \Ct + \betaW (6 - \betaW^2 + 4 \betaW \Ct + 4 \Ct^2)\big]
\nonumber\\
&+&4\betaW\DW(\Ct + 2\betaW + \betaW^2 \Ct)(\gAb + \gVb) \ssW^2
-8 \betaW^2 \DW^2 (\gAbb + \gVbb)\ssW^4
\Bigg\}
\nonumber\\
\nonumber\\
\ftuu_{7}[\Theta,\mWW]&=&
\frac{2\sqrt{2}\fWW\St}{3(1-\betaW^2)^{3/2}\DW}
\Bigg\{1 + \Ct + 
\betaW \Big(4 + 6 \Ct - 2 \betaW^4 \Ct - 4 \Ct^2 - 3 \betaW^3 (1 + \Ct)
\nonumber\\
&-& 4 \betaW^2 (1 - \Ct + \Ct^2) + 2 \betaW (3 - \Ct + 2 \Ct^2 - 2 \Ct^3)\Big)
\nonumber\\
&+&2\betaW \DW \Big[4 \betaW^3 + 4 \betaW (\Ct-2) + 
  \betaW^4 \Ct + (\Ct-2) (1 + 2 \Ct)
\nonumber\\
&-& 2 \betaW^2 (1 -\Ct + \Ct^2) \Big](\gAb + \gVb) \ssW^2
-8 \betaW^2 \DW^2 \Big[ 2(\betaW^2-2) \gAb \gVb + 
   \Ct (\gAbb + \gVbb)\Big] \ssW^4
\Bigg\}
\nonumber\\
\nonumber\\
\ftuu_{8}[\Theta,\mWW]&=&
\frac{\fWW}{3\sqrt{3}(1-\betaW^2)^2\DW}
\Bigg\{
1 + 6 \Ct + 3 \Ct^2 - \betaW^6 (1 + 6 \Ct + \Ct^2) -
4 \betaW^5 (3 + 3 \Ct + 3 \Ct^2 - \Ct^3)
\nonumber\\
&+&  8 \betaW^3 (3 + \Ct - 3 \Ct^2 - \Ct^3) - 
4 \betaW (3 - 5 \Ct - 9 \Ct^2 + 3 \Ct^3)
\nonumber\\
&+&  \betaW^2 (21 + 6 \Ct + \Ct^2 + 24 \Ct^3 - 12 \Ct^4) - 
\betaW^4 (11 + 6 \Ct + 3 \Ct^2 + 24 \Ct^3 - 4 \Ct^4)
\nonumber\\
&+&4 \betaW \DW \Big[3 - 
  2 \betaW^2 \St^2 (3 + \Ct) + 2 \betaW^3 (3 +6\Ct - \Ct^2)
  - \Ct (5 +9\Ct-3\Ct^2)
\nonumber\\
  &+& 2 \betaW (5 + 6\Ct  -3\Ct^2 ) 
-\betaW^4 (3 +3 \Ct +3\Ct^2 - \Ct^3)\Big] (\gAb + \gVb) \ssW^2
\nonumber\\
&+&8 \betaW^2 \DW^2 \Big[\big[5 - 3 \Ct^2 + \betaW^2 (\Ct^2-3)\big]
  (\gVbb+ \gAbb) +
   12 (1 - \betaW^2) \Ct \gAb \gVb \Big] \ssW^4
\Bigg\}\, .
\eea
The elements $\gtuu_{a}$ are identical: $\gtuu_{a}=\ftuu_{a}$. 

The elements $\htdd_{ab},~ \ftdd_{a},~ \gtdd_{a}$ can be obtained from the following transformations
\bea
A^{d \bar d}&=&A^{u \bar u}\Big\{ 
\gVb \to -\gVbD~~,~~\gAb \to -\gAbD~~,~~ \betaW\to -\betaW \Big\}\nonumber\\
\htdd_{ab}&=& \htuu_{ab} \Big\{ 
\gVb \to -\gVbD~~,~~\gAb \to -\gAbD~~,~~ \betaW\to -\betaW \Big\}\nonumber\\
\ftdd_{a}&=& \ftuu_{a} \Big\{ 
\gVb \to -\gVbD~~,~~\gAb \to -\gAbD~~,~~ \betaW\to -\betaW \Big\}\nonumber\\
\gtdd_{a}&=& \gtuu_{a} \Big\{ 
\gVb \to -\gVbD~~,~~\gAb \to -\gAbD~~,~~ \betaW\to -\betaW \Big\}\, ,
\eea
where in this case the angle $\Theta$ is the angle between the antidown quark
$\bar{d}$ and the $W^+$ momenta. The effective couplings $\bar{g}_{V,A}^{u,d}$ are defined in \eq{effgva}.

\subsection{Polarization density matrix for $q\,\bar{q}\to ZZ$}
We write below the coefficients $\Aqq [\Theta,\mVV]$, $\ftqq_a[\Theta,\mVV],\, \gtqq_a[\Theta,\mVV]$, and $\htqq_{ab}[\Theta,\mVV]$,
appearing in the polarization density matrix for
$q\, \bar{q}\to ZZ$. The angle $\Theta$ is the scattering angle in the CM frame from the anti-quark and one of the $Z$ momenta. Our convention is that the $Z$ is in this case the one with momentum parallel to the $\hat{k}$ unit vector of  the spin right-handed basis in \eq{basis}.  Results below will be given for a generic quark $q$. 
 
\bea
A^{q \bar q}=|\xbar{{\cal M}}_{\Z\Z}^{\;q\bar q}|^2
\eea
where the expression for the unpolarized square amplitude $|\xbar{{\cal M}}_{\Z\Z}^{\; q \bar q}|^2$ is
given in \eq{M2ZZ}.  Throughout the following expressions we use $\Ct\equiv \cos{\Theta}$, $\St\equiv \sin{\Theta}$ and $\DZ$, $\fZZ$ which are given in \eq{fZZ}.

The non-vanishing elements $\htqq_{ab}$ ($\htqq_{ba}=\htqq_{ab}$),
are given by
\vspace{0.5cm}
\bea
\htqq_{11}[\Theta,\mZZ]&=&
\fZZ(1 - \betaZ^2)
\Big\{
(1 + \Ct^2) (\gAAAA + 6 \gAA \gVV + \gVVVV)+ 8 \Ct \gA \gV (\gAA + \gVV) 
   \Big\}
\nonumber\\
\nonumber\\
\htqq_{15}[\Theta,\mZZ]&=&
\fZZ\sqrt{2}\sqrt{1-\betaZ^2}\St
\Big\{
\Ct (\gAAAA + 6 \gAA \gVV + \gVVVV)+4 \gA \gV (\gAA + \gVV) 
\Big\}
\nonumber\\
\nonumber\\
\htqq_{16}[\Theta,\mZZ]&=&
\fZZ(1-\betaZ^2)\St^2
\Big\{\gAAAA + 6 \gAA \gVV + \gVVVV
\Big\}
\nonumber\\
\nonumber\\
\htqq_{22}[\Theta,\mZZ]&=&
\frac{\fZZ (1-\betaZ^2)}{\DZ}
\Bigg\{
-8 \Ct \big[3 + 2 \betaZ^2 - \betaZ^4 - 4 \Ct^2\big] \gA \gV (\gAA + \gVV)
\nonumber\\
&+&\Big[(1 + \betaZ^2)^2 - (7 + 10 \betaZ^2 - \betaZ^4) \Ct^2 + 
  4 (2 + \betaZ^2) \Ct^4\Big] (\gVVVV + 6 \gAA \gVV +\gAAAA)
\Bigg\}
\nonumber\\
\nonumber\\
\htqq_{23}[\Theta,\mZZ]&=&
\frac{\fZZ 2\sqrt{2}\sqrt{1-\betaZ^2}\St}{\DZ}
\Bigg\{ \Big[\Ct (1 + \betaZ^2 + (\betaZ^2-3) \Ct^2)\Big]
(\gAAAA + 6 \gAA \gVV + \gVVVV)\nonumber\\
&+&\Big[2 (1 + \betaZ^2)^2 - 2 (5 - 2 \betaZ^2 + \betaZ^4) \Ct^2\Big]
\gA \gV (\gAA + \gVV)
\Bigg\}
\eea
\bea
\htqq_{24}[\Theta,\mZZ]&=&
\frac{\fZZ\sqrt{2}\sqrt{1-\betaZ^2}\St}{\DZ}
\Bigg\{ \Big[ (3 - \betaZ^2) (1 + \betaZ^2) \Ct  -
 4 \Ct^3 \Big](\gAAAA + 6 \gAA \gVV + \gVVVV)\nonumber\\
&+&\Big[4 (1 + \betaZ^2)^2 -  8 (1 + \betaZ^4) \Ct^2\Big]
\gA \gV (\gAA + \gVV)
\Bigg\}
\nonumber\\
\nonumber\\
\htqq_{27}[\Theta,\mZZ]&=&-
\frac{\fZZ (1-\betaZ^2)\St^2}{\DZ}
\Bigg\{\Big[(1 + \betaZ^2)^2 + 4 (\betaZ^2-2) \Ct^2\Big]
(\gAAAA + 6 \gAA \gVV + \gVVVV) 
\Bigg\}
\nonumber\\
\nonumber\\
\htqq_{28}[\Theta,\mZZ]&=&
\frac{\fZZ 2\sqrt{2}\sqrt{1-\betaZ^2}\St}{\sqrt{3}\DZ}
\Bigg\{
\Big[ 2 (1 + \betaZ^2)^2 (1 + \Ct^2)-8 (1 + \betaZ^2) \Ct^2\Big]
\gA \gV (\gAA + \gVV)
\nonumber\\
&+& \Big[2 (1 - 3 \betaZ^2) \Ct^3 + (1 + \betaZ^2)
  (3 \betaZ^2 + \Ct^2-2) \Ct \Big]
(\gAAAA + 6 \gAA \gVV + \gVVVV)
\Bigg\}
\nonumber\\
\nonumber\\
\htqq_{33}[\Theta,\mZZ]&=&-
\frac{\fZZ}{\DZ}
\Bigg\{8 \Ct \Big[2 + \betaZ^2
  + \betaZ^6 + (-3 + 2 \betaZ^2 - 3 \betaZ^4) \Ct^2\Big]\gA \gV (\gAA + \gVV)
\nonumber\\
&+&\Big[(\betaZ + 
  \betaZ^3)^2 + (7 - 5 \betaZ^2 - 3 \betaZ^4 + \betaZ^6) \Ct^2
\nonumber\\
  &-& (9 - 
      10 \betaZ^2 + 5 \betaZ^4) \Ct^4\Big] (\gAAAA + 6 \gAA \gVV + \gVVVV)
\Bigg\}
\nonumber\\
\nonumber\\
\htqq_{34}[\Theta,\mZZ]&=&
\frac{\fZZ(1-\betaZ^2)\St^2}{\DZ}
\Bigg\{\Big[2 (3 + \betaZ^2)\Ct^2-(1 + \betaZ^2)^2  \Big]
(\gAAAA + 6 \gAA \gVV + \gVVVV)
\nonumber\\
&+&8 (1 +\betaZ^2)\gA \gV (\gAA + \gVV)
\Bigg\}
\nonumber\\
\nonumber\\
\htqq_{37}[\Theta,\mZZ]&=&-
\frac{\fZZ \sqrt{2}(1-\betaZ^2)^{3/2}\Ct\St}{\DZ}
\Bigg\{
3 (1 + \betaZ^2 - 2 \Ct^2)(\gAAAA + 6 \gAA \gVV + \gVVVV)
\nonumber\\
&+&4 (1 - \betaZ^2) \Ct \gA \gV (\gAA + \gVV)
\Bigg\}
\nonumber\\
\nonumber\\
\htqq_{38}[\Theta,\mZZ]&=&
\frac{\fZZ}{\sqrt{3}\DZ}
\Bigg\{\Big[2 + 3 \betaZ^2 - 
  \betaZ^6 - (9 - 9 \betaZ^2 - \betaZ^4 + \betaZ^6) \Ct^2
\nonumber\\
&+& (9 - 18 \betaZ^2 + 
5 \betaZ^4) \Ct^4\Big] (\gAAAA + 6 \gAA \gVV + \gVVVV)
\nonumber\\
&+&8 \Ct \Big[2 + \betaZ^2 + \betaZ^6 - (3 - 2 \betaZ^2 + 3 \betaZ^4)
\Ct^2\Big]
\gA \gV (\gAA + \gVV)
\Bigg\}
\eea
\bea
\htqq_{44}[\Theta,\mZZ]&=&
\frac{2\fZZ\St^2}{\DZ}
\Bigg\{\Big[2 (1 + \betaZ^4) \Ct^2-(1 + \betaZ^2)^2\Big]
(\gAAAA + 6 \gAA \gVV + \gVVVV)
\nonumber\\
\nonumber\\
\htqq_{47}[\Theta,\mZZ]&=&
\frac{\fZZ \sqrt{2}\sqrt{1-\betaZ^2}\St}{\DZ}
\Bigg\{\Ct\Big[(\betaZ^2-3) (1 + \betaZ^2)  +4 \Ct^2\Big]
(\gAAAA + 6 \gAA \gVV + \gVVVV)
\nonumber\\
&+&4\Big[(1 + \betaZ^2)^2-2 (1 + \betaZ^4) \Ct^2\Big]
\gA \gV (\gAA + \gVV)
\Bigg\}
\nonumber\\
\nonumber\\
\htqq_{48}[\Theta,\mZZ]&=&
\frac{\fZZ(1-\betaZ^2)\St^2}{\sqrt{3}\DZ}
\Bigg\{
\Big[(1 + \betaZ^2)^2 - 2 (3 + \betaZ^2) \Ct^2\Big]
(\gAAAA + 6 \gAA \gVV + \gVVVV)
\nonumber\\
&+&24 (1 + \betaZ^2) \Ct\gA \gV (\gAA + \gVV)
\Bigg\}
\nonumber\\
\nonumber\\
\htqq_{55}[\Theta,\mZZ]&=&
\fZZ 2(1-\betaZ^2)\St^2 \Big[\gAAAA + 6 \gAA \gVV + \gVVVV\Big]
\nonumber\\
\nonumber\\
\htqq_{56}[\Theta,\mZZ]&=&-
\fZZ\sqrt{2}\sqrt{1-\betaZ^2}\St
\Big\{
\Ct(\gAAAA + 6 \gAA \gVV + \gVVVV)-4 \gA \gV (\gAA + \gVV)
\Big\}
\nonumber\\
\nonumber\\
\htqq_{66}[\Theta,\mZZ]&=&
\fZZ(1-\betaZ^2)\Big\{
(1 + \Ct^2)(\gAAAA + 6 \gAA \gVV + \gVVVV)
-8 \Ct \gA \gV (\gAA + \gVV)
\Big\}
\nonumber\\
\nonumber\\
\htqq_{77}[\Theta,\mZZ]&=&
\frac{\fZZ(1-\betaZ^2)}{\DZ}
\Bigg\{
8 \Ct \Big[3 + 2 \betaZ^2 - \betaZ^4 - 4 \Ct^2\Big]
\Ct \gA \gV (\gAA + \gVV)
\nonumber\\
&+&\Big[(1 + \betaZ^2)^2 - (7 + 10 \betaZ^2 - \betaZ^4) \Ct^2 +
  4 (2 + \betaZ^2) \Ct^4\Big](\gAAAA + 6 \gAA \gVV + \gVVVV)
\Bigg\}
\nonumber\\
\nonumber\\
\htqq_{78}[\Theta,\mZZ]&=&
\frac{\fZZ\sqrt{2}\sqrt{1-\betaZ^2}}{\sqrt{3}\DZ}
\Bigg\{
\Ct\Big[1 + 4 \betaZ^2 + 3 \betaZ^4 -2 (3 + \betaZ^2) \Ct^2\Big]
(\gAAAA + 6 \gAA \gVV + \gVVVV)
\nonumber\\
&+&4\Big[(9 - 2 \beta^2 + \betaZ^4) \Ct^2-2 (1 + \betaZ^2)^2 \Big]
\gA \gV (\gAA + \gVV)
\Bigg\}
\nonumber\\
\nonumber\\
\htqq_{88}[\Theta,\mZZ]&=&
\frac{\fZZ}{3\DZ}
\Bigg\{
\Big[(1 + \betaZ^2)^2 (4 + \betaZ^2) + (3 + 3 \betaZ^2 - 7 \betaZ^4 + 
  \betaZ^6) \Ct^2
\nonumber\\
  &-& (9 + 6 \betaZ^2 + 5 \betaZ^4) \Ct^4\Big]
(\gAAAA + 6 \gAA \gVV + \gVVVV)
\nonumber\\
&-&24 \Ct \Big[2 + \betaZ^2 + \betaZ^6 - (3 - 2 \betaZ^2 + 3 \betaZ^4)
  \Ct^2\Big]
\gA \gV (\gAA + \gVV)
\Bigg\}
\eea

\newpage
The non-vanishing elements $\ftqq_{a}$ are given by
\vspace{0.5cm}
\bea
\ftqq_{2}[\Theta,\mZZ]&=&
\frac{\fZZ 2\sqrt{2}\sqrt{1-\betaZ^2}}{3\DZ}
\Bigg\{\Ct \Big[2 \betaZ^2 + 3 \betaZ^4 -4 \betaZ^2 \Ct^2-1\Big]
 (\gAAAA + 6 \gAA \gVV + \gVVVV)
\nonumber\\
&+&4\Big[(1 + \betaZ^2)^2  + 4 \betaZ^2 (\beta^2-2) \Ct^2\Big]
\gA \gV (\gAA + \gVV)
\Bigg\}
\nonumber\\
\nonumber\\
\ftqq_{3}[\Theta,\mZZ]&=&
\frac{\fZZ}{3\DZ}
\Bigg\{\Big[(1 + \betaZ^2)^3 + (15 \betaZ^2 - 13 \betaZ^4 + \betaZ^6-3) \Ct^2
\nonumber\\
&+&   4 \betaZ^2 (\betaZ^2-3) \Ct^4\Big]
(\gAAAA + 6 \gAA \gVV + \gVVVV)
\nonumber\\
&+&8 \Ct \Big[1 + 3 \betaZ^4 - \betaZ^6 + \betaZ^2 (5 - 8 \Ct^2)\Big]
\gA \gV (\gAA + \gVV)
\Bigg\}
\nonumber\\
\nonumber\\
\ftqq_{4}[\Theta,\mZZ]&=&
\frac{\fZZ 2(1-\betaZ^2)\St^2}{3\DZ}
\Bigg\{\Big[1 + \betaZ^4 + \betaZ^2 (2 + 4 \Ct^2)\Big]
(\gAAAA + 6 \gAA \gVV + \gVVVV)
\Bigg\}
\nonumber\\
\nonumber\\
\ftqq_{7}[\Theta,\mZZ]&=&
\frac{\fZZ 2\sqrt{2}\sqrt{1-\betaZ^2}}{3\DZ}
\Bigg\{\Ct\Big[1 - 2 \betaZ^2 - 3 \betaZ^4 +4 \betaZ^2 \Ct^2\Big]
(\gAAAA + 6 \gAA \gVV + \gVVVV)
\nonumber\\
&+&4\Big[(1 + \betaZ^2)^2 +4 \betaZ^2 (\betaZ^2-2) \Ct^2\Big]
\gA \gV (\gAA + \gVV)
\Bigg\}
\nonumber\\
\nonumber\\
\ftqq_{8}[\Theta,\mZZ]&=&-
\frac{\fZZ}{3\sqrt{3}\DZ}
\Bigg\{\Big[(1 + \betaZ^2)^3 + (15 \betaZ^2 - 13 \betaZ^4 + \betaZ^6-3) \Ct^2
\nonumber\\
  &+& 4 \betaZ^2 (\betaZ^2-3) \Ct^4\Big]
(\gAAAA + 6 \gAA \gVV + \gVVVV)
\nonumber\\
&+&24 \Ct \Big[ \betaZ^6 + \betaZ^2 (8 \Ct^2-5)-1 - 3 \betaZ^4 \Big]
\gA \gV (\gAA + \gVV)
  \Bigg\}\, .
\eea
The elements $\gtqq_{a}$ are identical: $\gtqq_{a}=\ftqq_{a}$.

   \subsection{Polarization density matrix for $u\,\bar{d}\to W^+Z$}  

We write below the expressions for the coefficients
$\Aud [\Theta,\mVV]$, $\ftud_a[\Theta,\mVV],\, \gtud_a[\Theta,\mVV]$, and $\htud_{ab}[\Theta,\mVV]$, appearing in the polarization density matrix for $u\,\bar{d}\to W^+Z$ in the limit $M_W=M_Z=M_V$, where $\mVV$ is
the invariant mass of $WZ$ system in this approximation. The angle $\Theta$ is the scattering angle in the CM frame from the anti-down quark and $W^+$ momenta.  Our convention for the polarization matrix is that the  momentum of $W^+$ is chosen parallel to the $\hat{k}$ unit vector of  the spin right-handed basis in \eq{basis}.
\bea
A^{u \bar d}=|\xbar{{\cal M}}_{\W\Z}^{\; u\bar d}|^2
\eea
where the expression for the unpolarized square amplitude $|\xbar{{\cal M}}_{\W\Z}^{\; u \bar d}|^2$ is given in \eq{M2WZ}.  Throughout the following expressions we use $\Ct\equiv \cos{\Theta}$, $\St\equiv \sin{\Theta}$ and $\DV$, $\fWZ$ which are given in \eq{fWZ}.

The  non-vanishing elements $\htud_{ab}$ ($\htud_{ba}=\htud_{ab}$)
are given by
\vspace{0.5cm}
\bea
\htud_{11}[\Theta,\mVV]&=&
\fWZ (1+\Ct)^2
\Bigg\{(1 - \betaV^2) (3 + \betaV^2)^2
-2\, (3 + \betaV^2) \Big[3 - 
  \betaV \Big(3 - 15 \Ct + 2\betaV
  \nonumber\\
  &+& \betaV^2 (3 + \betaV + 9 \Ct)\Big)\Big] \ccW^2
+ \Big[9 - 
  \betaV \Big(18 - 33 \betaV + 24 \betaV^2 + 41 \betaV^3
  + 6 \betaV^4 + \betaV^5
\nonumber\\
  &+& 6\, (3 + \betaV^2) \big(-5 + 6 \betaV + 3\betaV^2\big) \Ct + 
       72 \betaV (-3 + \betaV^2) \Ct^2\Big)\Big] \ccW^4
\Bigg\}
\nonumber\\
\nonumber\\
\htud_{15}[\Theta,\mVV]&=&
\fWZ \sqrt{2}\sqrt{1-\betaV^2}(1+\Ct)
\Bigg\{(3 + \betaV^2)^2
+
(3 + \betaV^2) \Big[\betaV \big(3 + 3 \betaV^2-2\betaV
  - 30 \Ct\big)-6\Big]\ccW^2
\nonumber\\
&+&\Big[9 - \betaV \Big(9 - 24 \betaV + 12 \betaV^2 + 17 \betaV^3
  + 3 \betaV^4 + 
       6 (3 \betaV-5) (3 + \betaV^2) \Ct - 216 \betaV \Ct^2\Big)\Big]\ccW^4
\Bigg\}
\nonumber\\
\nonumber\\
\htud_{16}[\Theta,\mVV]&=&
\fWZ(1 - \betaV^2) \St^2
\Bigg\{(3 + \betaV^2)^2
-2\, (3 + \betaV^2) \Big[3 + \betaV^2 + 15 \betaV \Ct\Big] \ccW^2
\nonumber\\
&+&\Big[9 + 24 \betaV^2 - 17 \betaV^4 + 30\betaV(3+\betaV^2)\Ct
  + 216 \betaV^2 \Ct^2
  \Big] \ccW^4
\Bigg\}
\nonumber\\
\nonumber\\
\htud_{22}[\Theta,\mVV]&=&
\frac{\fWZ(1 +\Ct)^2)  }{\DV}
\Bigg\{(1 - \betaV^2) (3 + \betaV^2)^2
\Big[(1 + \betaV^2)^2 - 8 (1 + \betaV^2) \Ct + 4 (2 + \betaV^2) \Ct^2\Big]
\nonumber\\
&+&2\, (3 + \betaV^2)(1 + \betaV^2) \Big[(1 + \betaV^2) (
-3 - 15 \betaV + 2 \betaV^2 + 9 \betaV^3 + \betaV^4)
\nonumber\\
&+&  \big(24 + 3 \betaV - 16 \betaV^2 + 6 \betaV^3 - 8 \betaV^4 + 3 \betaV^5\big) \Ct
\nonumber\\
&+&
4\, \big(-6 + 36 \betaV + \betaV^2 - 9 \betaV^3 + 4 \betaV^4 - 21 \betaV^5
+ \betaV^6\big)\Ct^2
   \nonumber\\
   &+& 
   12 \betaV \big(-12 + 7 \betaV^2 + 3 \betaV^4\big) \Ct^3\Big]\ccW^2
\nonumber\\
&+&\Big[(1 + \betaV^2)^2 \big(9 + 90 \betaV + 33 \betaV^2 - 24 \betaV^3
  - 41 \betaV^4 - 18 \betaV^5 - \betaV^6\big)
  \nonumber\\
  &-&2\, (1 + \betaV^2) (3 + \betaV^2) \big(
    12 + 3 \betaV - 98 \betaV^2 + 6 \betaV^3 + 50 \betaV^4 + 3 \betaV^5
    \big)\Ct
    \nonumber\\
&+&4\, \big(18 - 216 \betaV - 105 \betaV^2 - 18 \betaV^3 - 121 \betaV^4
    + 144 \betaV^5 + 65 \betaV^6 + 42 \betaV^7 - \betaV^8  \big)\Ct^2
    \nonumber\\
    &+&24 \betaV (3 + \betaV^2) \big(12 - 36 \betaV - 7 \betaV^2 + 30 \betaV^3
    - 3 \betaV^4\big)\Ct^3
    \nonumber\\
    &+&288 \betaV^2 (9 - 6 \betaV^2 - \betaV^4) \Ct^4
    \Big]\ccW^4
\Bigg\}
\eea
\bea
\htud_{23}[\Theta,\mVV]&=&
\frac{\sqrt{2}\,\fWZ\sqrt{1-\betaV^2}(1+\Ct) \St}{\DV}
\Bigg\{
9 + 24 \betaV^2 + 22 \betaV^4 + 8 \betaV^6 + \betaV^8
\nonumber\\
&+& \big( 9 + 6 \betaV^2 - 8 \betaV^4 - 6 \betaV^6 - \betaV^8 \big)\Ct
- \big(54 + 18 \betaV^2 - 6 \betaV^4 - 2 \betaV^6 \big)\Ct^2
\nonumber\\
&-&\Big[(1 +\betaV^2)^2
  \big(18 - 9 \betaV + 12 \betaV^2 + 6 \betaV^3 + 2 \betaV^4 + 
  3 \betaV^5\big)
  \nonumber\\
  &+&  2\, \big(9 + 108 \betaV + 6 \betaV^2 + 144 \betaV^3 - 8 \betaV^4
  + 36 \betaV^5 - 6 \betaV^6 - \betaV^8\big)\Ct
  \nonumber\\
&-&  4\, \big(27 - 27 \betaV + 9 \betaV^2 - 3 \betaV^4 + 21 \betaV^5 - \betaV^6 + 
  6 \betaV^7\big)\Ct^2- 72\, \big( 9\betaV - \betaV^5 \big)\Ct^3
  \Big]\ccW^2
\nonumber\\
&+&\Big[(1 + \betaV^2)^2
  \big(9 - 9 \betaV - 48 \betaV^2 + 6 \betaV^3 + 19 \betaV^4 + 
  3 \betaV^5\big)\nonumber\\
  &+& \big( 9 + 216 \betaV - 48 \betaV^2 + 288 \betaV^3 - 26 \betaV^4 + 72 \betaV^5 +  48 \betaV^6 + 17 \betaV^8\big)\Ct
  \nonumber\\
  &-&2\, \big(27 - 54 \betaV - 477 \betaV^2 - 327 \betaV^4 + 42 \betaV^5 + 
  89 \betaV^6 + 12 \betaV^7\big)\Ct^2
  \nonumber\\
  &-&36 \betaV \big(18 - 9 \betaV + 6 \betaV^3 - 2 \betaV^4 + 3 \betaV^5\big)\Ct^3
 +648 \betaV^2 (\betaV^2-3)\Ct^4\Big]\ccW^4
\Bigg\}
\nonumber\\
\nonumber\\
\htud_{24}[\Theta,\mVV]&=&
\frac{\sqrt{2}\,\fWZ\sqrt{1-\betaV^2}(1+\Ct) \St}{\DV}
\Bigg\{
(3 + \betaV^2)^2 \Big[(1 + \betaV^2)^2 + 2 (1 - \betaV^4) \Ct - 4 \Ct^2\Big]
\nonumber\\
&-&(3 + \betaV^2)\, \Big[
(1 + \betaV^2)^2 \big(6 - 9 \betaV + 2 \betaV^2 + 3 \betaV^3\big)\nonumber\\
  &-&2\, (1 + \betaV^2) \big(
  6+27\betaV - 4\betaV^2 +3 \betaV^3  - 2 \betaV^4\big) \Ct
  \nonumber\\
  &-&4\, \big(6 - 18 \betaV + 2 \betaV^2 + 3 \betaV^3 + 9 \betaV^5\big)\Ct^2
+24 \betaV (-6 + \betaV^2) \Ct^3\Big]\ccW^2
\nonumber\\
&+&\Big[(1 + \betaV^2)^2 \big(9 -27\betaV - 12 \betaV^2 + 19 \betaV^4 +
  3 \betaV^5\big)
\nonumber\\
&+&2\,
(1 + \betaV^2) (3 + \betaV^2)
\big(3 + 27 \betaV - 29\betaV^2 + 3\betaV^3 + 8 \betaV^4\big)\Ct
\nonumber\\
&-&4\, \big(9 - 54 \betaV - 156 \betaV^2 - 9 \betaV^3 - 89 \betaV^4
+ 30 \betaV^5 + 36 \betaV^6 + 9 \betaV^7\big)\Ct^2
\nonumber\\
&-&24 \betaV (3 + \betaV^2)  \big(6 -9\betaV -\betaV^2 + 6 \betaV^3\big) \Ct^3
+432\, \betaV^2 (\betaV^2-3) \Ct^4
\Big]\ccW^2
\Bigg\}
\nonumber\\
\nonumber\\
\htud_{27}[\Theta,\mVV]&=&-
\frac{\fWZ(1-\betaV^2)\St^2}{\DV}
\Bigg\{
(3 + 4 \betaV^2 + \betaV^4)^2 + 4 (\betaV^2-2) (3 + \betaV^2)^2 \Ct^2
\nonumber\\
&-&2 (3 + \betaV^2) \Big[(1 + \betaV^2)^2 (3 + \betaV^2)  
  + 3 \betaV (15 - \betaV^2) (1 + \betaV^2) \Ct
\nonumber\\
  &+& 4 \big(\betaV^2 + \betaV^4-6\big) \Ct^2
   + 12 \betaV (5 \betaV^2-12) \Ct^3
   \Big] \ccW^2
\nonumber\\
&+&\Big[(1 + \betaV^2)^2 \big(9 - 84 \betaV^2 + 19 \betaV^4\big)
+6 \betaV (15 - \betaV^2) (1 + \betaV^2) \big(3 + \betaV^2\big) \Ct
 \nonumber\\
 &+& 4\, \big(321 \betaV^2 + 238 \betaV^4 - 53 \betaV^6-18\big) \Ct^2
 +  24\, \betaV (3 + \betaV^2) \big(5 \betaV^2-12\big) \Ct^3
\nonumber\\
&-&
864\, \betaV^2 (3 - \betaV^2) \Ct^4
\Big]\ccW^4
\Bigg\}
\eea

\newpage
\bea
\htud_{28}[\Theta,\mVV]&=&
\frac{\sqrt{2}\,\fWZ\sqrt{1-\betaV^2}\St}{\sqrt{3}\DV}
\Bigg\{(3 + \betaV^2)^2 
\Big[1 + 2 \betaV^2 + \betaV^4 - \big(4 - 2 \betaV^2 - 6 \betaV^4\big) \Ct
\nonumber\\
&-& (3 + 2 \betaV^2 - \betaV^4) \Ct^2 + \big(6 - 10 \betaV^2\big) \Ct^3\Big]
-(3 + \betaV^2) \Big[(1 + \betaV^2)^2
  \big(6  +21 \betaV +2 \betaV^2 - 9 \betaV^3\big)
\nonumber\\
&+&
  (1 + \betaV^2) (3 +  \betaV^2) \big(15 \betaV + 12 \betaV^2 + 3\betaV^3-8\big) \Ct
\nonumber\\
&-&  2 \big(
9 + 96 \betaV + 9 \betaV^2 - 42 \betaV^3 - \betaV^4 - 78 \betaV^5 - \betaV^6\big) \Ct^2
\nonumber\\
&-& 4 (3 + \betaV^2) \big(9 \betaV + 5 \betaV^2-3\big) \Ct^3 + 
24 \betaV (9 - 13 \betaV^2) \Ct^4\Big] \ccW^2
\nonumber\\
&+&\Big[(1 + \betaV^2)^2\big( 
9 + 63 \betaV + 24 \betaV^2 - 6 \betaV^3 + 19 \betaV^4 - 9 \betaV^5\big)
\nonumber\\
&+&(1 + \betaV^2)\big(-36 + 135 \betaV + 498 \betaV^2 + 117 \betaV^3
+ 32 \betaV^4 + 33 \betaV^5 -  102 \betaV^6 + 3 \betaV^7\big) \Ct
\nonumber\\
&+&\big(-27 - 576 \betaV + 450 \betaV^2 + 60 \betaV^3 + 336 \betaV^4 + 552 \betaV^5 + 
22 \betaV^6 + 156 \betaV^7 + 19 \betaV^8\big) \Ct^2
\nonumber\\
&+&2\, \big(27 - 162 \betaV - 1107 \betaV^2 - 108 \betaV^3 + 333 \betaV^4 - 18 \betaV^5 + 
571 \betaV^6\big) \Ct^3
\nonumber\\
&+& 12\, \big(54 \betaV - 81 \betaV^2 - 60 \betaV^3 - 26 \betaV^5 - 3 \betaV^6\big)+216\, \betaV^2 (9 - 11 \betaV^2) \Ct^5 \Big] \ccW^4
\Bigg\}
\nonumber\\
\nonumber\\
\htud_{33}[\Theta,\mVV]&=&
\frac{\fWZ(1+\Ct)}{\DV}
\Bigg\{(3+\betaV^2)^2\Big[\betaV^2 + 2 \betaV^4 + \betaV^6
  + \big(4 + \betaV^2 - 2 \betaV^4 + \betaV^6\big) \Ct
\nonumber\\
&+&  \big(3 - 6 \betaV^2 - \betaV^4\big) \Ct^2
+ \big(10 \betaV^2 - 5 \betaV^4-9\big) \Ct^3 \Big]
\nonumber\\
&-&2 (3 + \betaV^2)
\Big[\betaV (1 + \betaV^2)^2  (-6 + 3 \betaV  + 6 \betaV^2 + \betaV^3)
\nonumber\\
&+& (1 + 
\betaV^2) (12 -12 \betaV -5 \betaV^2 + 30 \betaV^3 + 6 \betaV^4 + \betaV^5) \Ct
\nonumber\\
&+& \big (9 + 102 \betaV - 15 \betaV^2 - 12 \betaV^3 - 9 \betaV^4 - 
  54 \betaV^5 - \betaV^6 + 12 \betaV^7 \big) \Ct^2
\nonumber\\
&-& (3 + \betaV) \big(9 - 21 \betaV + 32 \betaV^3 - 9 \betaV^4 + 5 \betaV^5\big) \Ct^3  - 6 \betaV \big(27 - 28 \betaV^2 + 9 \betaV^4\big) \Ct^4
\Big]\ccW^2
\nonumber\\
&+&\Big[2 \betaV (1 + \betaV^2)^2
\big(-18 + 27 \betaV + 12 \betaV^2 - 6 \betaV^3 + 6 \betaV^4 + 5 \betaV^5\big)
\nonumber\\
&-&4 (1 + \betaV^2)\big(-9 + 18 \betaV + 84 \betaV^2 - 39 \betaV^3 - 28 \betaV^4
- 24 \betaV^5 -  39 \betaV^6 - 3 \betaV^7 + 2 \betaV^8\big) \Ct
\nonumber\\
&+&\big(27 + 612 \betaV - 468 \betaV^2 + 132 \betaV^3 + 210 \betaV^4 - 348 \betaV^5 + 
492 \betaV^6 - 36 \betaV^7 - 37 \betaV^8 + 24 \betaV^9\big) \Ct^2
\nonumber\\
&-&\big(81 - 324 \betaV - 2412 \betaV^2 + 468 \betaV^3 + 534 \betaV^4 + 228 \betaV^5 + 
1172 \betaV^6 + 12 \betaV^7 - 175 \betaV^8\big) \Ct^3
\nonumber\\
&-&12\, \big(81 \betaV - 81 \betaV^2 - 57 \betaV^3 + 126 \betaV^4 - \betaV^5 - 9 \betaV^6 + 
9 \betaV^7\big) \Ct^4
\nonumber\\
&+& 36 \betaV^2 (3 - \betaV^2) (17 \betaV^2-27) \Ct^5\Big]\ccW^4
\Big\}
\eea
\bea
\htud_{34}[\Theta,\mVV]&=&-
\frac{\fWZ(1-\betaV^2)\St^2}{\DV}
\Bigg\{
(3 + 4 \betaV^2 + \betaV^4)^2 - 2 (1 + \betaV^2) \big(3 + \betaV^2\big)^2 \Ct - 
2 \big(3 + \betaV^2\big)^3 \Ct^2
\nonumber\\
&-&2 (3 + \betaV^2)\Big[
  (1 + \betaV^2)^2 (3 + 3 \betaV + \betaV^2)
  -(1 + \betaV^2) \big(6 - 33 \betaV +2 \betaV^2- 9 \betaV^3\big) \Ct
\nonumber\\
&-&2\, \big(9 + 18 \betaV + 6 \betaV^2 + 12 \betaV^3 + \betaV^4\big) \Ct^2 - 
  12 \betaV \big(9 + 2 \betaV^2\big) \Ct^3
  \Big]\ccW^2
\nonumber\\
&+& \Big[
  (1 + \betaV^2)^2 \big(9 + 18 \betaV -30 \betaV^2 + 6 \betaV^3 + \betaV^4
\big)
\nonumber\\
&+&2\, (1 + \betaV^2) (3 + \betaV^2)
\big( 33 \betaV + 17 \betaV^2 + 9 \betaV^3 -3\big) \Ct
\nonumber\\
&-&2\, \big(27 + 108 \betaV - 405 \betaV^2 + 108 \betaV^3 - 387 \betaV^4 + 24 \betaV^5 - 35 \betaV^6\big) \Ct^2
\nonumber\\
&-&24 \betaV (3 + \betaV) (3 + 2 \betaV) \big(3 + \betaV^2\big) \Ct^3 - 
  216 \betaV^2 \big(9 + \betaV^2\big) \Ct^4
  \Big]\ccW^4
\Bigg\}
\nonumber\\
\nonumber\\
\htud_{37}[\Theta,\mVV]&=&-
\frac{\sqrt{2}\,\fWZ\sqrt{1-\betaV^2}\St}{\DV}
\Bigg\{
(1 - \betaV^2) (3 + \betaV^2)^2 \Big[3 + \betaV^2 (3 - \Ct) + 
  \Ct (1 - 6 \Ct)\Big] \Ct
\nonumber\\
&+&2\, (3 + \betaV^2) \Big[
  3\, (2 \betaV + 3 \betaV^3 - \betaV^7)
  -3\, (1 - \betaV^4) \big(3 -2 \betaV  + \betaV^2\big) \Ct
\nonumber\\
&+&
\big( 3 + 75 \betaV - 5 \betaV^2 - 6 \betaV^3 + \betaV^4 - 45 \betaV^5 + \betaV^6\big) \Ct^2
\nonumber\\  
&-& 6\, (1 - \betaV^2) \big(3 -3 \betaV + \betaV^2 + \betaV^3\big) \Ct^3 -
12 \betaV \big(9 - 8 \betaV^2\big) \Ct^4
\Big]\ccW^2
\nonumber\\
&+&\Big[
6\, \betaV (1 + \betaV^2)^2 \big(6 \betaV + \betaV^2 + \betaV^4-6\big)
\nonumber\\
&+&3\, (1 + \betaV^2) \big(9 - 12 \betaV - 99 \betaV^2 + 8 \betaV^3 - 5 \betaV^4 + 4 \betaV^5 + 23 \betaV^6\big) \Ct
\nonumber\\
&+&\big(9 + 450 \betaV - 228 \betaV^2 + 114 \betaV^3 - 146 \betaV^4 - 282 \betaV^5 + 
76 \betaV^6 - 90 \betaV^7 + \betaV^8\big) \Ct^2
\nonumber\\
&+&6\, \big(-9 + 18 \betaV +291 \betaV^2 - 18 \betaV^3 - 19 \betaV^4
- 2 \betaV^5 -119 \betaV^6 + 2 \betaV^7 \big) \Ct^3
\nonumber\\
&-&12\, \big(54 \betaV - 27 \betaV^2 - 30 \betaV^3 + 18 \betaV^4 - 16 \betaV^5
- 3 \betaV^6
\big) \Ct^4 + 216 \betaV^2 (7 \betaV^2 -9) \Ct^5
\Big]\ccW^4
\Bigg\}
\nonumber\\
\nonumber\\
\htud_{38}[\Theta,\mVV]&=&
\frac{\fWZ(\Ct-1)}{\sqrt{3}\DV}
\Bigg\{
(3 + \betaV^2)^2 \Big[1 + \betaV^2 (1 - \Ct) + 3 \Ct\Big]
\Big[\betaV^4 + 3 \Ct^2 -
    \betaV^2 (1 + 4 \Ct + 5 \Ct^2)-2\Big]
\nonumber\\
&+&2 (3 + \betaV^2)\Big[
  (1 + \betaV^2)^2 \big(6 - 6 \betaV - \betaV^2 + 6 \betaV^3 - \betaV^4\big)
  \nonumber\\
  &+&(1 + \betaV) (1 + \betaV^2) \big( 18 + 36 \betaV - 33 \betaV^2 + 3 \betaV^3 - \betaV^4 + \betaV^5  \big) \Ct
    \nonumber\\
&-&\big(9 - 144 \betaV - 39 \betaV^2 - 30 \betaV^3 - 17 \betaV^4 +
 48 \betaV^5 - \betaV^6 - 18 \betaV^7\big) \Ct^2
 \nonumber\\
 &-&\big(27 + 54 \betaV - 45 \betaV^2 - 216 \betaV^3 - 3 \betaV^4
 + 18 \betaV^5 + 5 \betaV^6\big) \Ct^3 -6 \betaV (3 - \betaV^2) (9 - 13 \betaV^2) \Ct^4
 \Big]\ccW^2
\nonumber\\
&+&\Big[2 (1 + \betaV^2)^2 \big( 18 \betaV + 30 \betaV^2 - 12 \betaV^3 - 25 \betaV^4 -   6 \betaV^5 + 14 \betaV^6 -9\big)
  \nonumber\\
&-&  2 \big(27 + 162 \betaV - 135 \betaV^2 + 126 \betaV^3 - 90 \betaV^4 - 66 \betaV^5 +  160 \betaV^6 - 30 \betaV^7 + 75 \betaV^8 - 13 \betaV^{10}\big) \Ct
  \nonumber\\
&+&\big(  27 - 864 \betaV - 1404 \betaV^2 - 468 \betaV^3 - 54 \betaV^4 + 228 \betaV^5 + 
  628 \betaV^6 - 12 \betaV^7 - 253 \betaV^8 - 36 \betaV^9\big) \Ct^2
  \nonumber\\
&+&\big(  81 + 324 \betaV - 3348 \betaV^2 - 1188 \betaV^3 + 270 \betaV^4 - 324 \betaV^5 + 
  1308 \betaV^6 + 36 \betaV^7 - 391 \betaV^8\big) \Ct^3
  \nonumber\\
&+&  12\, (81 \betaV + 81 \betaV^2 - 117 \betaV^3 - 270 \betaV^4 - 9 \betaV^5 + 
  57 \betaV^6 + 13 \betaV^7) \Ct^4
  \nonumber\\
  &+& 108 (27 \betaV^2 - 42 \betaV^4 + 11 \betaV^6) \Ct^5
\Big]\ccW^4
  \Bigg\}
  \eea

  \bea
\htud_{44}[\Theta,\mVV]&=&
\frac{2\fWZ\St^2}{\DV}
\Bigg\{
 2 (3 + \betaV^2)^2 \big(1 + \betaV^4\big) \Ct^2-(3 + 4 \betaV^2 + \betaV^4)^2 
\nonumber\\
&+&2\, (3 + \betaV^2) \Big[(1 + \betaV^2)^2 (3 + \betaV^2) + 
   3 \betaV (1 + \betaV^2) \big(7 + \betaV^4\big) \Ct - 
   2\, (3 + \betaV^2) \big(1 + \betaV^4\big) \Ct^2
\nonumber\\
   &-& 12\, \betaV \big(3 - \betaV^2 + 2 \betaV^4\big) \Ct^3\Big]\ccW^2
+\Big[(1 + \betaV^2)^2 \big( 3 \betaV^2 - 19 \betaV^4 + 9 \betaV^6 -9 \big)
\nonumber\\
  &-&   6 \betaV (1 + \betaV^2) (3 + \betaV^2) \big(7 + \betaV^4\big) \Ct
+ 2 \big(9 - 210 \betaV^2 - 44 \betaV^4 + 42 \betaV^6 - 53 \betaV^8\big) \Ct^2
\nonumber\\
&+&  24 \betaV \big(9 + 5 \betaV^4 + 2 \betaV^6\big) \Ct^3 + 
     72 \betaV^2 \big(9 - 6 \betaV^2 + 5 \betaV^4\big) \Ct^4\Big] \ccW^4
\Bigg\}
\nonumber\\
\nonumber\\
\htud_{47}[\Theta,\mVV]&=&
\frac{\sqrt{2}\fWZ\sqrt{1-\betaV^2}(1-\Ct)\St}{\DV}
\Bigg\{(3 + \betaV^2)^2 \Big[(1 + \betaV^2)^2 - 2 \big(1 - \betaV^4\big) \Ct - 4 \Ct^2\Big]
\nonumber\\
&-&(3 + \betaV^2) \Big[
  (1 + \betaV^2)^2 \big(6 +9 \betaV +2 \betaV^2- 3 \betaV^3\big)
  \nonumber\\
&+&2\, (1 + \betaV^2) \big(27 \betaV + 4 \betaV^2 + 3 \betaV^3 + 2 \betaV^4-6\big) \Ct 
  \nonumber\\
&-&4\, \big(6 + 18 \betaV + 2 \betaV^2 - 3 \betaV^3 - 9 \betaV^4\big) \Ct^2  
  - 24\, \big( 6 \betaV - \betaV^3\big) \Ct^3\Big]\ccW^2
\nonumber\\
&+&\Big[(1 + \betaV^2)^2 \big(9 +27 \betaV - 12 \betaV^2 + 19 \betaV^4
  - 3 \betaV^5\big)
  \nonumber\\
  &-&2\, (1 + \betaV^2) (3 + \betaV^2) \big(3 -27 \betaV  -29 \betaV^2 -3 \betaV^3
  + 8 \betaV^4\big) \Ct \nonumber\\
  &-&4\big( 9 + 54 \betaV - 156 \betaV^2 + 9 \betaV^3 - 89 \betaV^4
  - 30 \betaV^5 + 36 \betaV^6 - 9 \betaV^7\big) \Ct^2
  \nonumber\\
  &-& 24\, (3 + \betaV^2) \big(6 \betaV +9 \betaV^2 - \betaV^3 - 6 \betaV^4\big) \Ct^3
  -432 \betaV^2 \big(3 - \betaV^2\big) \Ct^4
  \Big]\ccW^4
\Bigg\}
\nonumber\\
\nonumber\\
\htud_{48}[\Theta,\mVV]&=&
\frac{\fWZ(1-\betaV^2)\St^2}{\sqrt{3}\DV}
\Bigg\{(3 + 4 \betaV^2 + \betaV^4)^2 + 6 (1 + \betaV^2) (3 + \betaV^2)^2 \Ct - 
2\, (3 + \betaV^2)^3 \Ct^2
\nonumber\\
&-&2\, (3 + \betaV^2) \Big[
  (1 + \betaV^2)^2\big(3 -9 \betaV + \betaV^2\big)
+ 3\, (1 + \betaV^2) \big(6 + 11 \betaV + 2 \betaV^2 + 3 \betaV^3\big) \Ct
\nonumber\\
&-& 2\, \big( 9 - 54 \betaV + 6 \betaV^2 - 36 \betaV^3 + \betaV^4 \big) \Ct^2 - 
 12 \betaV (9 + 2 \betaV^2) \Ct^3
 \Big]\ccW^2
\nonumber\\
&+&\Big[(1 + \betaV^2)^2 \big(9 -54 \betaV -30 \betaV^2 -18 \betaV^3 + \betaV^4 \big)
\nonumber\\
&+&6\, (1 - \betaV) (1 + \betaV^2) (3 + \betaV^2)
\big(3 +14 \betaV - 3 \betaV^2\big) \Ct
\nonumber\\
&+&2\,\big(324 \betaV + 405 \betaV^2 + 324 \betaV^3 + 387 \betaV^4 + 72 \betaV^5 + 35 \betaV^6 -27\big) \Ct^2
\nonumber\\
&-&24 \betaV (3 + \betaV^2) \big(9 -27 \betaV + 2 \betaV^2\big) \Ct^3
-216 \betaV^2 (9 + \betaV^2) \Ct^4
\Big]\ccW^4
\Bigg\}
\nonumber\\
\nonumber\\
\htud_{55}[\Theta,\mVV]&=&
2\fWZ(1-\betaV^2)\St^2
\Bigg\{(3 + \betaV^2)^2
-2\, (3 + \betaV^2) \Big[3 + \betaV^2 + 3 \betaV \big(5 - \betaV^2\big) \Ct\Big] \ccW^2
\nonumber\\
&+&\Big[9 + 15 \betaV^2 - 17 \betaV^4 + 
  9 \betaV^6
+ \betaV\big(90  +12 \betaV^2 -6 \betaV^4 \big) \Ct + 
 72 \betaV^2 \big(3 - \betaV^2\big) \Ct^2\Big]\ccW^4
\Bigg\}~~~~~~~
\eea
\bea
\htud_{56}[\Theta,\mVV]&=&
\sqrt{2}\fWZ\sqrt{1-\betaV^2}(1-\Ct)
\Bigg\{(3 + \betaV^2)^2
- (3 + \betaV^2) \Big[6 + 3 \betaV + 2 \betaV^2
  + 3 \betaV^3 + 30 \betaV \Ct\Big] \ccW^2
\nonumber\\
&+&\Big[9 + 9 \betaV + 24 \betaV^2 + 12 \betaV^3
  - 17 \betaV^4 + 3 \betaV^5 + 
  6 \betaV (5 + 3 \betaV) \big(3 + \betaV^2\big) \Ct
  + 216 \betaV^2 \Ct^2\Big] \ccW^4
\Bigg\}
\nonumber\\
\nonumber\\
\htud_{66}[\Theta,\mVV]&=&
\fWZ(1-\Ct)^2
\Bigg\{
(1 - \betaV^2) (3 + \betaV^2)^2
\nonumber\\
&-& 2 (3 + \betaV^2)\Big[
  3 + 3 \betaV - 2 \betaV^2 + 3 \betaV^3 - \betaV^4
  + \betaV\big(15 - 9 \betaV^2\big) \Ct
  \Big] \ccW^2
\nonumber\\
&+&\Big[9 + 18 \betaV + 33 \betaV^2 + 24 \betaV^3 - 41 \betaV^4  
+  6 \betaV^5 - \betaV^6
  \nonumber\\
 &+&  6 \betaV (5 +6 \betaV - 3 \betaV^2) (3 + \betaV^2) \Ct + 
 72 \betaV^2 (3 - \betaV^2) \Ct^2\Big]\ccW^4
\Bigg\}
\nonumber\\
\nonumber\\
\htud_{77}[\Theta,\mVV]&=&
\frac{\fWZ(1-\Ct)^2}{\DV}
\Bigg\{(1 - \betaV^2) (3 + \betaV^2)^2
\Big[1 + \betaV^4 + 8 \Ct (1 + \Ct) + 2 \betaV^2 (1 + 4 \Ct + 2 \Ct^2)\Big]
\nonumber\\
&+&2 (3 + \betaV^2)\Big[(1 + \betaV^2)^2
  \big(15 \betaV + 2 \betaV^2 - 9 \betaV^3 + \betaV^4-3\big)
  \nonumber\\
  &+&(1 + \betaV^2) \big(3 \betaV + 16 \betaV^2 + 6 \betaV^3 + 8 \betaV^4 + 3 \betaV^5 -24
  \big)\Ct
  \nonumber\\
  &+&4\,\big( \betaV^2 + 9 \betaV^3 + 4 \betaV^4 + 21 \betaV^5 + \betaV^6
-6 - 36 \betaV  \big) \Ct^2
  \nonumber\\
  &+&12 \betaV \big(7 \betaV^2 + 3 \betaV^4 - 12\big) \Ct^3
  \Big]\ccW^2
\nonumber\\
&+&\Big[
(1 + \betaV^2)^2\big( 
  9 - 90 \betaV + 33 \betaV^2 + 24 \betaV^3 - 41 \betaV^4 + 18 \betaV^5 - \betaV^6 \big)
  \nonumber\\
  &-&2\, (1 + \betaV^2) (3 + \betaV^2) \big(
  3 \betaV + 98 \betaV^2 + 6 \betaV^3 - 50 \betaV^4 + 3 \betaV^5 -12
  \big) \Ct
  \nonumber\\
  &+&4\,\big(18 + 216 \betaV - 105 \betaV^2 + 18 \betaV^3
  - 121 \betaV^4 - 144 \betaV^5 + 65 \betaV^6 - 42 \betaV^7 - \betaV^8
  \big) \Ct^2
  \nonumber\\
  &+&24 \betaV (3 + \betaV^2)\big(12 + 36 \betaV - 7 \betaV^2
  - 30 \betaV^3 - 3 \betaV^4\big) \Ct^3
  +288 \betaV^2 \big(9 - 6 \betaV^2 - \betaV^4\big) \Ct^4
  \Big] \ccW^4
  \Bigg\}
\nonumber\\
\nonumber\\
\htud_{78}[\Theta,\mVV]&=&-
\frac{\sqrt{2}\fWZ\sqrt{1-\betaV^2}\,\St}{\sqrt{3}\DV}
\Bigg\{(3 + \betaV^2)^2\big(1 + \betaV^2 - 2 \Ct\big)
\Big[2 + 2 \betaV^2 +3 (1 - \betaV^2) \Ct - (3 + \betaV^2) \Ct^2\Big]
\nonumber\\
&-&2 (3 + \betaV^2) \Big[
  (1 + \betaV^2)^2 \big(6 -3 \betaV + 2 \betaV^2\big)
  \nonumber\\
&+&(1 + \betaV^2) ( 63 \betaV - 10 \betaV^2 + 6 \betaV^3 - 3 \betaV^4 + 3 \betaV^5 -3  \big) \Ct
  \nonumber\\
&-&  \big(  27 + 33 \betaV + 3 \betaV^2 + 66 \betaV^3 + \betaV^4 + 21 \betaV^5 + \betaV^6
  \big) \Ct^2
  \nonumber\\
  &+&2\, \big(  9 - 81 \betaV + 6 \betaV^2 + 18 \betaV^3 + \betaV^4 - 9 \betaV^5\big) \Ct^3
  +12 \betaV \big(9 + 2 \betaV^2\big) \Ct^4
  \Big]\ccW^2
\nonumber\\
&+&\Big[
  2\, (1 + \betaV^2)^2 \big(9 - 9 \betaV - 30 \betaV^2 - 3 \betaV^3 + 19 \betaV^4\big)
  \nonumber\\
&+&(1 + \betaV^2)\big(378 \betaV - 105 \betaV^2 + 162 \betaV^3 - 19 \betaV^4 + 30 \betaV^5 - 
  3 \betaV^6 + 6 \betaV^7 -9 \big) \Ct
  \nonumber\\
  &-&\big(81 + 198 \betaV - 1584 \betaV^2 + 462 \betaV^3 - 1110 \betaV^4
  + 258 \betaV^5 + 184 \betaV^6 + 42 \betaV^7 - 35 \betaV^8\big) \Ct^2
  \nonumber\\
&-&2\,\big(486 \betaV + 405 \betaV^2 + 54 \betaV^3 + 495 \betaV^4 + 18 \betaV^5 +   71 \betaV^6 + 18 \betaV^7 -27 \big) \Ct^3
  \nonumber\\
  &+& 12 \betaV \big(54 - 243 \betaV + 30 \betaV^2 + 54 \betaV^3 + 4 \betaV^4 - 15 \betaV^5
  \big) \Ct^4 +216 \betaV^2 (9 + \betaV^2) \Ct^5
  \Big] \ccW^4
\Bigg\}
  \eea
  \bea
\htud_{88}[\Theta,\mVV]&=&
\frac{\fWZ}{3\DV}
\Bigg\{
(4 + \betaV^2) \big(3 + 4 \betaV^2 + \betaV^4\big)^2
-  6\, (3 + \betaV^2)^2 \big(2 + \betaV^2 + \betaV^6\big) \Ct
\nonumber\\
&+& (3 + \betaV^2)^2 (3 + 
3 \betaV^2 - 7 \betaV^4 + \betaV^6) \Ct^2
+ 6\, (27 + 18 \betaV^4 + 16 \betaV^6 + 3 \betaV^8) \Ct^3
\nonumber\\
&-& (3 + \betaV^2)^2 (9 + 
6 \betaV^2 + 5 \betaV^4) \Ct^4
\nonumber\\
&+&\Big[-2 (1 + \betaV^2)^2 \big(36 + 54 \betaV + 33 \betaV^2 - 36 \betaV^3 + 10 \betaV^4 - 
  18 \betaV^5 + \betaV^6\big)
\nonumber\\
&+&12\, \big(18 - 33 \betaV + 21 \betaV^2 - 59 \betaV^3 + 8 \betaV^4 - 43 \betaV^5 + 
   10 \betaV^6 - 21 \betaV^7 + 6 \betaV^8 - 4 \betaV^9 + \betaV^{10}\big) \Ct
\nonumber\\
&+&2\, \big( 810 \betaV - 45 \betaV^2 + 324 \betaV^3 + 42 \betaV^4 - 
   144 \betaV^5 + 30 \betaV^6 + 108 \betaV^7 + \betaV^8 + 54 \betaV^9 - 
   \betaV^{10} -27 \big) \Ct^2
\nonumber\\
&-&12\, \big(27 + 36 \betaV + 36 \betaV^3 + 18 \betaV^4 - 52 \betaV^5 + 16 \betaV^6 - 
   20 \betaV^7 + 3 \betaV^8 \big) \Ct^3
\nonumber\\
&+&2\, \big(81 - 972 \betaV + 108 \betaV^2 + 324 \betaV^3 + 90 \betaV^4 - 
324 \betaV^5 + 36 \betaV^6 - 180 \betaV^7 + 5 \betaV^8\big) \Ct^4
\nonumber\\
&+&12 \betaV (3 + \betaV^2)^2 \big(9 + \betaV^2\big) \Ct^5
\Big] \ccW^2
\nonumber\\
&+&\Big[
  2 (1 + \betaV^2)^2 \big(18 + 54 \betaV + 21 \betaV^2 - 36 \betaV^3
  +   32 \betaV^4 - 18 \betaV^5 - 13 \betaV^6\big)
   \nonumber\\
&-&   6\, \big(18 - 66 \betaV - 123 \betaV^2 - 118 \betaV^3 - 64 \betaV^4 - 86 \betaV^5 + 
   154 \betaV^6 - 42 \betaV^7 + 78 \betaV^8 - 8 \betaV^9 + \betaV^{10}\big) \Ct
   \nonumber\\
&+& \Big(27 - 1620 \betaV + 1044 \betaV^2 - 648 \betaV^3 + 426 \betaV^4 + 
   288 \betaV^5 + 384 \betaV^6
\nonumber\\
&-& 216 \betaV^7 + 539 \betaV^8 - 108 \betaV^9 + 
    28 \betaV^{10} \Big) \Ct^2
    \nonumber\\
    &+&6\, \big(27 + 72 \betaV - 972 \betaV^2 + 72 \betaV^3 +  162 \betaV^4
    - 104 \betaV^5 + 340 \betaV^6 - 40 \betaV^7 - 69 \betaV^8\big) \Ct^3    
\nonumber\\
&+& \big(1944 \betaV + 1512 \betaV^2 - 648 \betaV^3 + 882 \betaV^4 + 
    648 \betaV^5 - 1944 \betaV^6 + 360 \betaV^7 - 257 \betaV^8 -81 \big) \Ct^4
    \nonumber\\
&-&12 \betaV \big(81 - 486 \betaV + 63 \betaV^2 + 324 \betaV^3 + 15 \betaV^4 - 
    126 \betaV^5 + \betaV^6\big) \Ct^5
\nonumber\\
 &-&  108 \betaV^2 \big(27 + 6 \betaV^2 - 5 \betaV^4\big) \Ct^6  
\Big] \ccW^4
 \Bigg\}
 \eea
 
\newpage
The  non-vanishing elements $\ftud_{a}$ are given by
\bea
\ftud_{2}[\Theta,\mVV]&=&
\frac{2\sqrt{2}\fWZ\sqrt{1-\betaV^2}\,\St}{3\DV}
\Bigg\{(3 + \betaV^2)^2 \Big[
  1 + 2 \betaV^2 + \betaV^4
\nonumber\\
  &+& \big(2 \betaV^2 + 3 \betaV^4-1\big) \Ct
  + \big( \betaV^2 + 4 \betaV^4-8\big) \Ct^2 - 4 \betaV^2 \Ct^3
  \Big]
\nonumber\\
&-&
\Big[
  2 (1 + \betaV^2)^2 (3 + \betaV^2) (3 + 6 \betaV + \betaV^2)
  \nonumber\\
  &-&2 (1 + \betaV^2) (3 + \betaV^2)\big(
  3 - 9 \betaV - 8 \betaV^2 - 30 \betaV^3 - 3 \betaV^4 + 3 \betaV^5 \big) \Ct
  \nonumber\\
  &-&8 \betaV (3 + \betaV^2)\big(6 + 6 \betaV - 6 \betaV^2 - \betaV^3
  - 6 \betaV^4 - \betaV^5\big) \Ct^2
  \nonumber\\
  &-&8 \betaV^2 (3 + \betaV^2)\big( 3 + 27 \betaV + \betaV^2 - 9 \betaV^3 \big) \Ct^3
  + 96 \betaV^3 (3 + \betaV^2) \Ct^4
  \Big] \ccW^2
\nonumber\\
&+&\Big[
  (1 + \betaV^2)^2 \big(9 + 36 \betaV + 78 \betaV^2 + 12 \betaV^3 - 35 \betaV^4\big)
  \nonumber\\
 &-& (1 + \betaV^2)\big(9 - 54 \betaV - 273 \betaV^2 - 198 \betaV^3 - 125 \betaV^4 - 42 \betaV^5 - 
  3 \betaV^6 + 6 \betaV^7 \big) \Ct
  \nonumber\\
  &-&8 \betaV \big(18 + 9 \betaV - 12 \betaV^2 - 48 \betaV^3 - 24 \betaV^4 - 65 \betaV^5 - 
  6 \betaV^6 + 4 \betaV^7 \big) \Ct^2
  \nonumber\\
  &-& 8 \betaV^2
  \big(72 + 81 \betaV - 6 \betaV^2 - 22 \betaV^4 - 9 \betaV^5 \big) \Ct^3
  \nonumber\\
  &-&48\betaV^3\big(6 + 27 \betaV + 2 \betaV^2 - 6 \betaV^3 \big) \Ct^4
  - 432 \betaV^4 \Ct^5
  \Big]\ccW^4
\Bigg\}
\nonumber\\
\nonumber\\
\ftud_{3}[\Theta,\mVV]&=&
\frac{\fWZ (1-\Ct)}{3\DV}
\Bigg\{(3 + \betaV^2)^2\Big[(1 + 3 \betaV^2 + 3 \betaV^4 + \betaV^6)
\nonumber\\
&+& (3 + 13 \betaV^2 + 9 \betaV^4 - \betaV^6) \Ct
  +4 \betaV^2 (7  -  \betaV^2) \Ct^2
  +4 \betaV^2 (3 -  \betaV^4) \Ct^3
  \Big]
  \nonumber\\
  &-&\Big[  2\, (1 + \betaV^2)^2 (3 + \betaV^2)
  \big(3 + 6 \betaV + 4 \betaV^2 - 6 \betaV^3 + \betaV^4
  \big)
  \nonumber\\
&+&  2\, (1 + \betaV) (1 + \betaV^2) (3 + \betaV^2) \big(
  9 + 45 \betaV - 12 \betaV^2 - 12 \betaV^3 + 19 \betaV^4 - \betaV^5 \big)\Ct
  \nonumber\\
  &+&8\, \betaV (3 + \betaV^2)^2\big(6 + 7 \betaV + 12 \betaV^2 - \betaV^3\big) \Ct^2
  \nonumber\\
  &+&8\,\betaV^2 (3 + \betaV^2) \big(9 + 90 \betaV - 18 \betaV^3 - \betaV^4\big) \Ct^3
  +96 \betaV^3 (3 - \betaV^2) (3 + \betaV^2) \Ct^4
  \Big]\ccW^2
  \nonumber\\
&+&\Big[(1 + \betaV^2)^2 \big( 9 + 36 \betaV - 165 \betaV^2 - 24 \betaV^3 + 115 \betaV^4 - 12 \betaV^5 + \betaV^6\big)
  \nonumber\\
&+&(1 + \betaV^2)\big(
  27 + 324 \betaV + 216 \betaV^2 - 36 \betaV^3 - 162 \betaV^4 + 60 \betaV^5 
-  32 \betaV^6 + 36 \betaV^7 - \betaV^8\big) \Ct
  \nonumber\\  
&+&16 \betaV \big(  27 + 117 \betaV + 72 \betaV^2 + 51 \betaV^3 + 39 \betaV^4 - 47 \betaV^5 + 
  6 \betaV^6 + 11 \betaV^7\big) \Ct^2
\nonumber\\
&+&16 \betaV^2 \big( 108 + 135 \betaV + 144 \betaV^2 + 18 \betaV^3 + 6 \betaV^4 - 9 \betaV^5 + 
2 \betaV^6\big)
\nonumber\\
&+&48 \betaV^3 (18 + 81 \betaV - 15 \betaV^3 - 2 \betaV^4) \Ct^4 + 
 432 \betaV^4 (3 - \betaV^2) \Ct^5
\Big] \ccW^4  
  \Bigg\}
\nonumber\\
\nonumber\\
\ftud_{4}[\Theta,\mVV]&=&
\frac{2\fWZ (1-\betaV^2)\St^2}{3\DV}
\Bigg\{(3 + \betaV^2)^2 \Big[
(1 + 2 \betaV^2 + \betaV^4) + 4 \betaV^2 (3 + \betaV^2)^2 \Ct^2  \Big]
\nonumber\\
&-& 2\, (3 + \betaV^2) \Big[3 + 7 \betaV^2 + 5 \betaV^4 + \betaV^6
  +\big(24 \betaV + 24 \betaV^3\big) \Ct
  \nonumber\\
&+&4 \betaV^2 \big(3 + \betaV^2\big) \Ct^2
  +48 \betaV^3 \Ct^3
  \Big]\ccW^2
+\Big[
  (1 + \betaV^2)^2 \big(9 - 30 \betaV^2 + \betaV^4\big)
  \nonumber\\
&+&48 \betaV \big(3 + 4 \betaV^2 + \betaV^4\big) \Ct
+ 32 \betaV^2 (18 + 12 \betaV^2 - \betaV^4) \Ct^2
  \nonumber\\
&+&96 \betaV^3 \big(3 + \betaV^2\big) \Ct^3 + 432 \betaV^4 \Ct^4
\Big]\ccW^4
\Bigg\}
\eea
\bea
\ftud_{7}[\Theta,\mVV]&=&
\frac{2\sqrt{2}\fWZ \sqrt{1-\betaV^2}\St}{3\DV}
\Bigg\{(3 + \betaV^2)^2 \big(1 + \Ct + \betaV^4 (1 -3\Ct + 4 \Ct^2\big)
\nonumber\\
&+&   2 \betaV^2 \big(1 - \Ct - 4 \Ct^2 + 2 \Ct^3\big)
\nonumber\\
&-&\Big[2\, (3 -6\betaV + \betaV^2) (1 + \betaV^2)^2 (3 + \betaV^2) - 
 2 (1 + \betaV^2) (3 + \betaV^2)
 \nonumber\\
 &-&2\, (1 + \betaV^2) (3 + \betaV^2)\big(-3 - 9 \betaV + 8 \betaV^2 - 30 \betaV^3 + 3 \betaV^4 + 3 \betaV^5\big) \Ct
 \nonumber\\
 &+&8 \betaV (3 + \betaV^2) \big( 6 - 6 \betaV - 6 \betaV^2 + \betaV^3 - 6 \betaV^4 + \betaV^5
 \big) \Ct^2
 \nonumber\\
 &+&8 \betaV^2 (3 + \betaV^2) \big(3 - 27 \betaV + \betaV^2 + 9 \betaV^3\big) \Ct^3+96 \betaV^3 (3 + \betaV^2) \Ct^4
 \Big]\ccW^2
\nonumber\\
&+&\Big[(1 + \betaV^2)^2 \big[9  + \betaV (7 \betaV-6) \big(6 - 6 \betaV
  - 5 \betaV^2\big)\big]
  \nonumber\\
&+&(1 + \betaV^2)\big(9 + 54 \betaV - 273 \betaV^2 + 198 \betaV^3 - 125 \betaV^4 + 42 \betaV^5 -  3 \betaV^6 - 6 \betaV^7\big) \Ct
  \nonumber\\
  &+& 8 \betaV \big(18 - 9 \betaV - 12 \betaV^2
  + 48 \betaV^3 - 24 \betaV^4 + 65 \betaV^5 -  6 \betaV^6
  - 4 \betaV^7\big) \Ct^2
  \nonumber\\
  &+&8 \betaV^2 \big[72 - \betaV (81 + 6 \betaV + 22 \betaV^3 - 9 \betaV^4)\big] \Ct^3
  \nonumber\\
&+&48 \betaV^3 \big(6 -27 \betaV + 2 \betaV^2 + 6 \betaV^3\big) \Ct^4 + 432 \betaV^4 \Ct^5
\Big]\ccW^4
\Bigg\}
\nonumber\\
\nonumber\\
\ftud_{8}[\Theta,\mVV]&=&-
\frac{\fWZ}{3\sqrt{3}\DV}
\Bigg\{ (3 + \betaV^2)^2\Big[(1 + \betaV^2)^3 -
  6\, \big(1 + 5 \betaV^2 + 3 \betaV^4 - \betaV^6\big) \Ct
\nonumber\\
&-&  \big(3 - 15 \betaV^2 + 13 \betaV^4 - \betaV^6\big) \Ct^2 + 
 48 \betaV^2 \Ct^3 + 
 4 \betaV^2 \big(\betaV^2-3\big) \Ct^4\Big]
\nonumber\\
&+&\Big[
-2 (1 + \betaV^2)^2 \big(9 - 54 \betaV + 15 \betaV^2 + 36 \betaV^3 + 7 \betaV^4 + 
   18 \betaV^5 + \betaV^6\big)
   \nonumber\\
   &-&12\, \big(24 \betaV - 51 \betaV^2 + 20 \betaV^3 - 58 \betaV^4
   + 4 \betaV^5 - 14 \betaV^6 + 12 \betaV^7 + 3 \betaV^8 + 4 \betaV^9
   + \betaV^{10} - 9\big) \Ct
   \nonumber\\
   &+&2\, \big(27 + 162 \betaV - 117 \betaV^2 + 1296 \betaV^3 + 30 \betaV^4 + 
   900 \betaV^5 + 54 \betaV^6 + 7 \betaV^8 - 54 \betaV^9 - \betaV^{10}\big)\Ct^2
   \nonumber\\
  &+& 48 \betaV \big(9 - 18 \betaV - 21 \betaV^2 - 12 \betaV^3 + 7 \betaV^4 - 2 \betaV^5 + 
   5 \betaV^6\big) \Ct^3
   \nonumber\\
   &-&8 \betaV^2 \big(486 \betaV - 9 \betaV^2 + 108 \betaV^3 + 3 \betaV^4 - 
   18 \betaV^5 + \betaV^6 -27\big) \Ct^4 + 96 \betaV^3 \big(9 - \betaV^4\big) \Ct^5
   \Big]\ccW^2
\nonumber\\
&+&\Big[
(1 + \betaV^2)^2 \big(9 - 108 \betaV - 165 \betaV^2 + 72 \betaV^3 + 115 \betaV^4 + 
    36 \betaV^5 + \betaV^6\big)
\nonumber\\
&+&  6\, \big(48 \betaV - 195 \betaV^2 + 40 \betaV^3 - 130 \betaV^4 + 8 \betaV^5 + 
  130 \betaV^6 + 24 \betaV^7 + 75 \betaV^8 + 8 \betaV^9 + \betaV^{10} -9\big) \Ct
\nonumber\\
&-&\big(27 + 324 \betaV - 1629 \betaV^2 + 2592 \betaV^3 - 762 \betaV^4
+ 1800 \betaV^5 + 558 \betaV^6 - 209 \betaV^8 - 108 \betaV^9 - \betaV^{10}\big) \Ct^2
\nonumber\\
&-&48 \betaV \big(9 - 9 \betaV - 21 \betaV^2 + 93 \betaV^3 +
7 \betaV^4 +   53 \betaV^5 + 5 \betaV^6 - 9 \betaV^7\big) \Ct^3
\nonumber\\
&-&16 \betaV^2 \big(108 - 243 \betaV - 99 \betaV^2 - 54 \betaV^3 + 51 \betaV^4 + 
9 \betaV^5 + 2 \betaV^6\big) \Ct^4
\nonumber\\
&-&96 \betaV^3 (9 - 81 \betaV + 9 \betaV^3 - \betaV^4) \Ct^5 - 
432 \betaV^4 (3 - \betaV^2) \Ct^6
\Big]\ccW^4
\Bigg\}
 \eea
The elements $\gtud_{a}$ are identical: $\gtud_{a}=\ftud_{a}$.
}
  \clearpage



\end{document}